\documentclass[a4paper,11pt]{article}
\usepackage{jinstpub}
\usepackage{lineno}
\usepackage{float}
\usepackage{siunitx}
\usepackage{hepunits}
\usepackage{soul}
\usepackage{comment}
\usepackage{overpic}
\usepackage{subcaption}
\usepackage{lineno}
\title{Construction, commissioning, and performance of the ENUBET demonstrator}

\author[1]{F.~Acerbi} 
\author[21]{I.~Angelis}
\author[2,3]{L.~Bomben}
\author[3]{M.~Bonesini}
\author[3,4]{F.~Bramati}
\author[3,4]{A.~Branca}
\author[3,4]{C.~Brizzolari}
\author[3,4]{G.~Brunetti}
\author[5,10]{G.~T.~Burton}
\author[2,3]{S.~Capelli}
\author[6,24]{M.~Capitani}
\author[7,10]{S.~Carturan}
\author[8]{M.G.~Catanesi}   
\author[9]{S.~Cecchini}   
\author[6]{N.~Charitonidis}   
\author[9]{F.~Cindolo}   
\author[25]{J.~Cogan}   
\author[10]{G.~Cogo}   
\author[5,10]{G.~Collazuol}  
\author[5]{D.~D'Ago}
\author[5]{F.~Dal Corso}   
\author[5,10]{C.~Delogu}   
\author[11]{G.~De Rosa}   
\author[6]{S.~Dolan}
\author[3,4]{A.~Falcone} 
\author[5,10]{M.~Feltre}
\author[5,10]{C.~Forza}
\author[1]{A.~Gola}   
\author[3,4]{D.~Guffanti}   
\author[20]{L.~Halić}  
\author[5,10]{F.~Iacob}  
\author[16]{C.~Jollet}   
\author[17]{A.~Kallitsopoulou}
\author[20]{B.~Kliček}
\author[13]{Y.~Kudenko}  
\author[21]{Ch.~Lampoudis}   
\author[17]{P.~Legou} 
\author[5]{S.~Levorato}
\author[5,10]{A.~Longhin}   
\author[15]{L.~Ludovici}
\author[2,3]{E.~Lutsenko}   
\author[8,14]{L.~Magaletti}   
\author[9]{G.~Mandrioli}   
\author[3,4]{S.~Marangoni}   
\author[9]{A.~Margotti}   
\author[22,23]{V.~Mascagna}
\author[5,10]{M.~Mattiazzi}
\author[9,18]{N.~Mauri}   
\author[16]{J.~McElwee}  
\author[3,4]{L.~Meazza}   
\author[16]{A.~Meregaglia}   
\author[5]{M.~Mezzetto}
\author[6]{L.~Munteanu} 
\author[5,10]{M.~Pari}
\author[17]{T.~Papaevangelou}   
\author[4]{E.G.~Parozzi}   
\author[9,18]{L.~Pasqualini}   
\author[1]{G.~Paternoster}  
\author[9]{L.~Patrizii}
\author[25]{M.~Perrin-Terrin}   
\author[9]{M.~Pozzato}   
\author[2,3]{M.~Prest}   
\author[5]{F.~Pupilli}   
\author[8]{E.~Radicioni}   
\author[11]{A.C.~Ruggeri}   
\author[2,3]{G.~Saibene} 
\author[21]{D.~Sampsonidis}  
\author[10]{C.~Scian}  
\author[10]{L.~Silvestrin}  
\author[9]{G.~Sirri}  
\author[20]{M.~Stipčević}
\author[9]{M.~Tenti}
\author[3,4]{F.~Terranova} 
\author[3,4]{M.~Torti}  
\author[21]{S.E.~Tzamarias}  
\author[3]{E.~Vallazza}  
\author[25]{C.~Vall\'ee}

\affiliation[1]{Fondazione Bruno Kessler (FBK) and INFN TIFPA, Trento, Italy.}

\affiliation[2]{DiSAT, Università degli studi dell’Insubria, via Valleggio 11, Como, Italy.}

\affiliation[3]{INFN, Sezione di Milano-Bicocca, piazza della Scienza 3, Milano, Italy.}
 
\affiliation[4]{Phys. Dep. Università di Milano-Bicocca, piazza della Scienza 3, Milano, Italy.}

\affiliation[5]{INFN Sezione di Padova, via Marzolo 8, Padova, Italy.}

\affiliation[6]{CERN, Geneva, Switzerland.}
\affiliation[7]{INFN Laboratori Nazionali di Legnaro, Viale dell’Università, 2 - Legnaro (PD), Italy.}

\affiliation[8]{INFN Sezione di Bari, via E. Orabona 4, Bari, Italy.}

\affiliation[9]{INFN, Sezione di Bologna, viale Berti-Pichat 6/2, Bologna, Italy.}

\affiliation[10]{Phys. Dep. Università di Padova, via Marzolo 8, Padova, Italy.}

\affiliation[11]{INFN, Sezione di Napoli, via Cinthia, 80126, Napoli, Italy.}

\affiliation[12]{IPHC, Université de Strasbourg, CNRS/IN2P3, Strasbourg, France.}

\affiliation[13]{Institute of Nuclear Research of the Russian Academy of Science, Moscow, Russia. \\ National Research Nuclear University MEPhI,  115409 Moscow, Russia. \\ Moscow Institute of Physics and Technology (MIPT), 141701 Moscow.}

\affiliation[14]{Politecnico di Bari, Via E. Orabona 4, Bari, Italy.}

\affiliation[15]{INFN, Sezione di Roma 1, piazzale A. Moro 2, Rome, Italy.}

\affiliation[16]{LP2I Bordeaux, Universitè de Bordeaux, CNRS/IN2P3, 33175 Gradignan, France.}

\affiliation[17]{CEA, Centre de Saclay, Irfu/SPP, F-91191 Gif-sur-Yvette, France.}

\affiliation[18]{Dip. di Fisica e Astronomia ``Augusto Righi''
Viale Berti-Pichat 6/2, Bologna, Italy.}

\affiliation[19]{Phys. Dep. Università degli Studi di Napoli Federico II, via Cinthia, 80126, Napoli, Italy.}

\affiliation[20]{Center of Excellence for Advanced Materials and Sensing Devices, Ruder Boskovic Institute, HR-10000 Zagreb, KR
Center of Excellence for Advanced Materials and Sensing Devices, Ruđer Bo\v{s}kovi\'c Institute, 10000~Zagreb,~Croatia.}

\affiliation[21]{Aristotle University of Thessaloniki. Thessaloniki 541 24, Greece.}

\affiliation[22]{DII, Università degli studi di Brescia, via Branze 38, Brescia, Italy.}

\affiliation[23]{INFN, Sezione di Pavia, via Bassi 6, Pavia, Italy.}

\affiliation[24]{Department of Physics, University of Ioannina, Ioannina 451 10, Greece.}
\affiliation[25]{Aix Marseille Univ, CNRS/IN2P3, CPPM, Marseille, France.}

\emailAdd{andrea.longhin@pd.infn.it, leon.halic@irb.hr}
\abstract{

A final full-size prototype of the instrumented decay tunnel of ENUBET, referred to as the Demonstrator, was constructed and its performance was tested during the 2022, 2023 and 2024 beam test campaigns at the CERN East Area facility. We demonstrate the full scalability of this specialized longitudinal sampling calorimeter technology, based on iron absorbers and scintillator tiles readout by wavelength-shifting fibers and SiPMs. In addition, we evaluate the channel-by-channel response and the performance in terms of electron energy resolution, linearity, and particle identification capabilities, using charged particle beams with energies up to 5~GeV. A comparison with the predictions of a GEANT4 simulation of the prototype is also presented. 
The achieved energy resolution ($\sim 17$\% at 1 GeV and 10\% at 3 GeV) fully satisfies the requirements for positron identification.
The results meet the requirements for neutrino monitoring through lepton identification in the decay tunnel and validate the ENUBET full simulation. Although some limitations have been identified, they can be readily addressed in future developments.
}


\arxivnumber{2608.021640}

\begin{document}
\maketitle
\flushbottom

\section{Introduction}

Monitored neutrino beams are neutrino beams characterized by an unprecedented control of the flux at the source \cite{Longhin:2014yta,ENUBET_proposal}. This feature is essential to substantially improve the knowledge of neutrino cross sections, matching the requirements of next-generation long-baseline neutrino experiments. Flux monitoring is achieved by detecting, on an event by event basis, the charged leptons produced in the decay tunnel, which are proportional to the number of beam neutrinos, as both originate from the same meson decays ($\pi \rightarrow \mu \nu_\mu$, $K \rightarrow \mu \nu_\mu$, $K \rightarrow e \nu_e \pi^0$).

Neutrino beam monitoring is a major experimental challenge, since particle rates in the decay tunnel of a modern neutrino beam are forbidding. For this reason, monitored neutrino beams are designed without relying on magnetic horns and instead employ a meson focusing system based on quadrupoles and dipoles. These components can operate in DC mode and therefore do not require the primary protons to be extracted in short bunches. This technique, in turn, reduces the instantaneous particle rate by several orders of magnitude and paves the way for the use of moderate-granularity particle detectors capable of coping with pile-up over large areas. With a careful design of the beamline ENUBET managed to achieve rates of {$\mathcal{O}$}(100-1000)~kHz/cm$^2$ in the hottest region of the decay region instrumentation in typical working conditions~\cite{ENUBET:2023hgu}.

The detector technology for the instrumentation of the decay tunnel ("tagger") was investigated by the ENUBET collaboration from 2016 to 2022 \cite{Berra:2016thx,Berra:2017rsi,Ballerini:2018hus,Acerbi:2019wti,Acerbi:2020itd,Acerbi:2020nwd}, and it is also pivotal for the neutrino moni\-toring equipment of the recently proposed “neutrino SPS COmplex for Precision Experiments” (nuSCOPE)~\cite{Acerbi:2025wzo}. The prototyping activity 
culminated in the construction of a Demonstrator: a 1.7~m-long section of the instrumented tunnel validated with a charged particle beam at the CERN East Experimental Area \cite{Torti:2023hsc}.
The Demonstrator is a longitudinally segmented hadron calorimeter complemented by a photon veto. The most cost-effective option is based on iron–scintillator modules that sample energy deposits every 4.3 radiation lengths ($X_0$). The module cross-section, in the direction perpendicular to the beam, is $\sim 3\times 3$~\unit{\centi\metre\squared}. This granularity, combined with an electromagnetic energy resolution better than $25\%/\sqrt{E(\unit{\GeV})}$ and sensitivity to MIP-like deposits above $3\sigma$ over noise, provides sufficient information to achieve $e/\pi/\mu$ particle identification at the level required by ENUBET \cite{ENUBET:2023hgu}.
The detector is equipped with a photon veto made of scintillator layers located in the innermost part of the calorimeter. It must identify MIP-like particles with a time resolution better than 1~ns and distinguish single MIP-like deposits from zero or double-MIP (converted photon) signals with a signal-to-noise ratio above 3 \cite{Ballerini:2018hus,ENUBET:2023hgu}.

\par\medskip
The solution validated in this work represents a significant improvement over the 
earlier prototypes tested in the past~\cite{Ballerini:2018hus,Acerbi:2020nwd}.
Given its size (3.5 t, spanning several meters) and complexity (1275 electronic channels), this new detector demonstrates that the technology is mature and scalable to the dimensions foreseen for the final experiment.
\par\medskip
The first prototype tested \cite{Ballerini:2018hus} implemented a shashlik configuration in which the Silicon PhotoMultipliers (SiPM) were embedded in the calorimeter. Although elegant and effective, this scheme was abandoned to make the SiPMs accessible and, most importantly, to extend their lifetime, as they would otherwise have been severely damaged by neutron irradiation after only a fraction of the experiment’s lifetime~\cite{Acerbi:2019wti}.
\par\medskip
The following prototype \cite{Acerbi:2020nwd} was developed to test an improved solution aimed at transporting the light out of the calorimeter by means of lateral WLS fibers running along the short sides of the scintillator tiles. However, this scheme was difficult to scale to a geometry containing up to four radial units because the available space along the lateral sides is limited. We therefore introduced WaveLength Shifting (WLS) fibers on the front faces of the tiles.
\par\medskip
In addition, the prototype 
described in \cite{Acerbi:2020nwd}
was based on small, mechanically disconnected absorbers. By contrast, the demonstrator uses complete, self-supporting iron arches that also fulfil a structural role. Another substantial improvement is 
the implementation of a borated polyethylene (BPE) shield to reduce the neutron fluence to SiPMs \cite{ENUBET:2023hgu}
and its integration with the numerous WLS fibers emerging from the detector. In the previous prototypes, these fibers were essentially left “in air,” resulting in a fragile assembly that was difficult to handle. For the demonstrator, we developed an effective routing scheme based on grooves in the BPE slabs and 3D-printed plastic “concentrators,” making the system more robust, orderly, and reproducible.
\par\medskip
The system used to translate and tilt the detector is also significantly more sophisticated and effective than those employed in the previous prototypes. Other calorimeters based on a similar concept (scintillators + WLS fibers + SiPMs/MPPCs) have been described in the literature, including 
the CALICE-HCAL~\cite{Danilov2007_CALICE_AHCAL},
the T2K ECAL~\cite{Allan2013_T2K_ND280_ECal},
the SuperFGD~\cite{Blondel2018_FullyActiveDetector,
Blondel2020_SuperFGD_Prototype,
Abe2026_T2K_SuperFGD},
or BabyMIND~\cite{Antonova2017_BabyMIND}.
\par\medskip
Compared with these detectors, the ENUBET demonstrator is designed to meet the complex requirement of transporting light out of a highly irradiated decay region. It achieves this through an elegant solution based on fiber routing through dedicated grooves and a unique WLS-fiber bundling scheme.

Thanks to a full simulation of the ENUBET beamline, including realistic particle reconstruction, ENUBET has demonstrated that the systematic uncertainty on the neutrino flux can be reduced from about 10\% to below 1\% when information from the tunnel instrumentation is included in the standard flux estimators \cite{ENUBET:2023hgu,systematics,Bramati2026}. The goal of the Demonstrator is therefore to experimentally validate this finding by operating the apparatus in conditions that reproduce the ENUBET beam environment.

This validation campaign was carried out at the CERN East Experimental Area from 2022 to 2024 using a mixed beam of muons, charged pions, and electrons in the few-GeV energy range relevant to ENUBET. This paper summarizes the entire Demonstrator experimental program. In Sec.~\ref{sec:detector_structure}, we describe the Demonstrator design and implementation. The construction methods and detector assembly are detailed in Sec.~\ref{sec:assembly}. Sec.~\ref{sec:experiment} presents the beam test setup employed in 2022-24 for the validation of the Demonstrator. Data analysis and results are discussed in Sec.~\ref{sec:data_analysis_results}.
\section{Detector structure}
\label{sec:detector_structure}

The ENUBET Demonstrator described in this paper is the first large-scale prototype of the ENUBET decay tunnel instrumentation, and, as anticipated, it consists of two detectors: a modular sampling calorimeter and a photon veto (``$t_0$ layer''). 
\begin{figure}
\centering
\includegraphics[width=0.5\textwidth]{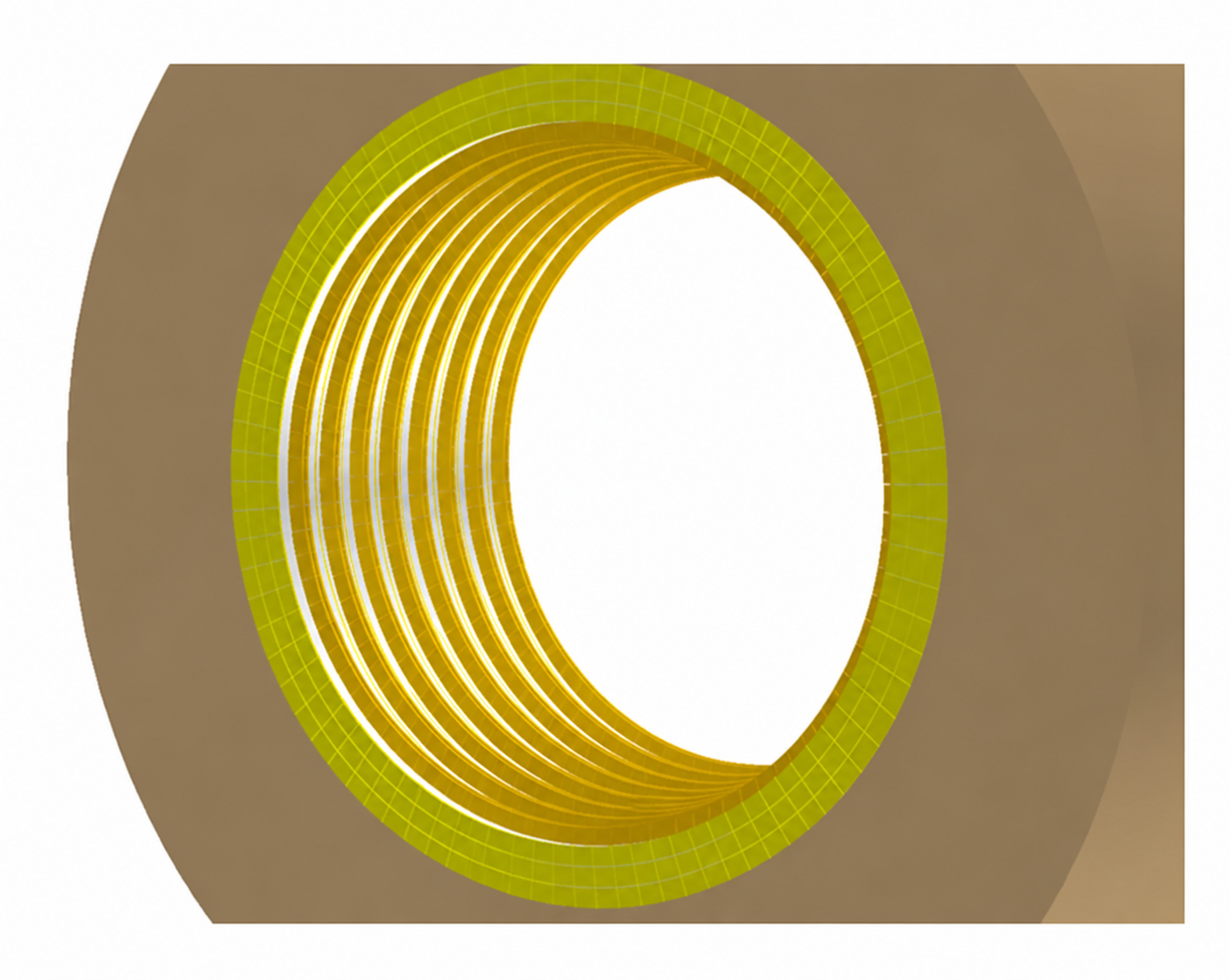}
\caption{\label{fig:tunnel_render}Rendering of the ENUBET instrumented decay tunnel. 
The tunnel instrumentation comprises, from inner to outer radii, the photon veto (orange), three radial layers of LCMs (yellow), a borated polyethylene shielding (in brown, which protects the photosensors from non-ionizing doses), and the SiPMs with their front-end electronics (located above the shielding and not shown in the figure).
}
\end{figure}

\begin{figure}
\centering
\includegraphics[width=.9\textwidth]{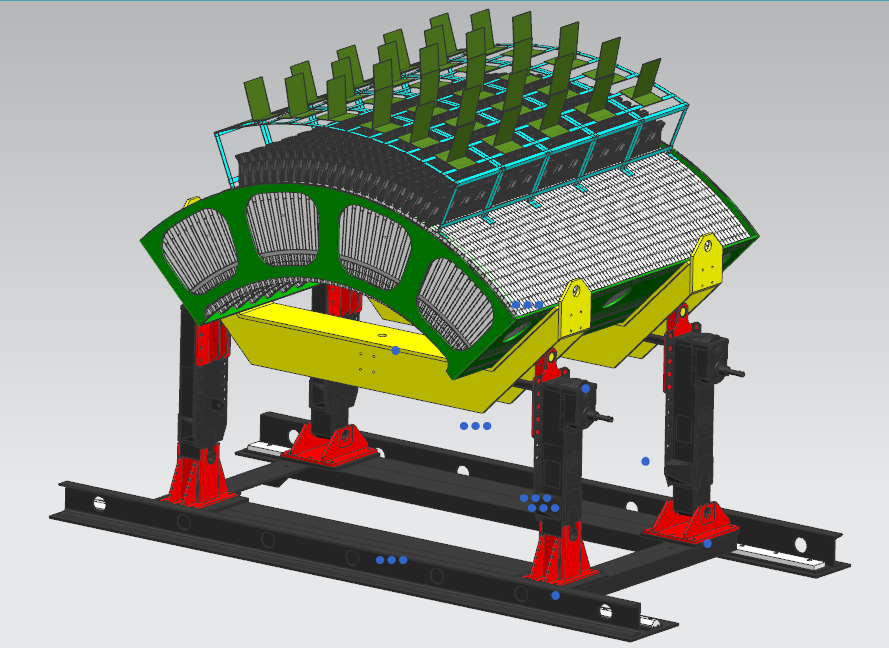}
\caption{\label{fig:3Ddemo}3D rendering of the ENUBET Demonstrator. See text for details.}
\end{figure}

The calorimeter will cover the tunnel walls, with three radial layers starting respectively at $R$~=~100, 103 and 106~cm from the tunnel axis, as shown in Fig. \ref{fig:tunnel_render}. Each radial layer will be further split in 11~cm-thick layers along the tunnel axis $z$, and in 200 sectors along the azimuthal direction $\phi$. The Demonstrator reproduces a reduced portion of this design, corresponding to 15 layers along $z$. Moreover, the detector instrumentation covers only 10 $\phi$ sectors in the first 8 $z$ layers, corresponding to $\sim 18 ^\circ$, and 25 $\phi$ sectors in the remaining 7 layers, corresponding to $\sim 45 ^\circ$ (see Fig.~\ref{fig:demoatCERN_bottom} and Fig.~\ref{fig:demo_layout_simple}). Former being constructed and tested in 2022, while the latter was constructed in 2023 and tested in 2023 and 2024. The instrumented portion was chosen to optimize cost effectiveness of the test while characterizing detector performance in terms of detection efficiency, response uniformity, and particle identification capabilities. A 3D rendering of the detector is shown in Fig.~\ref{fig:3Ddemo}.\\
\begin{figure}
\centering
\includegraphics[width=0.55\textwidth]{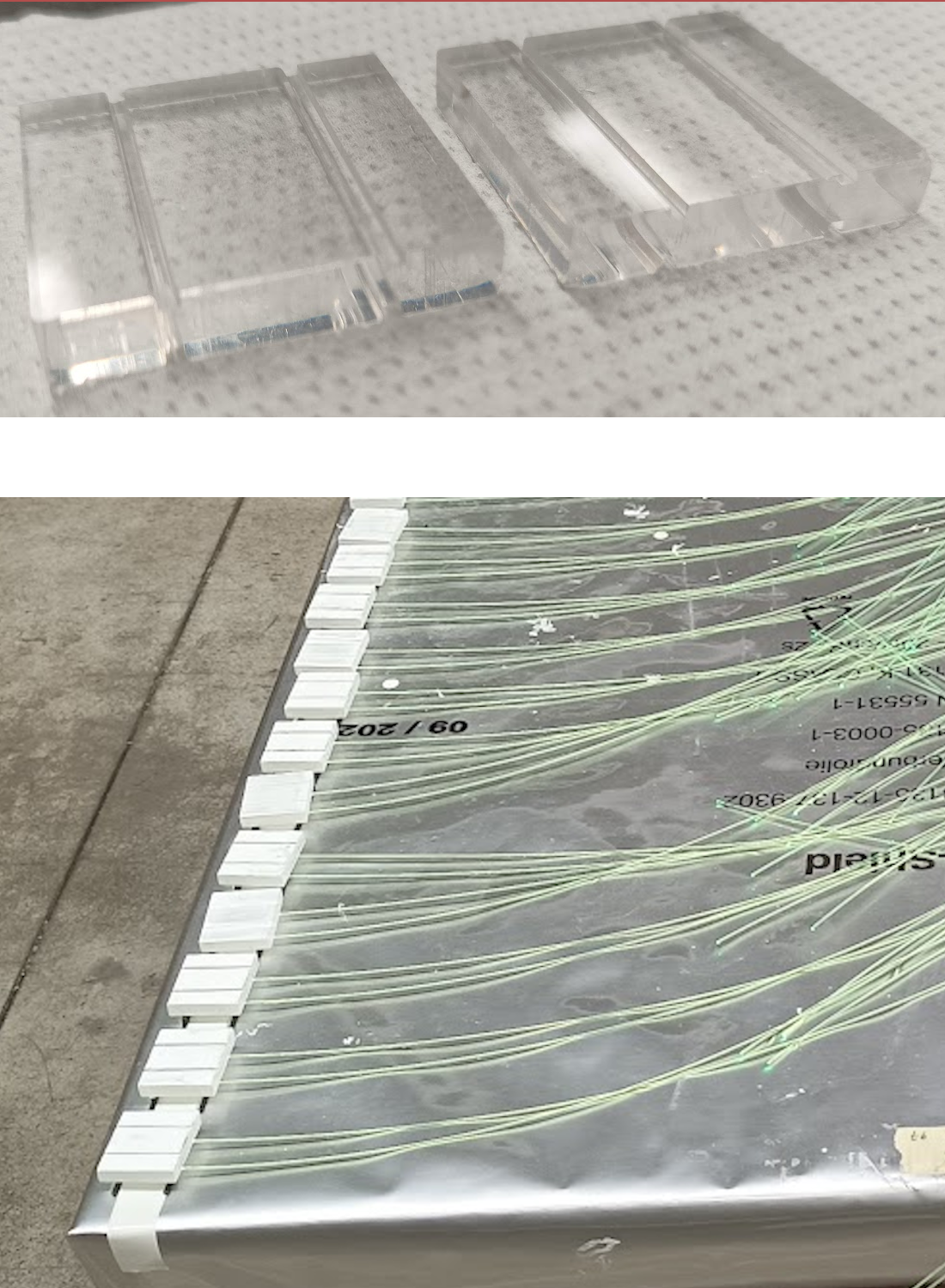} \hfill
\includegraphics[width=0.32\textwidth]{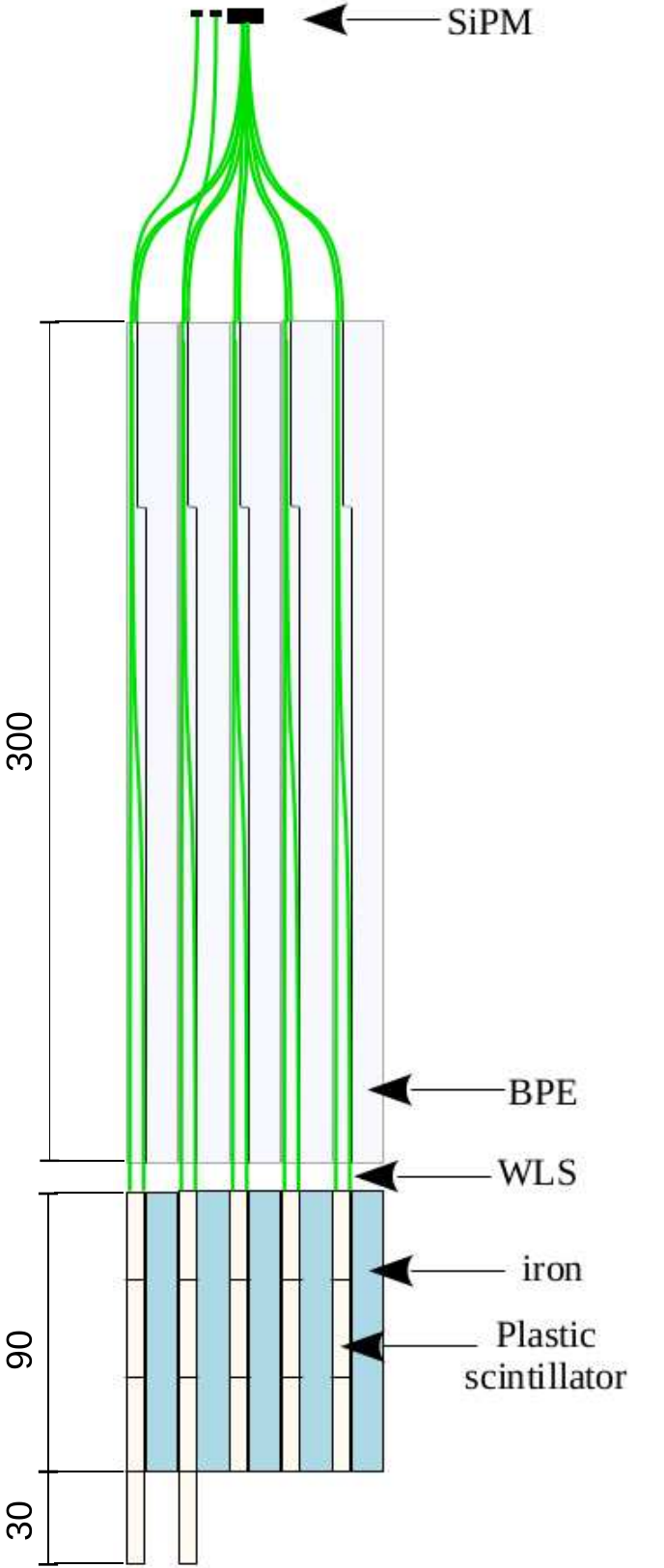}
\caption{\label{fig:demo_elements}Top left: plastic scintillator tiles. Bottom left: plastic scintillator tiles with coating and WLS fibers. Right: conceptual schematic of the main components of the Demonstrator: iron slabs, plastic scintillator tiles, fibers, borated polyethylene shielding, and SiPMs.}
\end{figure}
The unit module of the calorimeter (Lateral-readout Compact Module - LCM), corresponding to a single segment along each of the three directions ($R$, $\phi$, $z$), is made of five iron slabs interleaved along $z$ with tiles of plastic scintillator (Eljen EJ-200, Fig.~\ref{fig:demo_elements}). Each LCM covers an azimuthal angle of $\sim 31.4$~mrad, and the nominal slab and tile thicknesses are 1.4 and 0.75~cm, respectively (see Sec.~\ref{sec:assembly}). The tile surface is $\sim 3\times 3$~\unit{\centi\meter\squared}. The dimension of the LCM is a trade-off between the need for high-granularity modules for pile-up reduction and particle identification, and cost ($\lesssim 10$\% of the cost of the facility for the tagger). In the Demonstrator, the light produced by charged particles in the tiles is trapped by a diffusive coating (Eljen EJ-510) deposited on the tile surfaces. The only areas that are not covered by the diffuser are a pair of radial grooves. Two WLS optical fibers (Y11, Kuraray) are glued to the grooves using an Eljen optical cement (EJ-500) with a refraction index similar to the plastic scintillator. As a consequence, part of the light impinges on the WLS fibers and is re-emitted at $\lambda \sim 440$ inside the fiber, transported outside the calorimeter, and recorded by a Hamamatsu S14160-4050HS $4\times 4$~mm$^2$ Silicon Photomultiplier (SiPM).
All (ten) fibers belonging to the same LCM are grouped and optically connected to the same SiPM after crossing a
30 cm Borated PolyEthylene (BPE) shielding (see Figs.~\ref{fig:demo_elements} and \ref{fig:BPEgrooves}) that protects the photosensors from neutron irradiation in the tunnel.

\begin{figure}
\centering
\includegraphics[width=0.5\textwidth]{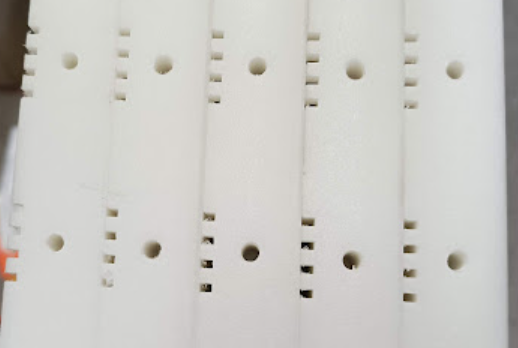}
\caption{\label{fig:BPEgrooves} Grooves for fiber routing on the top of the BPE shielding (see text for details).}
\end{figure}

The photon veto is made of a doublet of scintillator tiles, similar to the calorimeter ones, for each $z$ layer and $\phi$ sector. For each $\phi$ sector this scintillators doublet is positioned, radially, below the calorimeter, outside of the iron-filled volume (See Fig.~\ref{fig:demo_elements}, right). Along the beam coordinate, they are aligned with the two upstream tiles of the innermost LCM of the corresponding $\phi$-$z$ sector.\\
Each $t_0$ tile is read out by two WLS fibers and one Hamamatsu S14160-3050HS $3\times 3$ \unit{\milli\metre\squared} SiPM. Since each tile of the doublet is read out separately, the $t_0$ layer provides a detector to veto photons (no signal in any tile of the doublet) and converted photons (two MIP-like signals in one or two tiles of the doublet) against charged particles (one MIP-like signal per tile). The light produced in the tiles of the photon veto is transported by WLS fibers running through the corresponding LCM tiles of each layer by means of additional grooves. Thanks to the positioning of its tiles (see Fig. \ref{fig:demoatCERN_bottom}), the photon veto also provides the absolute time at which a charged particle (a positron or muon candidate) impinges on the tunnel wall. 
Testbeam data collected at CERN in 2018 and 2022 show a time resolution of 400~ps when the waveform is sampled by a 1~GS/s digitizer \cite{Acerbi:2020nwd}, well below the ENUBET specifications.

In total each $\phi$ sector then contains 17 trapezoidal scintillator tiles, 15 for the calorimeter ($z=1, 2, 3, 4, 5$ and $R=1, 2, 3$) and two $t_0$-layer scintillator tiles ($R=0$, $z=1,2$). The total number of WLS fibers is hence 34. The $t_0$-layer tiles are readout individually by two $3\times3$~mm$^2$ SiPMs each collecting a pair of WLS fibers while the 15 calorimeter tiles (5 in 3 radial layers) are readout in groups of five, corresponding to the same radius, by bundling the ten corresponding WLS fibers to three $4\times4$~mm$^2$ SiPMs. Each $\phi$ sector provides then five signals: two for the $t_0$-layer and three for the calorimeter radial layers.

Fig.~\ref{fig:BPEgrooves} shows the outer surface of the BPE shielding for two different $\phi$ sectors (top and bottom). In the picture the $z$ coordinate goes horizontally and the $\phi$ coordinate vertically. The rectangles are the grooves carved in each BPE slab that can host, each, a pair of WLS fibers. For each $z$ position there are four grooves that carry the WLS fibers coming from different radial layers with a routing that is achieved within in a suited pocket carved on the BPE slabs.

\begin{figure}
\centering
\includegraphics[width=0.6\textwidth]{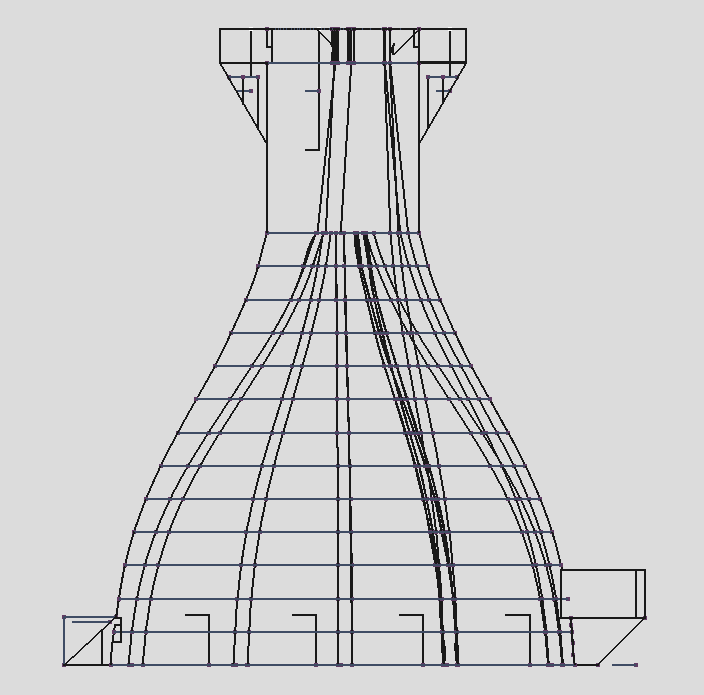}
\includegraphics[width=0.2\textwidth]{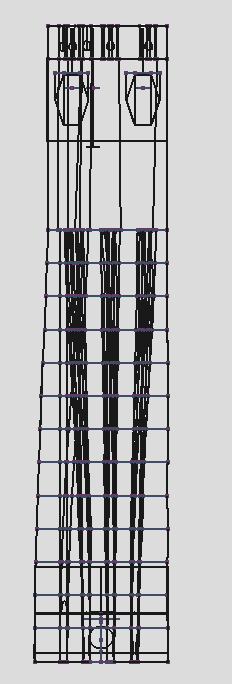}
\includegraphics[width=0.805\textwidth]{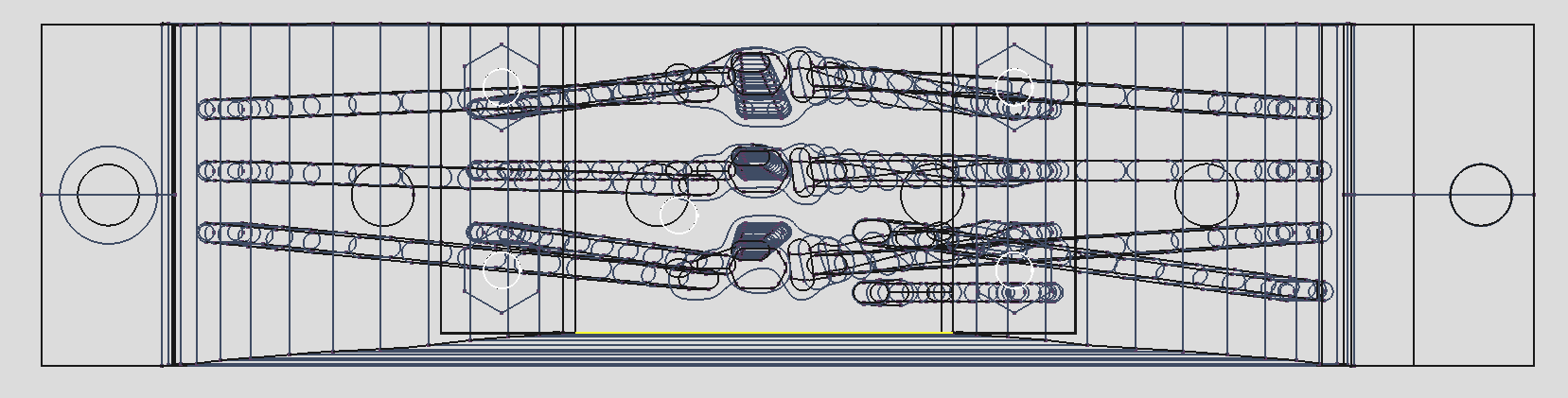}
\includegraphics[width=0.4\textwidth]{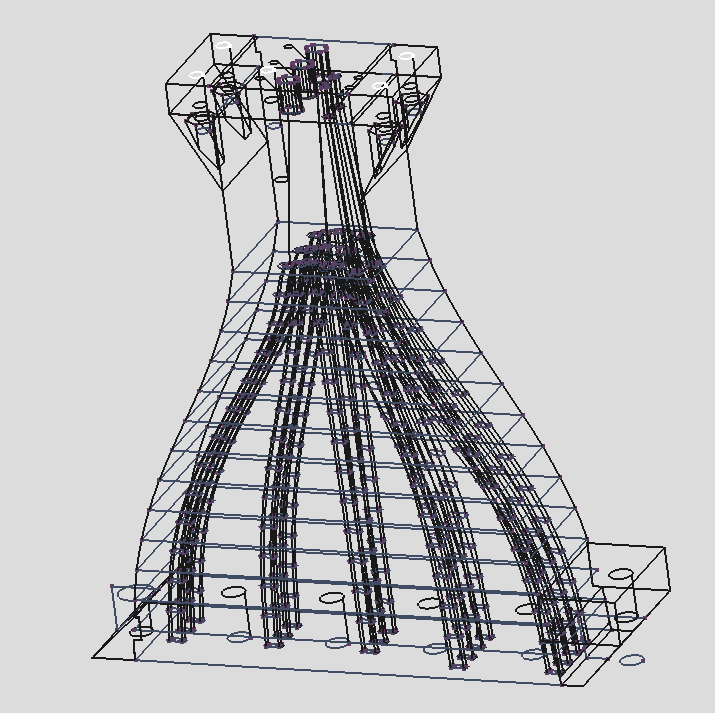}%
\includegraphics[width=0.4\textwidth]{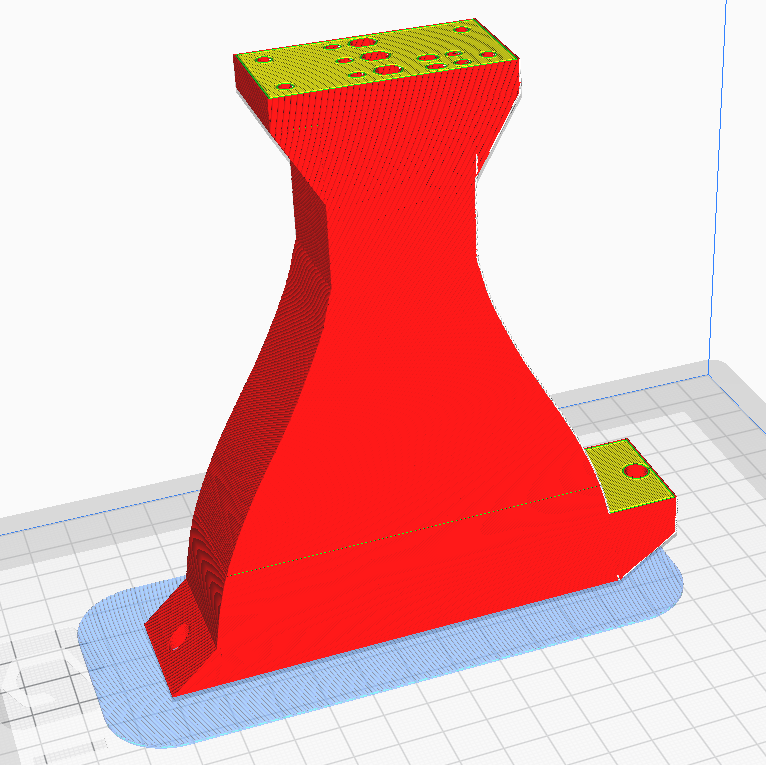}
\caption{\label{fig:fc}Fiber concentrators. Top left: $z-R$ view with transparency to show the channels carrying each a pair of WLS fibers. Top right: $\phi-R$ view. Middle: $z-\phi$ view. The bottom views show a transparent (left) and solid (right) 3D rendering.}
\end{figure}

\begin{figure}
\centering
\includegraphics[angle=90,width=0.245\textwidth]{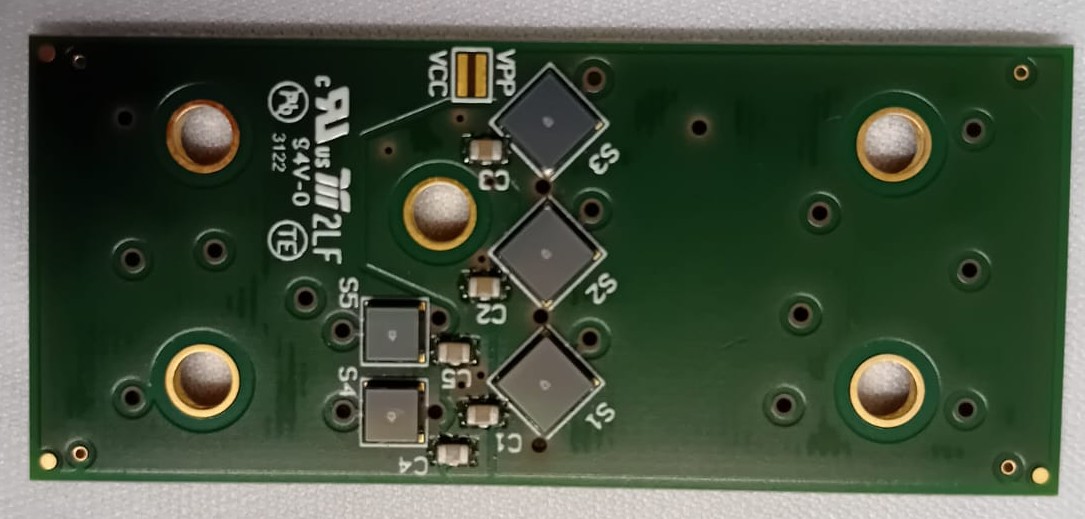} 
\includegraphics[width=0.262\textwidth]{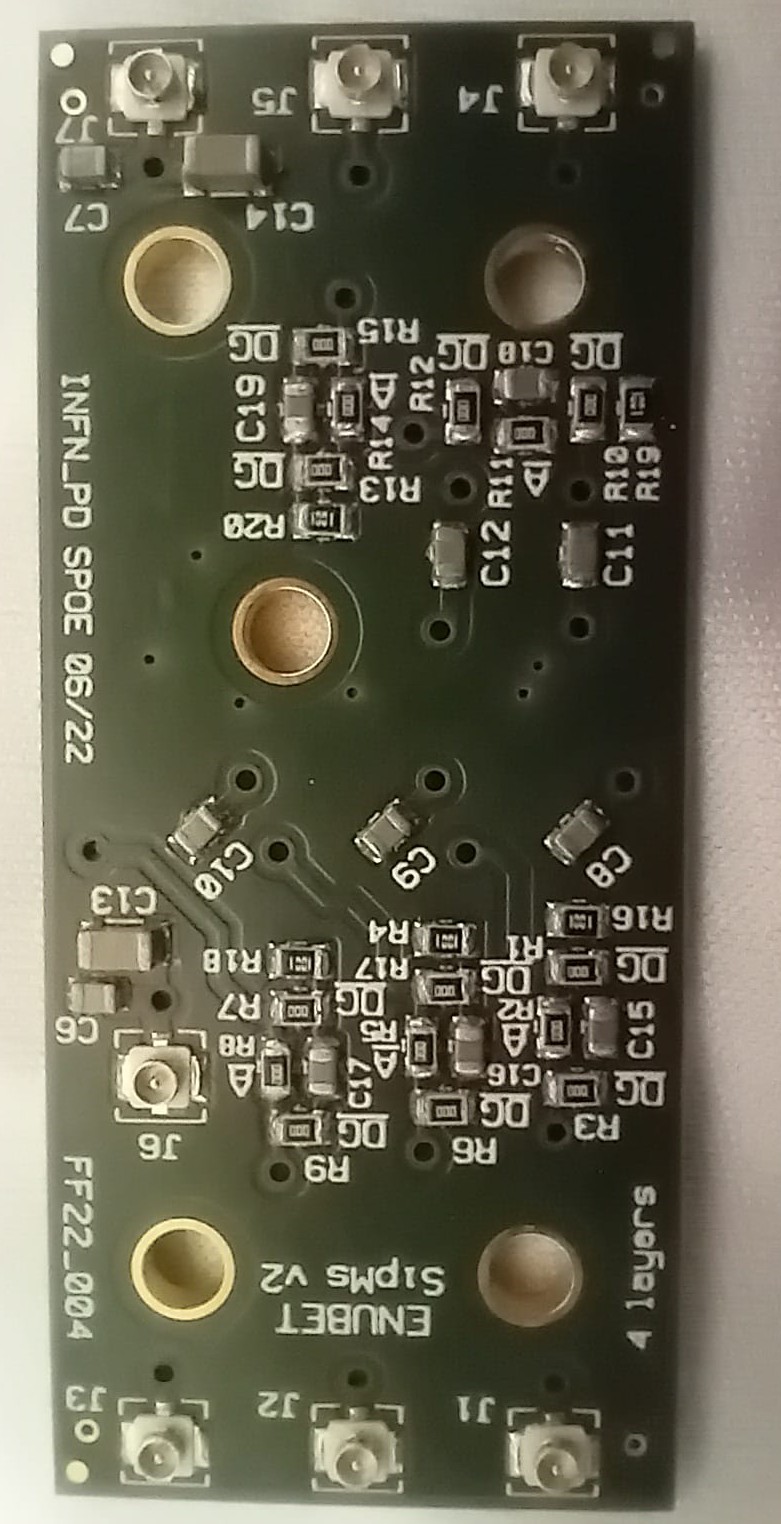} \hfill
\includegraphics[width=0.457\textwidth]{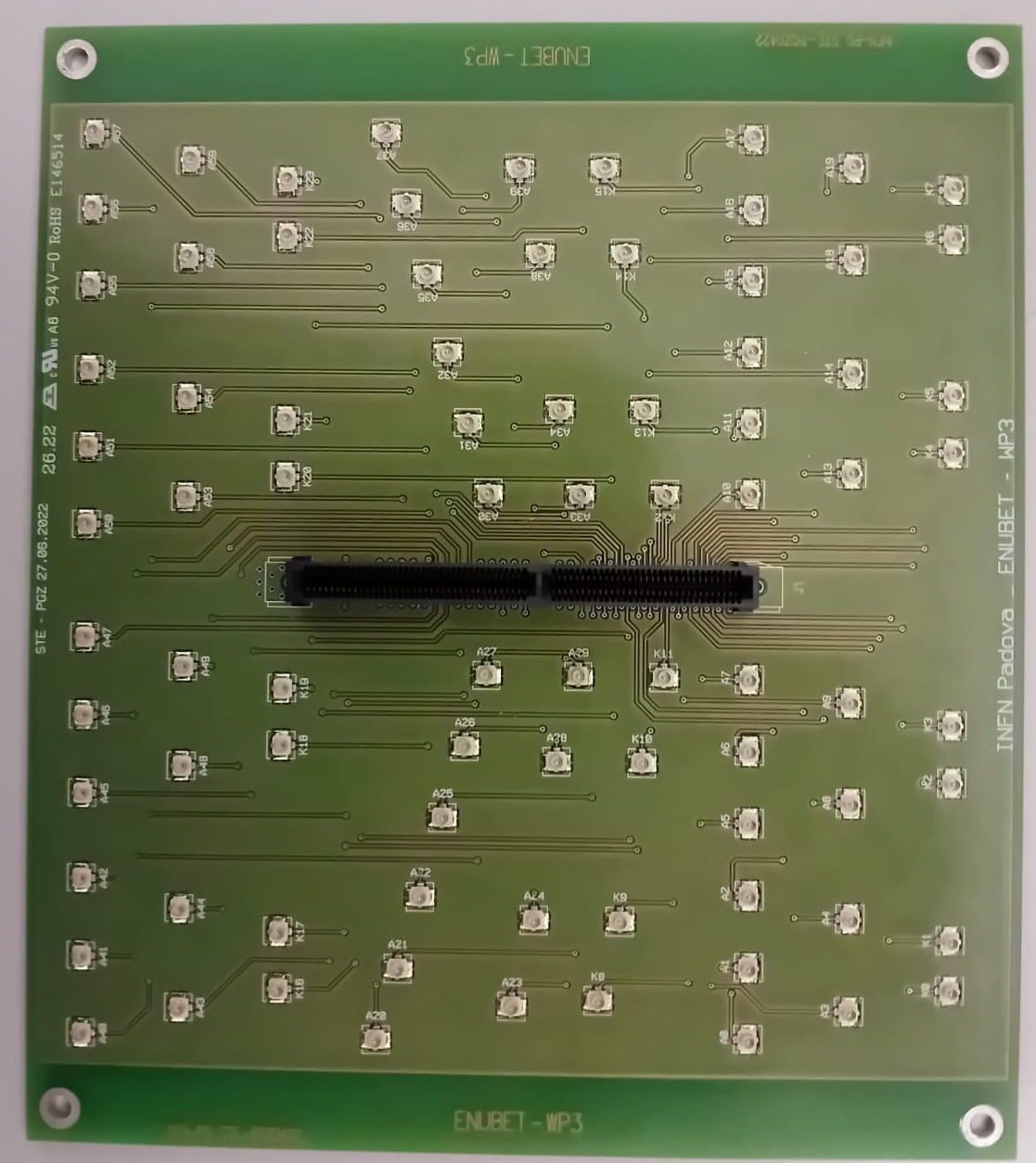}
\caption{\label{fig:boards_pictures}Pictures of the ENUBET Front End Boards (left and middle) and of the Inter-Connection board (right). Left: The three $4\times4$ mm$^2$ SiPM for calorimetric channels and two $3\times3$ mm$^2$ SiPM for $t_0$-layer channels, rotated by 45$^\circ$ with respect to each other, are visible. Middle: the connector side of the FEB. The seven HIROSE female micro-coaxial signal outputs are visible in the top, bottom and middle. Right: Inter-Connection board with 84 female HIROSE micro-coaxial connectors serving 12 FEB boards.}
\end{figure}

The coupling of the fibers to the SiPMs is achieved using plastic mechanical guides called fiber concentrators (FC), shown in Fig.~\ref{fig:fc}. Each FC is positioned on the outside surface of the borated polyethylene shielding, and collects the fibers from the LCM tiles of a given $\phi$-$z$ sector and those from the corresponding $t_0$ tiles. The FC routes the fibers so that they are properly grouped into the five corresponding readout channels (three LCMs and a photon-veto doublet) and coupled to the corresponding readout SiPMs. These SiPMs are in turn installed on dedicated Front-End Boards (FEBs), shown in Fig.~\ref{fig:boards_pictures}, left. 

Due to their geometrical complexity (see Fig.~\ref{fig:fc}) the FCs have been produced with 3D printing. The production of the 255 (10$\times$8+25$\times$7) FCs was achieved using a battery of five LONGER LK5 PRO commercial 3D printers installed at INFN-LNL with each printing taking about half a day.

\begin{figure}
\centering
\includegraphics[width=\textwidth]{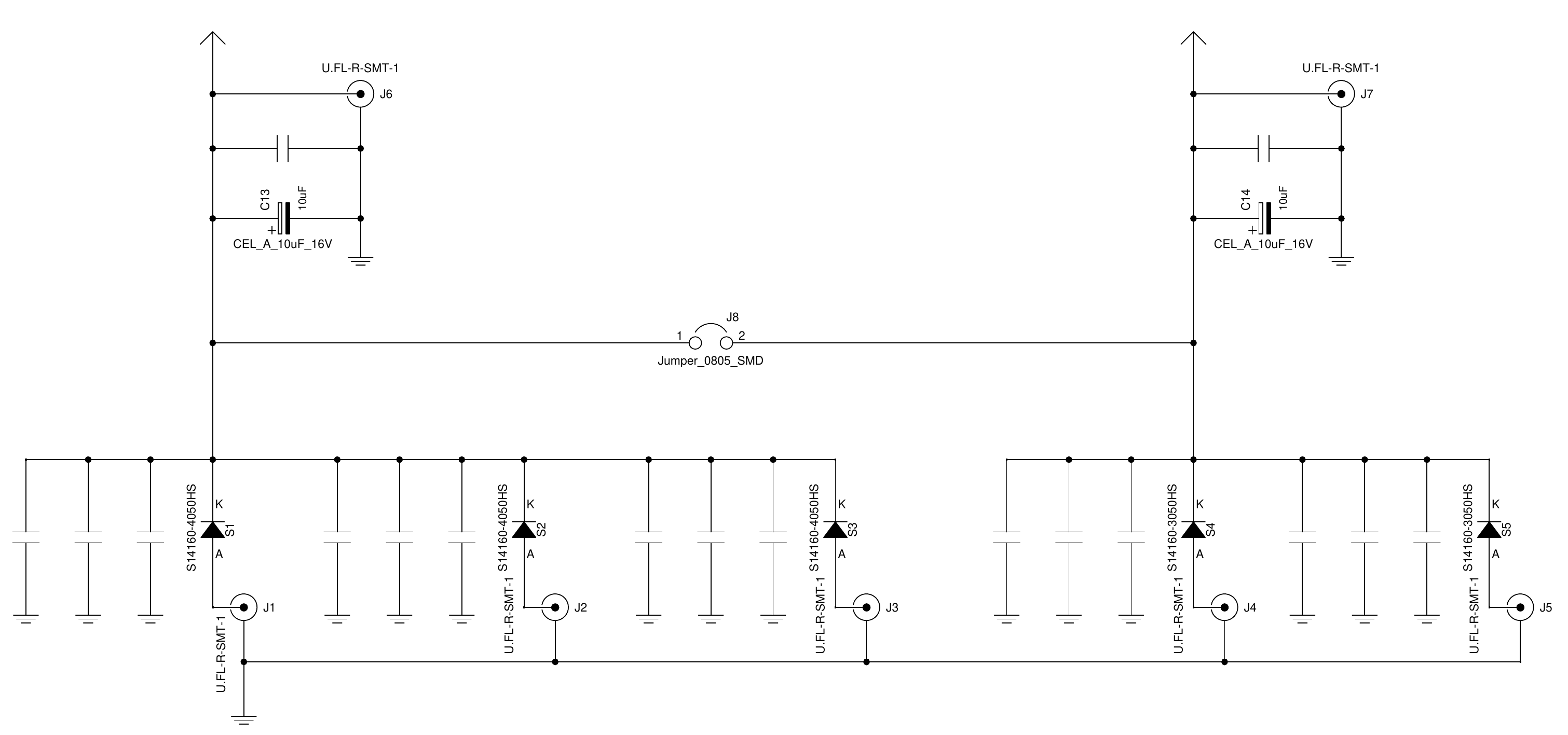}
\caption{\label{fig:FEB_circuit}Circuit schematic for a Front End Board. SIPMs are labeled as S1 to S5 (S14160-4050HS and S14160-3050HS) and are represented with the diode symbols with A and K indicating the Anode and Cathode. The labels J1, J2, J3, J4, J5, J6 (see also Fig.~\ref{fig:boards_pictures}) indicate the micro-coaxial HIROSE sockets from where the signal is taken. Diodes are polarized with a voltage bias of 40.7~V provided directly by a power supply embedded into the FERS readout boards.}
\end{figure}

\begin{figure}
\centering
\includegraphics[width=0.7\textwidth]{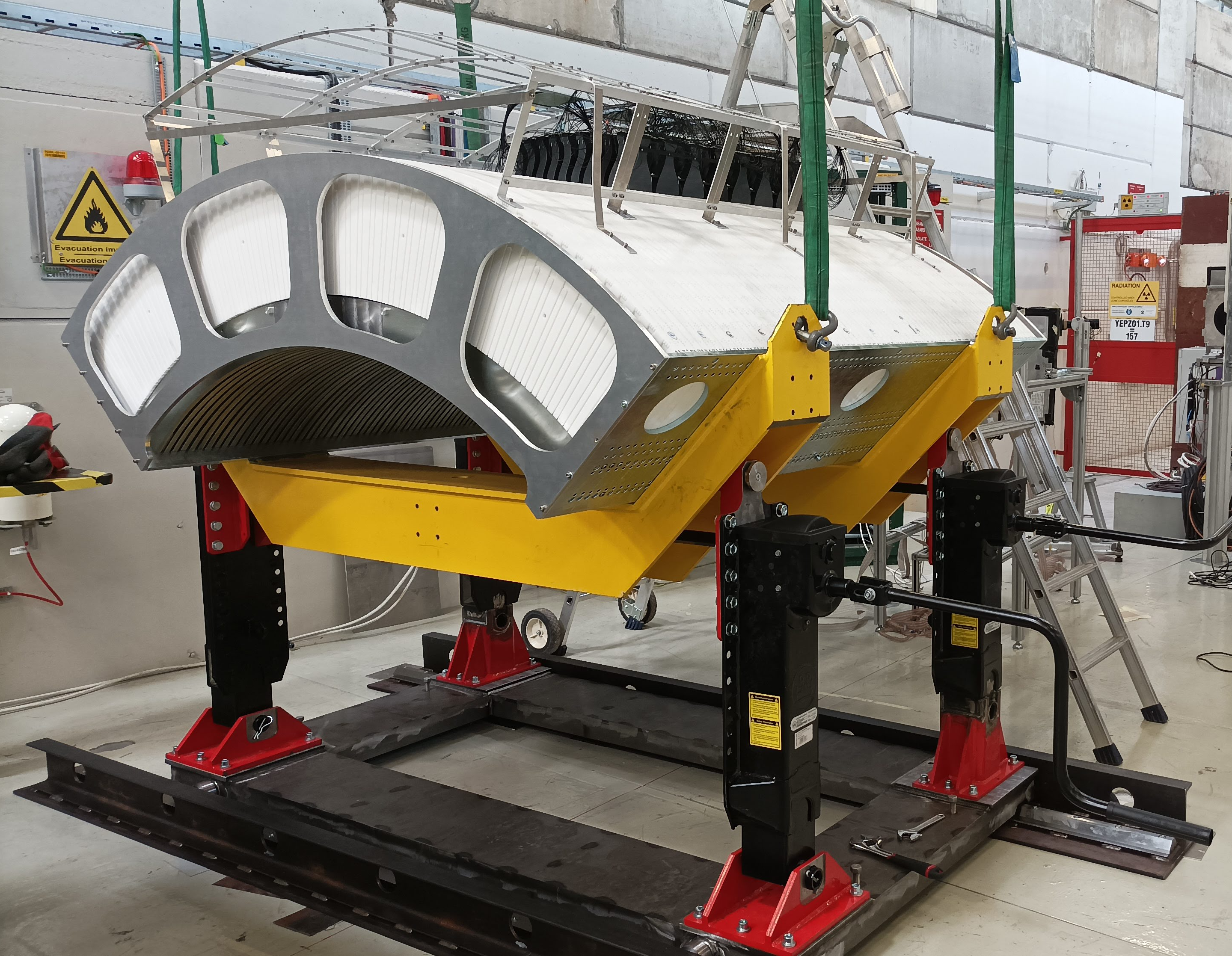}
\caption{\label{fig:demo_on_support} The ENUBET Demonstrator mounted on its mechanical support.}
\end{figure}

On a FEB, the cathodes of the three LCM SiPMs are connected together, and the same is done for the two $t_0$-doublet SiPMs. The seven electrodes (five anodes and two cathodes) are then routed to Hirose U.FL connectors on the opposite side of the board. A schematic of a FEB is shown in Fig.~\ref{fig:FEB_circuit}. 
The readout of the SiPMs is performed by CAEN A5202 boards. Each of these boards supports up to 64 channels, providing bias, shaping, sampling, and readout to the SiPMs. In the Demonstrator design, each A5202 board reads out 12 FEBs, for a total of 60 channels. The connection between the FEBs and the A5202 boards is performed via dedicated Inter-Connection (IC) boards. Each IC board (Fig.~\ref{fig:boards_pictures}, right) mounts 84 U.FL connectors, which connect via U.FL coaxial cables to the electrodes of 12 FEBs. On the IC boards, these electrodes are routed to a single Samtec HSEC8‐170‐01‐S‐DV socket, on which the CAEN A5202 board is installed.\\
In addition to coupling to the A5202 boards, the FEBs are prepared for external biasing and readout via waveform digitizers. This is implemented through a coupling RC circuit that can be selected using a jumper. This readout strategy was employed  in the 2022 beam test for a subset of the FEBs.\\
To perform the beam test measurements, the Demonstrator was mounted on a movement system shown in Fig.~\ref{fig:3Ddemo} and~\ref{fig:demo_on_support}. The system consists of four legs, usually employed for holding articulated lorries' trailers, and a rail system. A pair of legs is constrained to be vertical while the other can rotate with respect to the base frame at the bottom. The demonstrator is connected to the legs by pivots that allow for rotation. The legs length can be raised from 65 and 70 up to 95.5~cm, at the front, and 105.5~cm, at the back allowing  vertical angles between - 409 and 355 mrad. In addition, the rails allow the Demonstrator to be moved horizontally to its full extension. These degrees of freedom make it possible to scan all Demonstrator channels with the beam and to study their response to energy deposition.

\section{Detector Assembly}
\label{sec:assembly}

The ENUBET Demonstrator is composed of 75 absorbing (passive) and 75 active layers.
The absorbers are iron annulus sectors (``arcs'') with a 90$^\circ$ extension in $\phi$, a radial span of 11~cm (from $R_0=100$~cm to $R_1=111$~cm), and a thickness of 14~mm. They were obtained from iron slabs using high-pressure water cutting at INFN-LNL. A zinc coating was applied to prevent rusting.
They are inter-spaced by 8.5~mm to leave enough room for the scintillator layers.
The active layers are composed of EJ-200 scintillator tiles. Their nominal thickness is $\sim$7~mm, both for the t$_0$-layer and the calorimeter. In order to optimize geometrical coverage, the tiles are produced in four different trapezoidal shapes in the transverse plane, thus fully covering a $\phi$ sector at the four different radial positions. The transverse area of a tile is roughly 3$\times$3~cm$^2$. 

\begin{figure}
\centering
\includegraphics[width=0.9\textwidth]{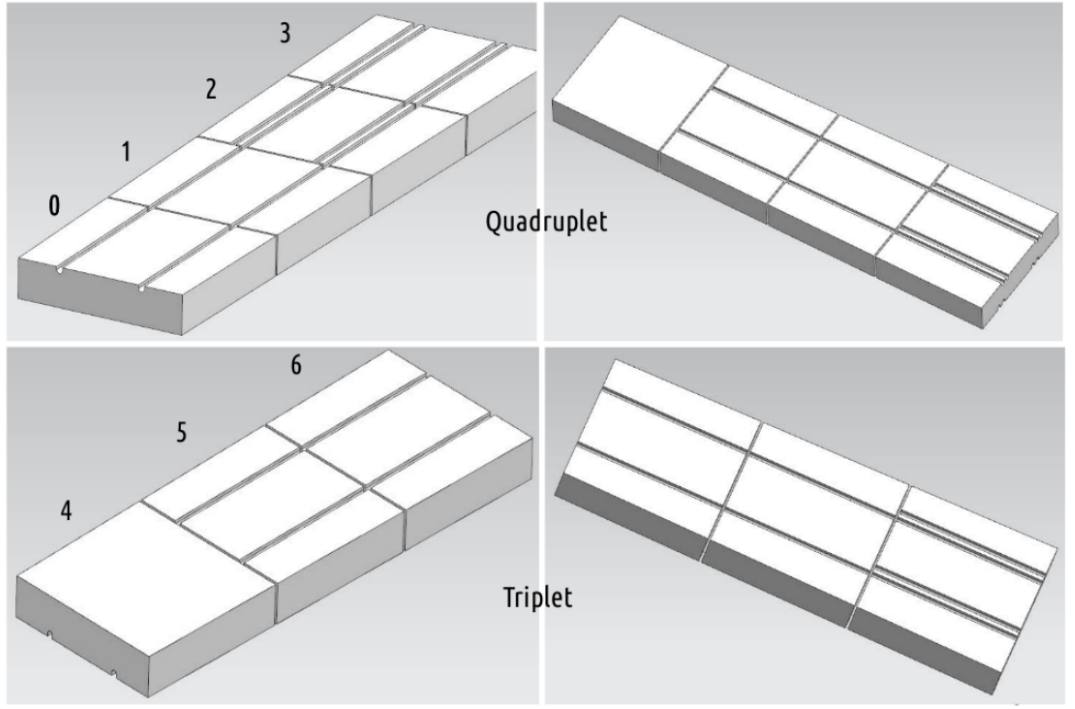}
\caption{\label{fig:tileshapes}Schematic view of the quadruplet (top) and triplet (bottom) of stacked scintillator tiles along the radial coordinate. The front (left) and back (right) sides of the tiles are shown. A quadruplet is composed of one $t_{0}$ tile (0) and three calorimetric tiles (1–2–3), instrumenting the first two scintillator layers along the beam axis within one azimuthal sector of channels. A triplet consists of three calorimetric tiles (4–5–6) and instruments the remaining three scintillator layers along the beam axis within one azimuthal sector of channels. The readout and routing grooves in each tile, running along the radial direction, are also indicated.}
\end{figure}

The tiles were produced by Stylplex~\cite{stylplex2026} through cutting and milling with numerical control machines, starting from large scintillator sheets produced by Scionix~\cite{scionix2026}. Non-uniformities in the sheets resulted in variations in tile thickness at the level of $0.1~\mathrm{mm}$ RMS, with outliers up to $\pm 0.5~\mathrm{mm}$. The measured average thickness was $6.7~\mathrm{mm}$.
The WLS fibers are hosted in grooves milled into their tiles. The grooves have a 1.2$\times$1.2~\unit{\milli\meter\squared} square cross-section.
Two types of grooves are used: readout grooves, hosting the WLS fibers of the tile, and “transit” grooves, allowing fibers from lower $R$ tiles to reach the FC without introducing gaps between active and absorbing layers. Transit grooves have to be optically isolated from the fibers they host. On the contrary, optimal transparency is needed for the readout grooves. Each tile has two readout grooves separated by 1.5~cm.
To avoid interference, the readout grooves at different $R$ are azimuthally shifted by 2~mm. Schematic drawings of the different type of tiles used and of their grooves are shown in Fig.~\ref{fig:tileshapes}. The number of transit grooves increases with $R$, as detailed in Tab.~\ref{tab:tiles_specs}.\\

\begin{table}
\begin{center}
\begin{tabular}{ c | c | c | c | c | c }
type & detector & $z$ position & inner radius & outer radius & transit grooves \\ \hline
0 & $t_0$-layer & 1, 2 & 97~cm & 100~cm & 0 \\
1 & calorimeter & 1, 2 & 100~cm & 103~cm & 2 \\
2 & calorimeter & 1, 2 & 103~cm & 106~cm & 4 \\
3 & calorimeter & 1, 2 & 106~cm & 109~cm & 6 \\
4 & calorimeter & 3, 4, 5 & 100~cm & 103~cm & 0 \\
5 & calorimeter & 3, 4, 5 & 103~cm & 106~cm & 2 \\
6 & calorimeter & 3, 4, 5 & 106~cm & 109~cm & 4 \\
\end{tabular}
\caption{\label{tab:tiles_specs}Specific geometrical parameters of the 7 types of scintillator tiles}
\end{center}
\end{table}

The preparation of the tiles consisted of the following steps.
\begin{itemize}
\item Cutting of the WLS fibers using scissors. The length was chosen to allow the fibers to reach the sensor outside the BPE shielding.
\item Polishing of the fibers. This task was performed using cylindrical ``cuffs'' to hold several tens of fibers in position, allowing perpendicular and uniform polishing. The polishing was done manually, using sandpaper of progressively finer granularity, with a final finish using standard copy paper.
\item Gluing of the WLS fiber pair into the readout grooves with Eljen EJ-500 optical cement. A planar support was used to keep the WLS fibers at the same height along their full length, ensuring that they remained parallel to the tiles planes. The glue was prepared by mixing 3 parts of resin to 1 part of hardener by volume
and was allowed to cure for 30 minutes. After gluing, the fibers and tiles were left to dry for one day.
\item Painting of the tiles with Eljen EJ-510 TiO$_2$ diffusive coating. The paint was applied manually, with two coats per tile. Particular care was taken as to avoid excess paint from clogging the transit grooves. 
\item Painting of the end surfaces of the WLS fibers with the same diffusive coating. This was done only on the ends that are not coupled to SiPMs, to limit light leakage. Other treatments (such as applying a mirroring surface with a glued aluminum foil or a sputtered Aluminum layer) were considered but finally discarded as unpractical.
\end{itemize}
Once the tiles were ready, they were assembled into arcs together with the iron layers and the BPE shielding.  
The BPE shielding consists of 150 arcs, each 22.5~mm thick. Each arc extends 30~cm in the radial direction, has an inner radius of 111~cm, and spans 90$^\circ$ in $\phi$. For each iron-scintillator layer pair, two BPE arcs are joined along $\phi$ and attached to the iron arc by means of six iron pegs. 
\\ 
Fiber  
routing pockets were machined into the BPE arcs, allowing the routing of the fibers from the tiles to the FCs.
\\
The assembly of the BPE, scintillator, and iron arcs proceeded as follows. First, each BPE arc was fixed to an iron arc by means of the six iron pins holding the iron and BPE arcs together. The scintillator tiles were then positioned on top of the iron arc in the required geometry. During the first construction phase (2022) for the instrumentation of 8 longitudinal layers, each arc consisted of 30 tiles (three radial layers and ten $\phi$ sectors) for arcs without a $t_{0}$ layer, and 40 tiles for arcs including the $t_{0}$ layer. During the second phase (2023) in which the remaining 7 longitudinal layers were instrumented, the number of $\phi$ sectors per arc was increased to 25, resulting in 75 tiles per arc (no $t_{0}$) or 100 tiles per arc (with $t_{0}$). In this case the array of $\phi$ sectors was separated in two regions divided by a non-instrumented $\phi$ sector (Fig.~\ref{fig:demo_layout_simple}). This layout, that was later faithfully simulated, was needed as the real $\phi$ extension of tile was in excess with respect to the nominal one due to the TiO$_2$ painting irregularities. This introduced a cumulative shift that prevented a match of the fibers with the carved pattern in the BPE layers for a stack exceeding about 15 $\phi$ sectors. After the tiles were in place, the WLS fibers were routed through the BPE routing pockets.
A sheet of paper was then placed on top of the tiles to protect the diffusive coating from friction during the stacking of the arcs.
Once prepared, the individual arcs were assembled together to form the final structure. The final cylindrical geometry was defined by an external iron support consisting of two lateral iron plates and two aluminum end caps as shown in Fig.~\ref{fig:demo_on_support}. The iron–scintillator–BPE arcs were inserted one at a time into the support and bolted on both sides to the lateral plates. Groups of five adjacent BPE arcs (corresponding to 30/75 LCMs) were additionally sandwiched together by four 11.25-cm long steel threaded rods inserted along the $z$ direction.

After all arcs were installed in the mechanical structure, thus forming the final detector geometry, the fibers were manually routed into the FCs positioned on top of the BPE shielding. The fibers were glued to the FCs by pouring EJ-500 optical cement into small recesses around the fiber exit points. The fibers were then cut flush and polished, ensuring optimal coupling conditions to the SiPMs. Finally, the FEBs were mounted on top of the FCs. An aluminium frame was then fixed on top of the demonstrator, and each IC board was mounted on the frame and connected through HIROSE micro-coaxial cables to twelve corresponding FEBs. This frame, together with the installed electronics, and final cabling, is shown in the photograph of the setup in Fig.~\ref{fig:TBsetup}.
\section{Beam test setup at CERN}
\label{sec:experiment}

The ENUBET Demonstrator was tested and characterized during three beam tests at the CERN PS East area facility, on the T9 secondary beam line. The detector was exposed to beams of electrons, hadrons, and muons with momenta ranging between 0.5 and 10~GeV$/c$, fully covering the expected momentum range inside the ENUBET decay tunnel.

Secondary beams in the T9 line are obtained by means of a primary 24~GeV/$c$ beam impinging on a fixed Be target (``target~1'' in~\cite{T09_userdoc}). To test the Demonstrator, two different beams were employed: ``electron enriched'' for high-purity electron beams obtained by switching on two sweeper magnets at the start of the line to dump all charged particles and select the photons to hit a 4~\unit{\milli\meter} Pb target (this converts to $e^{+}$/$e^{-}$ pairs and the downstream beamline then selects the charge and momentum) and ``hadron'' for mixed hadrons, electrons and muons beams obtained by switching off the sweeper magnets and letting the mixed beam go through the beamline. Additionally, muon beams with high purity were also obtained by
closing a CuCr1Zr and Inconel718 \cite{T09_userdoc} beam stopper.

\begin{figure}
\centering
\includegraphics[width=0.9\textwidth]{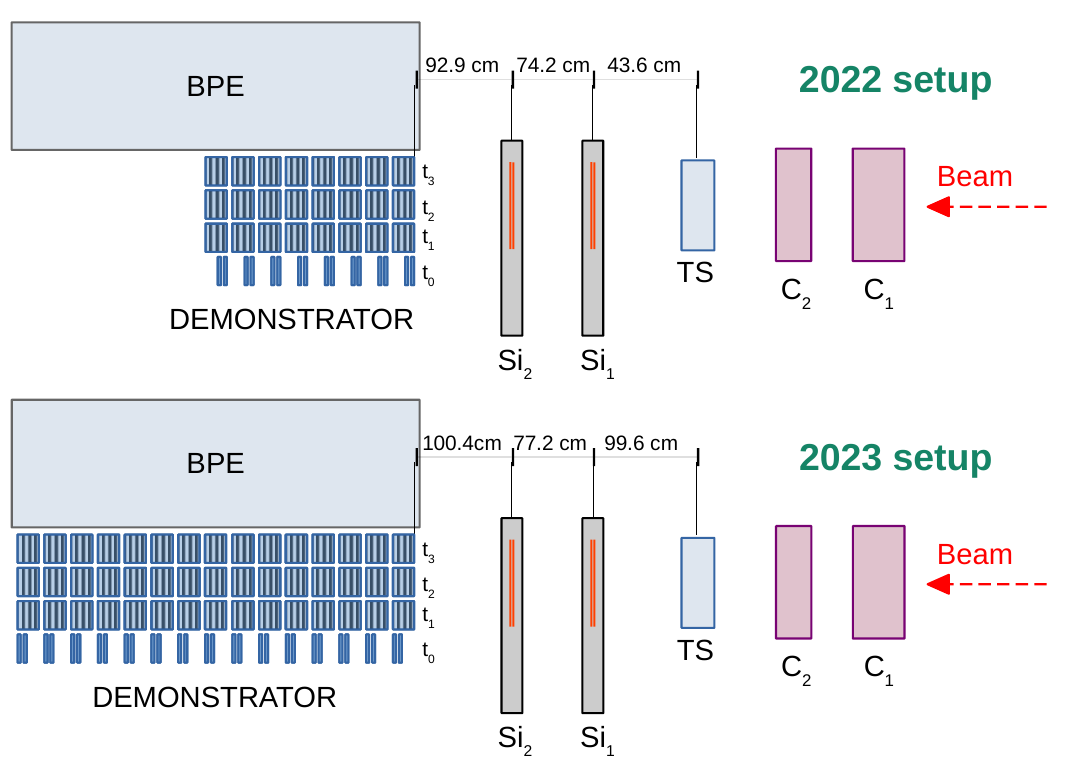}
\includegraphics[width=0.9\textwidth]{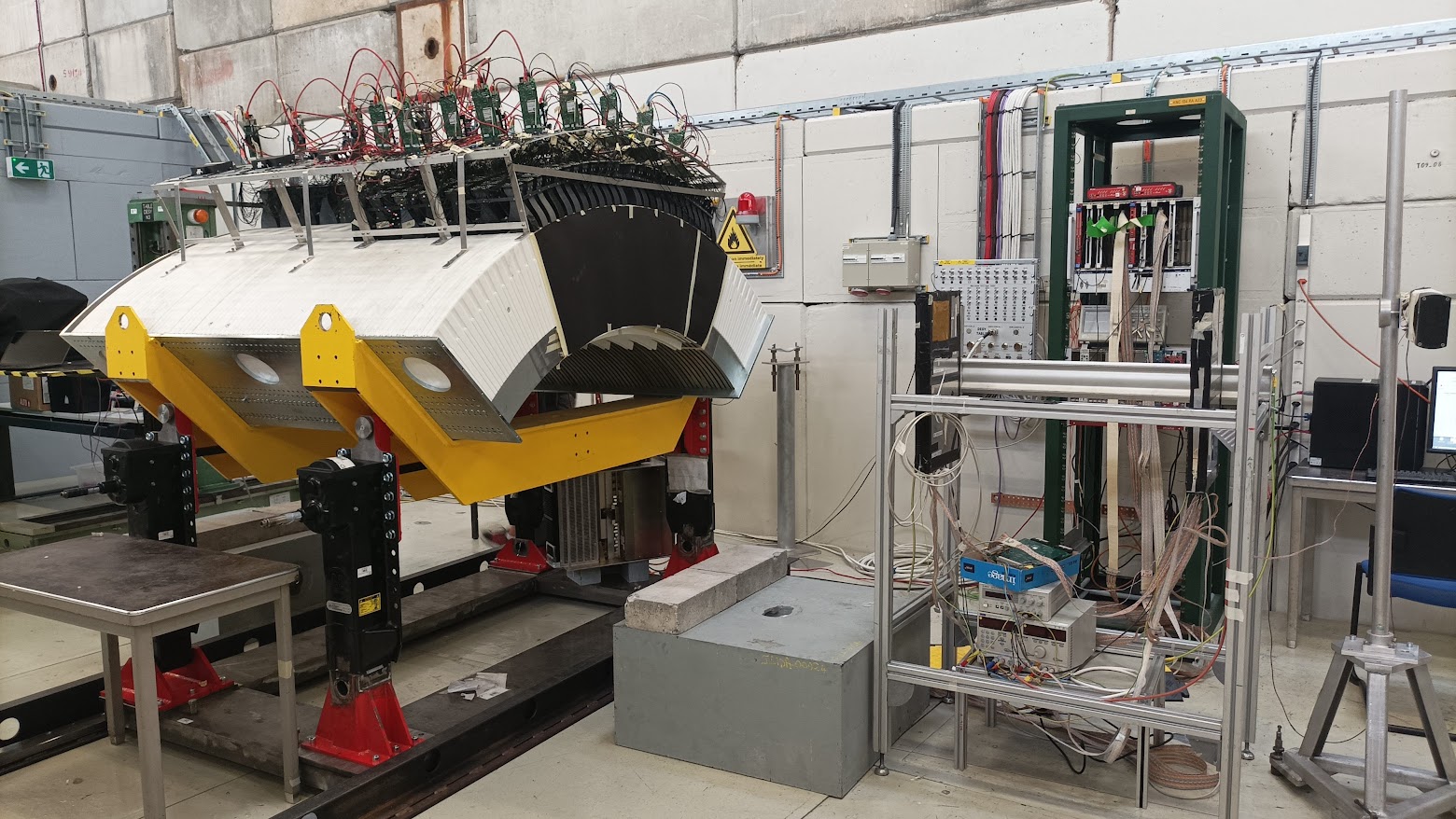}
\caption{\label{fig:TBsetup} Top: schematic of the beam test setup for the 2022 and 2023-24 tests of the ENUBET Demonstrator, including the \v{C}erenkov detectors (C$_1$, C$_2$), the trigger scintillator (TS), the silicon trackers (Si$_1$, Si$_2$) and the Demonstrator (ED) without its light shielding. In the 2023 and 2024 setup, the demonstrator was rotated by $180^{\circ}$ with respect to the 2022 configuration. The 2024 configuration is the same as that used in 2023, except for different detector distances. Drawing not to scale. Bottom: photograph of the 2023 beam test setup.}
\end{figure}

A schematic of the Demonstrator beam test setup is shown in Fig. \ref{fig:TBsetup}, along with a photograph.
The Demonstrator, mounted on its mechanical structure, was positioned on the beam line downstream of two silicon micro-strip trackers and a plastic scintillator. Each tracker consists of two 9.3$\times$9.3~\unit{\centi\meter\squared} high-resolution ($\sim 30$~\unit{\micro\meter}) silicon micro-strip detectors~\cite{AGILE_2003} (Si$_1$ and Si$_2$ in Fig. \ref{fig:TBsetup}), arranged in a $x$-$y$ configuration. Each tracker provides the hit position of a particle on its plane, which is oriented perpendicularly to the incoming beam. The combined information from the two trackers allows the reconstruction of the track of the incoming particles. A 10$\times$10~\unit{\centi\meter\squared} plastic scintillator (TS in Fig. \ref{fig:TBsetup}), read out by a photomultiplier tube, is placed on the beam line upstream of the first silicon tracker and serves as the trigger for the acquisition of the whole system. Two \v{C}erenkov threshold detectors (T09.XCET044 and T09.XCET048 in \cite{T09_userdoc}, C$_1$ and C$_2$ in Fig. \ref{fig:TBsetup}) from the T9 beamline are positioned further upstream. The operating pressure was adjusted for each beam momentum so that C$_2$ would only detect electrons, enabling their separation from heavier particles, while C$_1$ would detect both electrons and muons (except at 1 GeV/$c$ and below, where only electrons could be detected due to pressure limitations).

The Demonstrator, positioned downstream of the silicon trackers, was translated transversely with respect to the beam between runs in order to characterize the response of all LCMs and to measure the calorimeter performance for different particle species. 
The Data AcQuisition (DAQ) system for the tests consisted of two acquisition chains, both communicating with a Linux PC located in the experimental area:
\begin{itemize}
\item the acquisition chain handling the trackers, the \v{C}erenkovs detectors, and the trigger scintillator. 
The system, similar to the one used in~\cite{Acerbi:2020nwd}, is based on VME boards interfaced with a custom DAQ software through an optical link to a communication bridge (SBS Bit3 model 620 in 2022, CAEN V2718 in later setups).
The readout of the silicon trackers is performed by a chain of custom front-end and acquisition boards similar to those described in~\cite{lietti:trackers}). The signals of the \v{C}erenkov counters and of the trigger scintillator are digitized by a CAEN DT5730 digitizer.
\item the acquisition chain for the Demonstrator FEBs. This chain consists of several CAEN A5202 boards (eight in 2022, nine in 2023, and twenty-two in 2024 for the complete readout of the whole prototype), controlled via an Ethernet connection by a customized version of the open-source Janus software provided by CAEN. The software has been modified to allow the acquisition of the A5202 boards to be synchronized with the rest of the acquisition chain. 
\end{itemize}
Control of the DAQ system -- including starting and stopping runs, configuration of the boards, and pre-processing and preliminary monitoring of the data -- was performed by a second PC located in the control room, connected via Ethernet to the DAQ computer. In addition to the trigger scintillator signal, the data acquisition system also receives the $\sim$400~ms beam spill signal, which serves as an ``enable'' signal for the acquisition.

\begin{figure}
\centering
\includegraphics[width=0.8\textwidth]{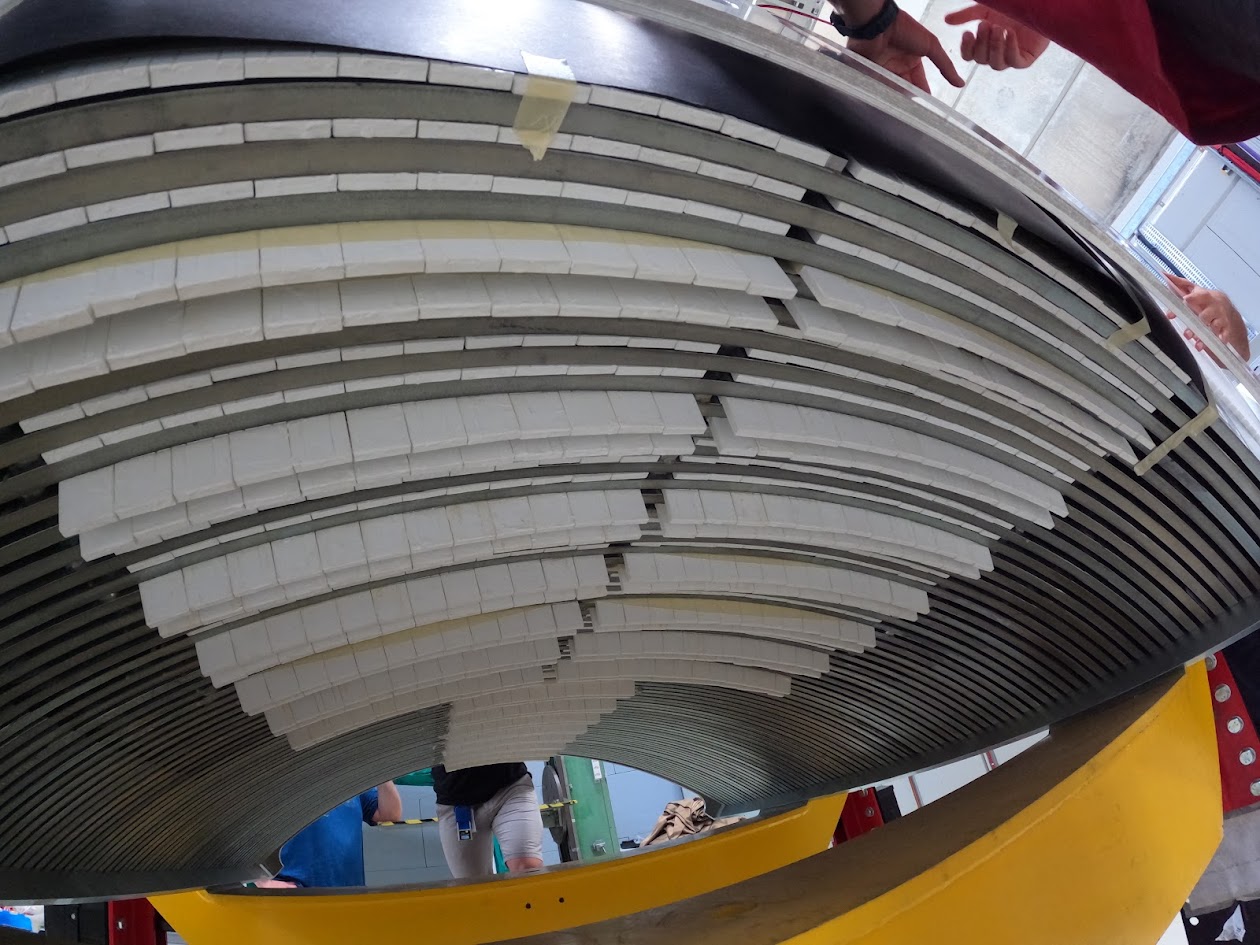}
\caption{\label{fig:demoatCERN_bottom}Bottom view of the Demonstrator on the T9 beamline at CERN, showing the photon-veto scintillator doublets located beneath the instrumented LCMs.}
\end{figure}

\begin{figure}
\centering
\includegraphics[width=0.9\textwidth]{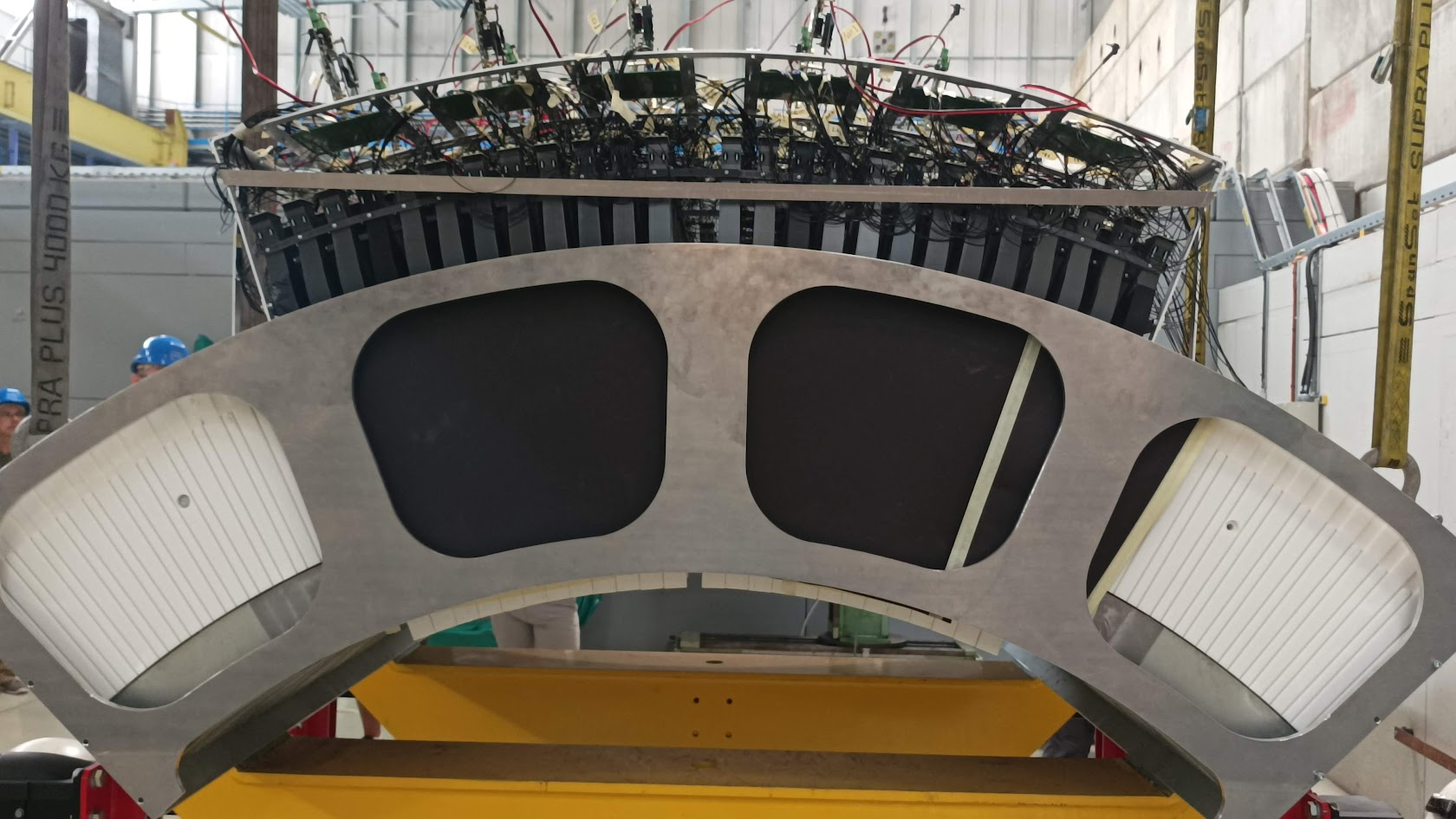}
\caption{\label{fig:demoatCERN_front}Front view of the Demonstrator on the T9 beamline at CERN.}
\end{figure}

\begin{figure}
\centering
\includegraphics[width=0.9\textwidth]{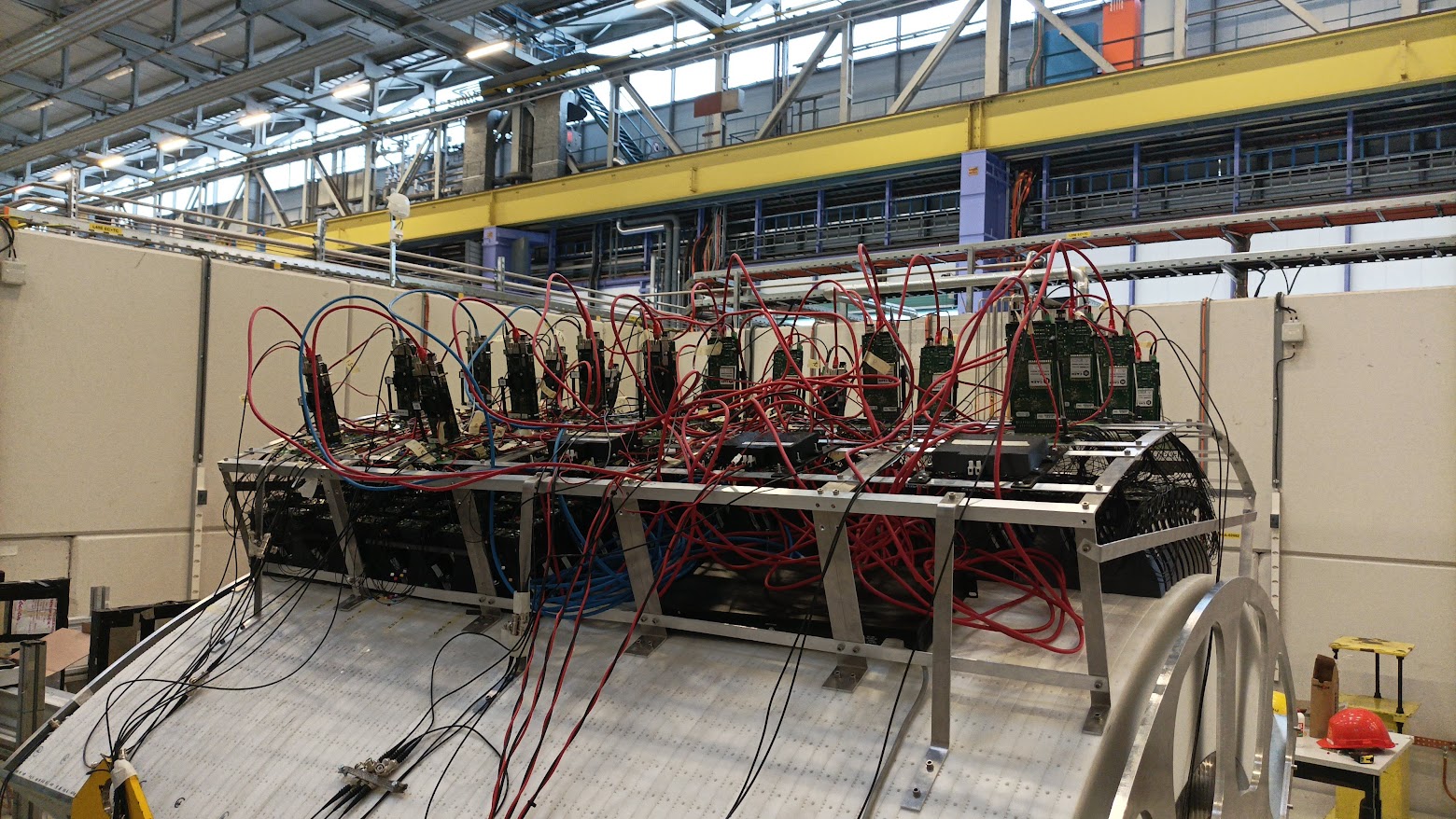}
\caption{\label{fig:demoatCERN_top}Top view of the Demonstrator on the T9 beamline at CERN, showing the  CAEN A5202 boards mounted on top of the detector.}
\end{figure}

\begin{figure} 
\centering
\includegraphics[width=0.9\textwidth]{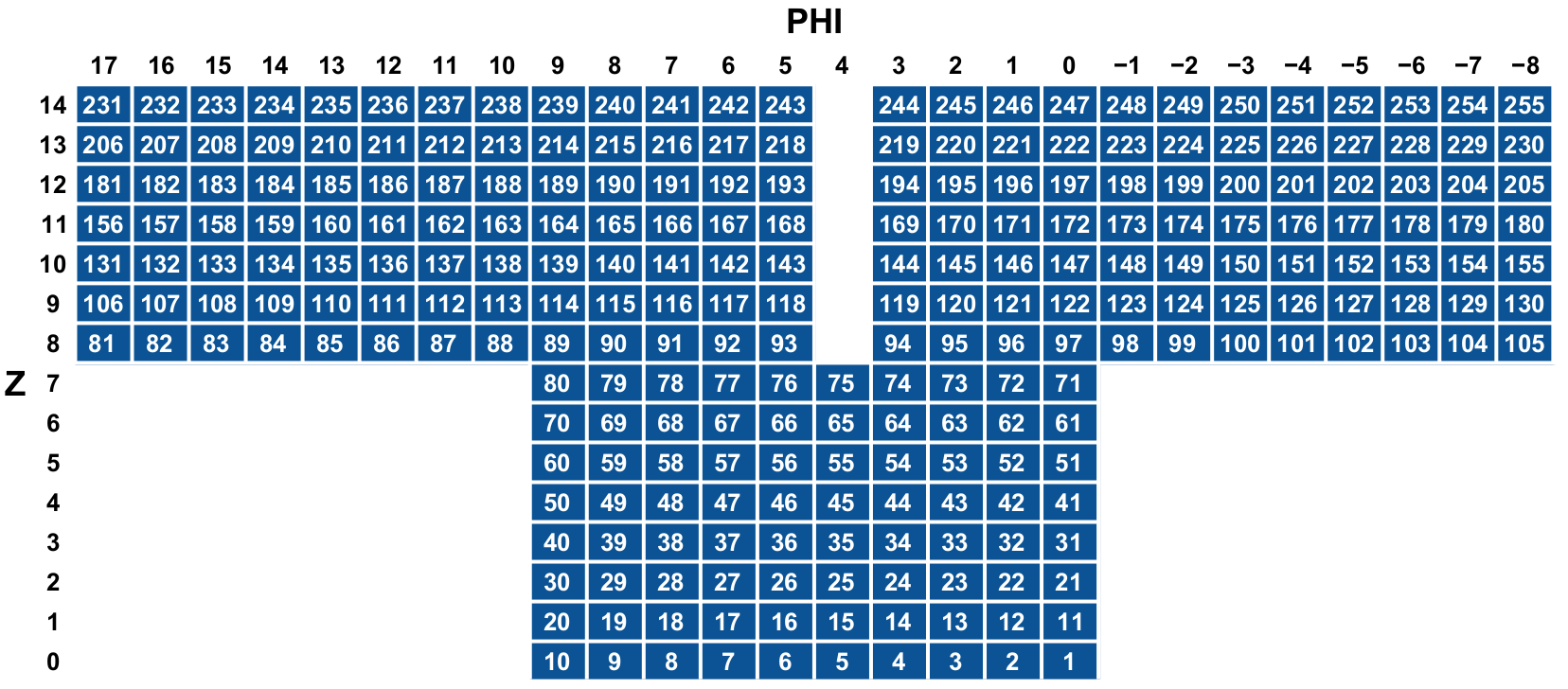}
\caption{Layout of the instrumented parts of the Demonstrator. Each cell shows an instrumented $\phi-z$ sector corresponding to a fiber concentrator. The origin of the discontinuity in $\phi$ for the upper part is explained in the text.}
\label{fig:demo_layout_simple}
\end{figure}

The layout of instrumented parts of the Demonstrator changed from 2022 to 2023, while the 2024 beam test used the same layout as in 2023. The final layout, shown in Figs. \ref{fig:demoatCERN_bottom}, \ref{fig:demoatCERN_front}, \ref{fig:demoatCERN_top}, and \ref{fig:demo_layout_simple}, consists of $15$ layers along the $z$-axis (beam axis) and $26$ $\phi$-axis layers. During the 2022 beam test, only $z$-axis layers $0-7$ were instrumented, with additional layers added in 2023. In the 2023 and 2024 beam tests, most measurements focused on $\phi \in [5,17]$ and $z \in [8,14]$. Additionally, in the 2022 beam test, the most upstream layer was $z=0$, whereas in 2023 and 2024 it was $z=14$.

\section{Data analysis and results}
\label{sec:data_analysis_results}

The purpose of the data analysis is to assess the capabilities of the Demonstrator and to compare the data gathered during the beam tests with the expectations from simulations. In this section, we describe the steps we took to analyze the beam test data, how we built a complete Demonstrator simulation, and the resulting analysis outcomes. 

Demonstrating that event-by-event lepton monitoring can provide a 1\% constraint on the neutrino flux under the beam conditions envisaged for ENUBET is a highly complex task that occupied the ENUBET Collaboration for several years. The outcome of these studies, which involved optimizing both the beam and the reconstruction in a high-particle-fluence environment, was reported in \cite{ENUBET:2023hgu}. In that paper, we described the reconstruction and PID algorithms used to obtain samples of reconstructed muons and electrons from $K$ decays with the achievable purity and efficiency. The algorithms and the simulation of the full ENUBET instrumented decay region (the ``tagger'') map directly onto the parameters of the ENUBET demonstrator, including the dimensions of the scintillators and absorbers and the light-collection system. The energy resolution that we measure is also the one achievable with a detector having the same sampling structure.

The goal of the present paper is therefore to validate the energy-deposition patterns on which the full ENUBET particle-identification chain is based. 
In principle, we could have developed dedicated PID algorithms for the demonstrator. However, such a study would have provided limited additional information because its results would not have been directly comparable with those expected under operational conditions. This is due both to the different beam and particle-flux conditions in a monitored neutrino beam compared with a PS test beam and to the limited size of the demonstrator.

The final study demonstrating that a monitored neutrino beam based on the detector technology validated in this paper can achieve the target systematic uncertainty on the neutrino flux will be published in 2027 and is beyond the scope of the present paper. Preliminary results are available in \cite{Bramati2026}.

\subsection{Demonstrator simulation}
\label{sec:simu}

\begin{figure} 
\centering
\includegraphics[trim=8cm 0cm 8cm 0cm, clip, width=0.9\textwidth]{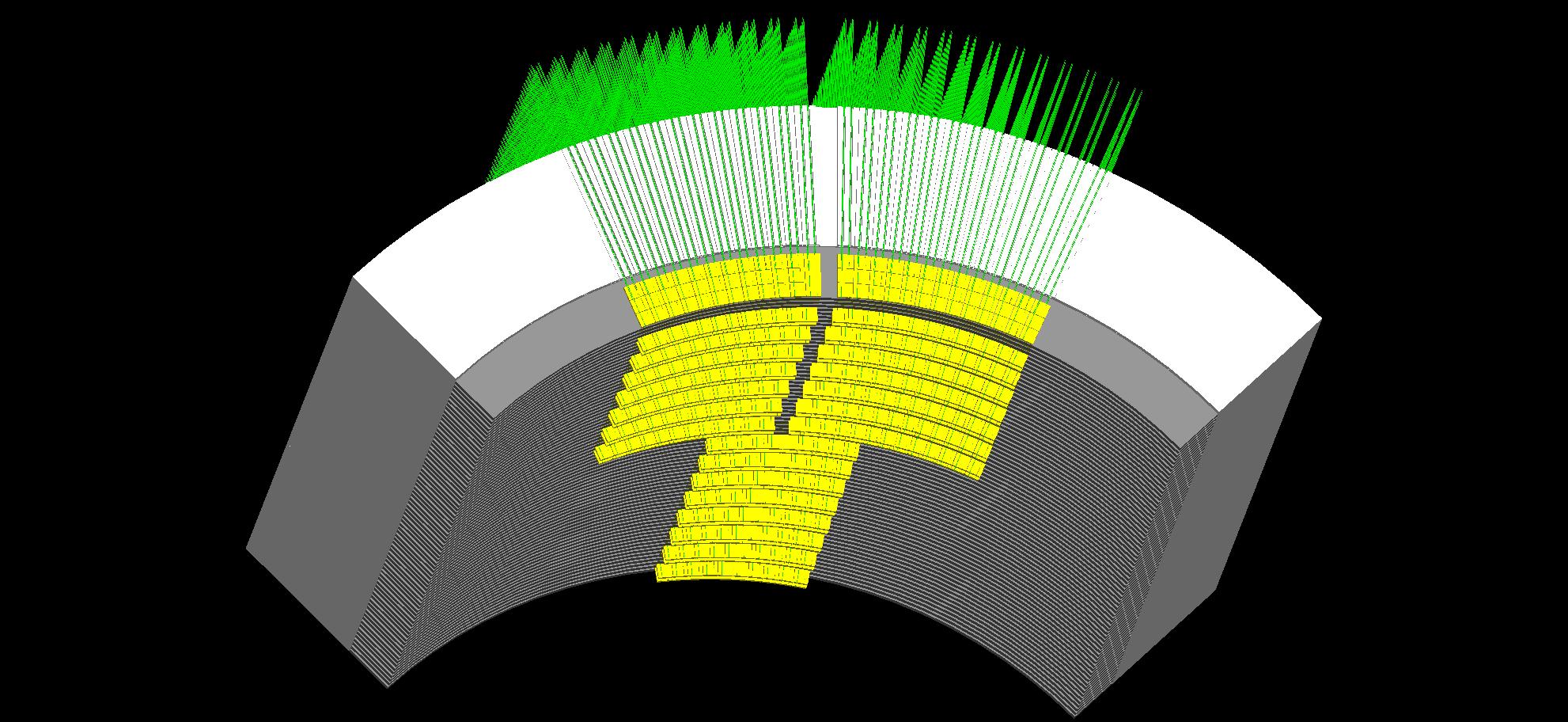}
\caption{Demonstrator geometry as implemented in the Geant4 simulation of the prototype, shown from the bottom and looking towards the tunnel walls. The instrumentation includes all channels tested during the different beam tests periods. The scintillator tiles (yellow) are displayed together with the iron arcs (grey), borated polyethylene (white) and WLS fibers (green).}
\label{fig:demo_sim}
\end{figure}

To better understand and validate the beam test results, a Geant4 simulation \cite{GEANT4:2002zbu,Allison:2006ve,Allison:2016lfl} of the ENUBET Demonstrator was developed. The simulation includes a full implementation of the detector geometry (see Fig. \ref{fig:demo_sim}) and the relevant elements of the beam test setup. The geometry was implemented by positioning 75 layers of iron and BPE arcs, each instrumented with the corresponding tiles. The tile shapes, positioning, and grooves for fiber routing reproduce the real ones, as described in Sec.~\ref{sec:assembly} and summarized in Tab.~\ref{tab:tiles_specs}. The WLS fibers are included in the simulation, positioned in the corresponding grooves of each scintillator tile. This allows to account for the full material budget and all geometrical details that may affect the calorimeter performance. The slight variations in the tile thickness discussed in Sec.~\ref{sec:assembly} were taken into account in the simulation by randomly sampling the tile widths from the measured distributions. Nevertheless, no significant impact on the visible energy was observed due to these thickness variations. Therefore, for the production of the simulation samples used in the data analysis, the width of all tiles was set to the average value of 6.7~mm. A radial and azimuthal gap between scintillator tiles can be configured when running the simulation. This enabled the study of the impact of dead regions on the energy resolution (see Sec.~\ref{sec:enereso} for details). Such dead regions are expected and are caused by the thickness of the TiO$_2$ coating. As already mentioned the operation of painting the tiles had to be performed manually.
This increased the variations of the thickness from tile to tile with respect to an industrial protocol. In particular the presence of regions with a thickened layer might happen due to the deposition of paint by gravity before the final drying which could only be mitigated but not completely avoided.

The signal in each Demonstrator channel is evaluated by measuring the total energy deposited in the scintillator tiles without simulating the propagation of photons within scintillators and WLS fibers. As such, it does not account for inefficiencies arising from scintillation light production, transport, wavelength shifting, or detection by the SiPM.

A detailed optical simulation was developed and used during the detector-design phase to study the absolute light yield and the uniformity of the light collection for the different scintillator geometries. However, this simulation was not used in the present data--MC comparison. A full optical simulation of the demonstrator would be computationally very demanding. More importantly, its predictive power at the individual-channel level would be limited by small variations in construction and optical coupling that cannot be reliably modelled for every channel.

We therefore adopted a data-driven approach in which the measured channel-equalization constants and saturation corrections effectively account for these effects.

To simulate the beam test conditions as closely as possible, the beam impinging on the demonstrator was modeled using the transverse beam profiles measured by the silicon trackers. Specifically, the initial particle positions and directions were sampled from the distributions of the beam hit coordinates ($x$, $y$) on the micro-strips and from the angular divergence inferred from the relative hit positions in the two tracking stations, respectively.

For the analysis presented here, simulated samples with statistics at least one order of magnitude larger than those available in data were produced.

\subsection{Channel equalization}
\label{sec:equalization}

\begin{figure}[h]
    \centering

        \centering
        \includegraphics[width=0.5\linewidth]{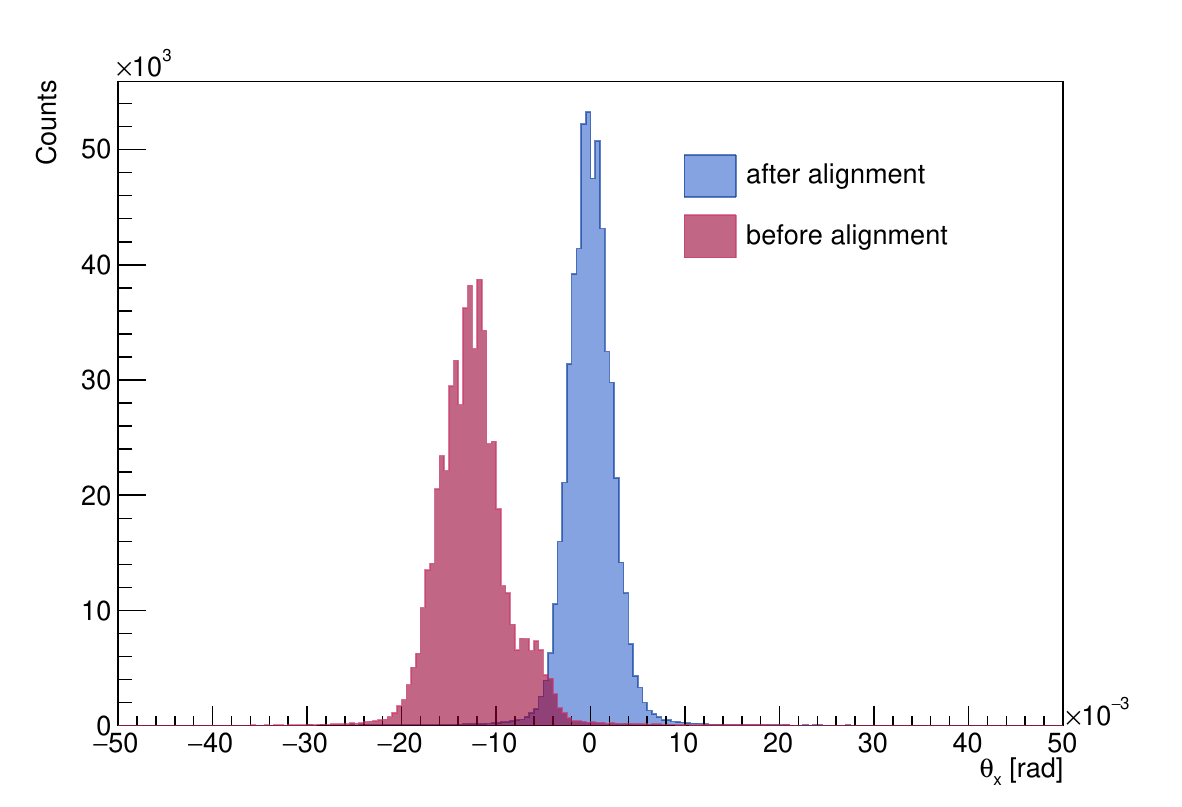}%
        \includegraphics[width=0.5\linewidth]{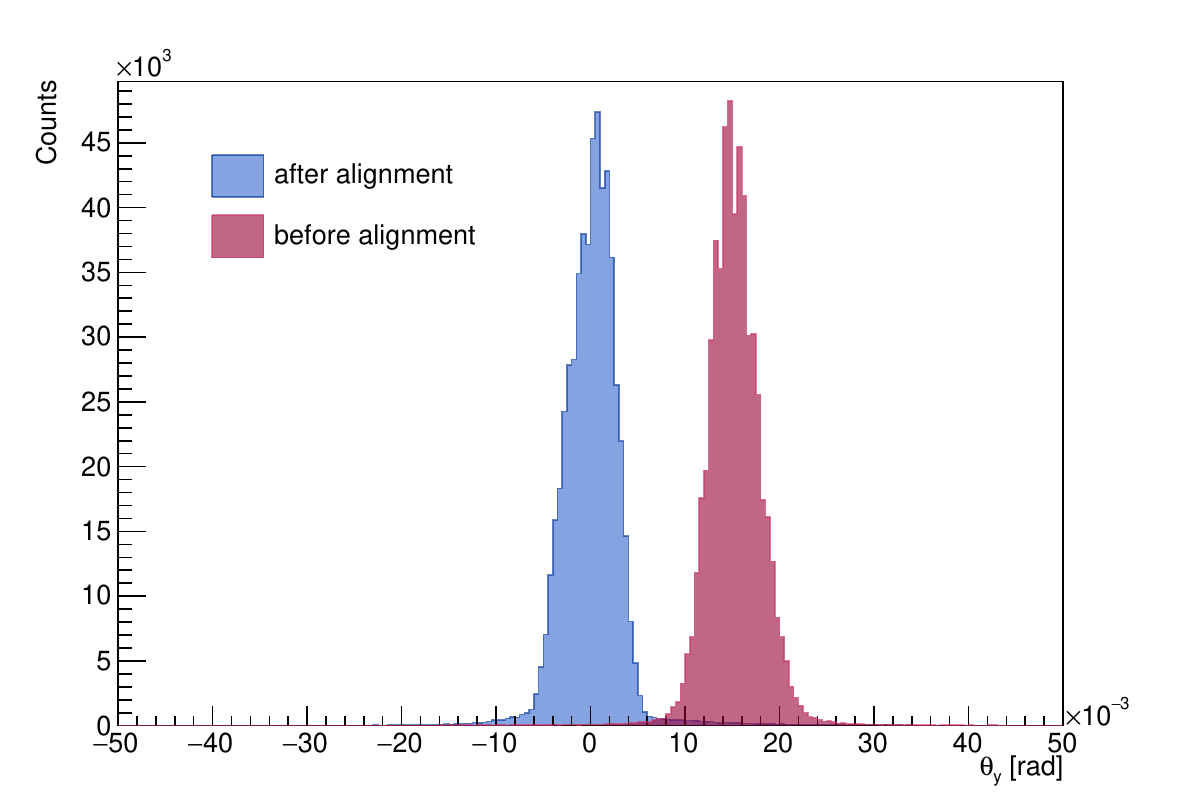}
    \caption{Distribution of the particle trajectory angles $\theta_x$ (left) and $\theta_y$ (right) between two silicon tracker detectors in the 2024 beam test. The magenta distribution shows the angles before the misalignment correction, while the blue one corresponds to the angles after correction.}
    \label{fig:tracker_misalignment}
\end{figure}

The first step in the Demonstrator beam test data analysis was the equalization of the channels, as the different LCM channels exhibit different responses to the same energy deposit. These differences arise from the intrinsic variations among the SiPMs, differences in the thickness of the scintillator tiles and possible inefficiencies in the transport and collection of light from the scintillators to the SiPMs.\\
For the equalization, the energy deposit of Minimum Ionizing Particles (MIPs) was used as a reference for all calorimeter channels. In all beam tests, dedicated calibration runs were performed to acquire the data necessary to estimate this response. In 2022, the calibration run was carried out with a 5 GeV$/c$ beam containing hadrons, electrons and muons. In 2023, two separate runs were performed. The first calibration run, covering the right side of the Demonstrator ($\phi \in [5,17]$ in Fig. \ref{fig:demo_layout_simple}), used a 10 GeV$/c$ beam of similar composition. The second calibration run, covering the left side ($\phi \in [3,-8]$), employed a 10 GeV/$c$ muon-enriched beam. In 2024, the calibration was primarily focused on the right side of the Demonstrator, using a 10 GeV$/c$ muon-enriched beam. Since most of the electron runs used for energy-resolution estimates were done on the right side of the Demonstrator during both the 2023 and 2024 beam tests, the equalization effort was mainly concentrated on that region.\\
Before the equalization, alignment of the Si micro-strip detectors was performed. The Si trackers were misaligned primarily due to imperfect positioning on the mounting rack. Since the rack was moved between some runs, the alignment was carried out on a run-by-run basis. The alignment procedure relied on the mean of particle trajectory angle between two Si trackers. If the trackers were perfectly aligned and assuming that the beam is perpendicular to the tracker, the expected mean of the particle angle is 0.0 rad. Using this assumption, each measurement in the downstream Si tracker was corrected by the mean misalignment. Fig. \ref{fig:tracker_misalignment} shows the distribution of the particle trajectory angle before and after the alignment.

\begin{figure}[htbp]
    \centering
    
    \begin{subfigure}{0.9\textwidth}
        \centering
        \includegraphics[width=\linewidth]{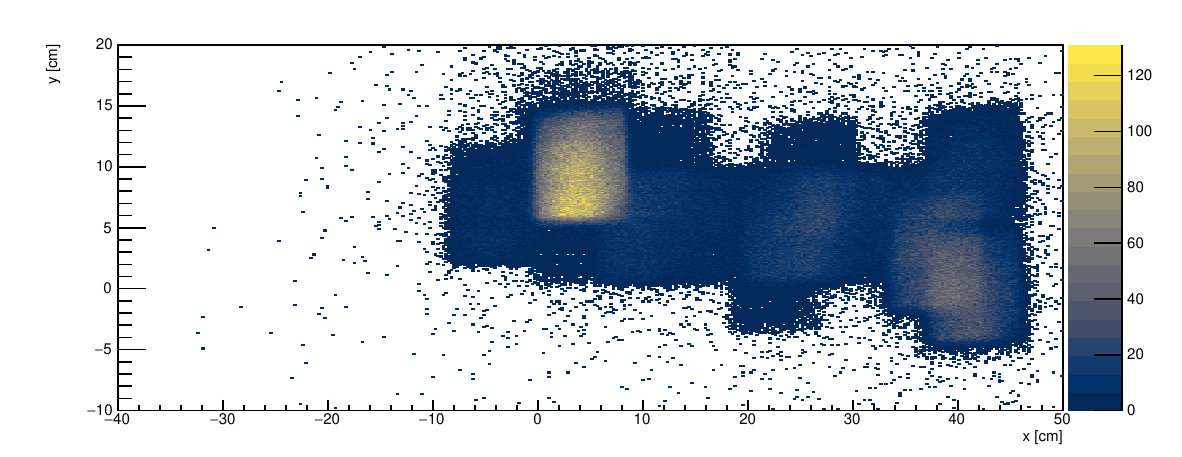}
        \caption{Projection of particle tracks from the Si trackers to the Demonstrator front face. The spots in yellow correspond to higher statistics runs.}
        \label{fig:efficiency_map_procedure_sub1}
    \end{subfigure}
    
    \vspace{1em} 

    \begin{subfigure}[t]{0.45\textwidth}
        \centering
        \includegraphics[width=\linewidth]{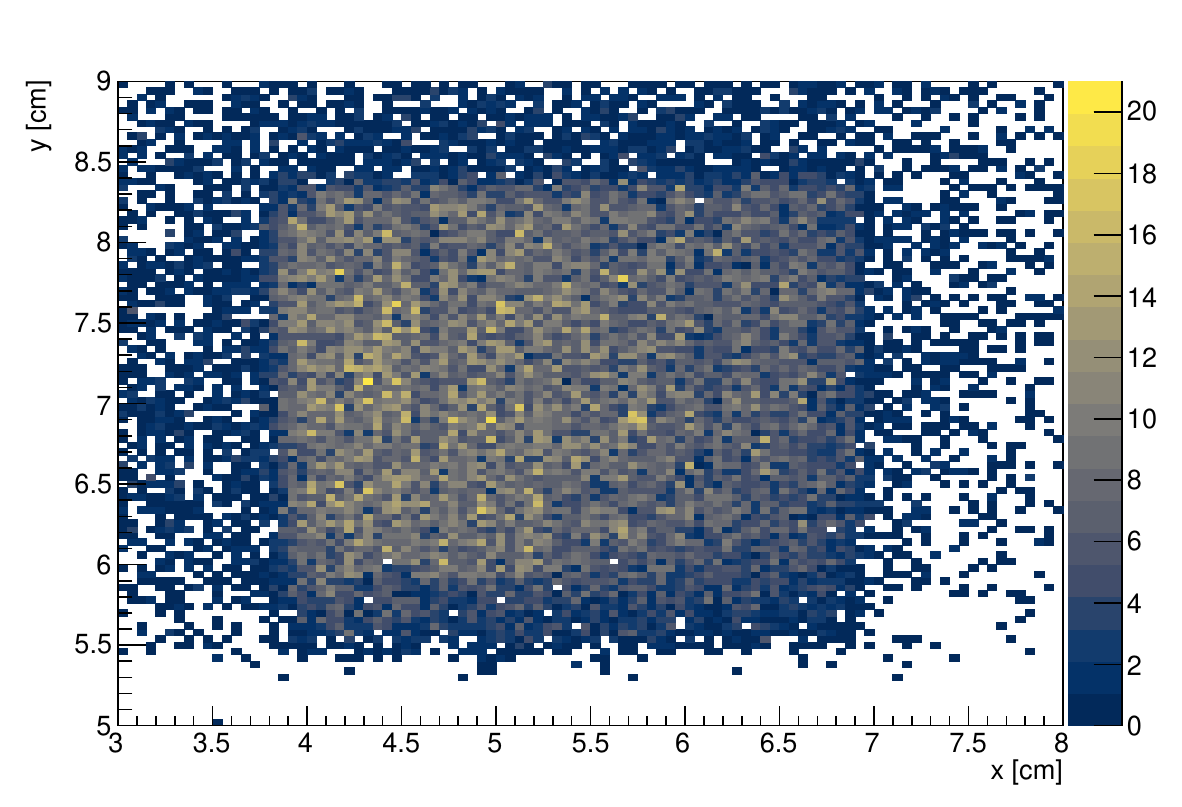}
        \caption{Projection of particle tracks from the Si trackers to the Demonstrator front face, only for particles that produce a signal above threshold in a given channel.}
        \label{fig:efficiency_map_procedure_sub2}
    \end{subfigure}
    \hfill
    \begin{subfigure}[t]{0.45\textwidth}
        \centering
        \includegraphics[width=\linewidth]{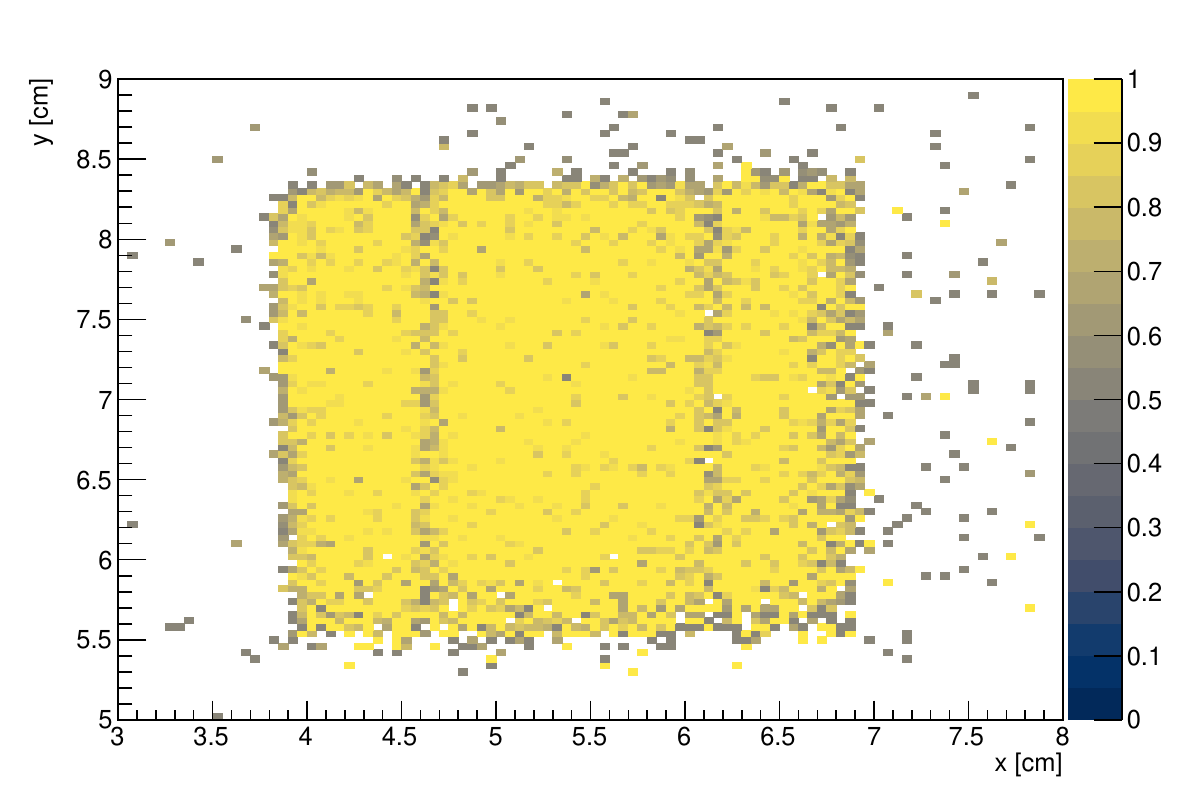}
        \caption{Efficiency map of one channel. The low-efficiency vertical stripes at $x = 4.5$ and $x = 6$ originates from the reduced material thickness due to the grooves for the optical fibers.}
        \label{fig:efficiency_map_procedure_sub3}
    \end{subfigure}

    \vspace{1em} 
    
    \begin{subfigure}{0.9\textwidth}
        \centering
        \includegraphics[width=\linewidth]{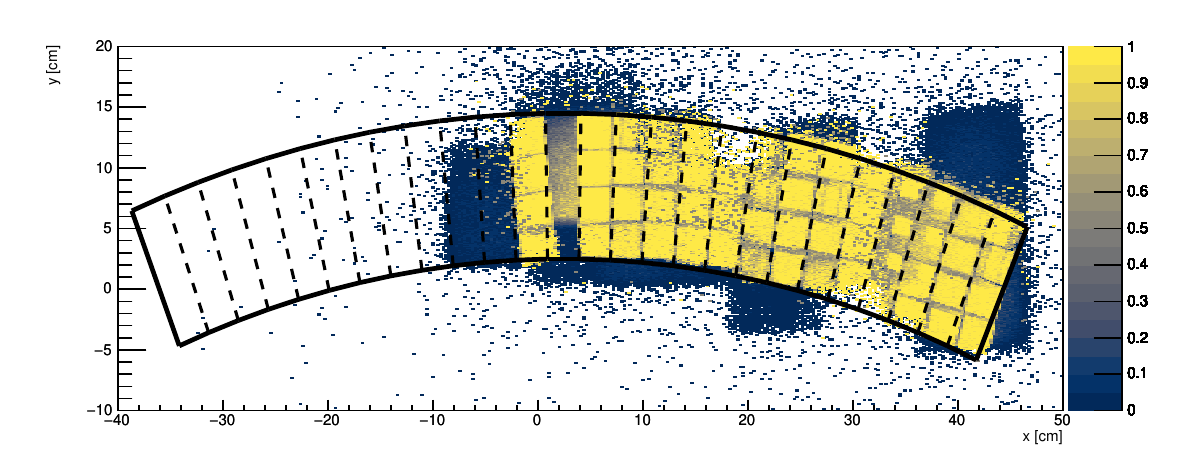}
        \caption{All efficiency maps overlayed on the histogram shown in Fig.  \ref{fig:efficiency_map_procedure_sub1}}
        \label{fig:efficiency_map_procedure_sub4}
    \end{subfigure}
    
    \caption{Procedure for obtaining the efficiency maps for each channel. The histogram in Fig.~\ref{fig:efficiency_map_procedure_sub2} is divided by the histogram in Fig.~\ref{fig:efficiency_map_procedure_sub1}, resulting in the efficiency map shown in Fig.~\ref{fig:efficiency_map_procedure_sub3}. The data used to create these figures were collected during the 2024 beam test.}
    \label{fig:efficiency_map_procedure}
\end{figure}

For a given channel, a sub-sample of events where a single particle impinged on its tiles was selected, in order to precisely estimate the MIP energy deposit.
The selection of this sub-sample relies on the definition of a fiducial area, determined based on the channel position. The channel position was estimated using the following procedure. The primary particle tracks measured by the Si micro-strip detectors are extrapolated to produce a 2D histogram $h_{h}$ (Fig. \ref{fig:efficiency_map_procedure_sub1}) of the primary-track hit positions ($x$, $y$) at the $z$ coordinate corresponding to the midpoint of the LCM. A second 2D histogram $h_{m}$ was then produced (Fig.~\ref{fig:efficiency_map_procedure_sub2}) using the same binning but including only the events where the considered channel recorded a signal larger than a fixed threshold corresponding to $300$\,ADC counts. A 2D ``detection efficiency'' map $h_{e}$ (Fig.~\ref{fig:efficiency_map_procedure_sub3}) was computed by taking bins ($i$, $j$) of the 2 histograms, and evaluating $h_{e,{ij}} = \frac{h_{m,{ij}}}{h_{h,{ij}}}$. The $h_{e}$ map (Fig.~\ref{fig:efficiency_map_procedure_sub3}) therefore corresponds to the fraction of events in which the primary particle impinges on a given location and produces a non-negligible signal in the Demonstrator channel. The $h_e$ map thus reproduces a projection of the LCM shape on the $x$-$y$ plane.

\begin{figure}[h!]
    \centering
        \centering
        \includegraphics[width=0.5\linewidth]{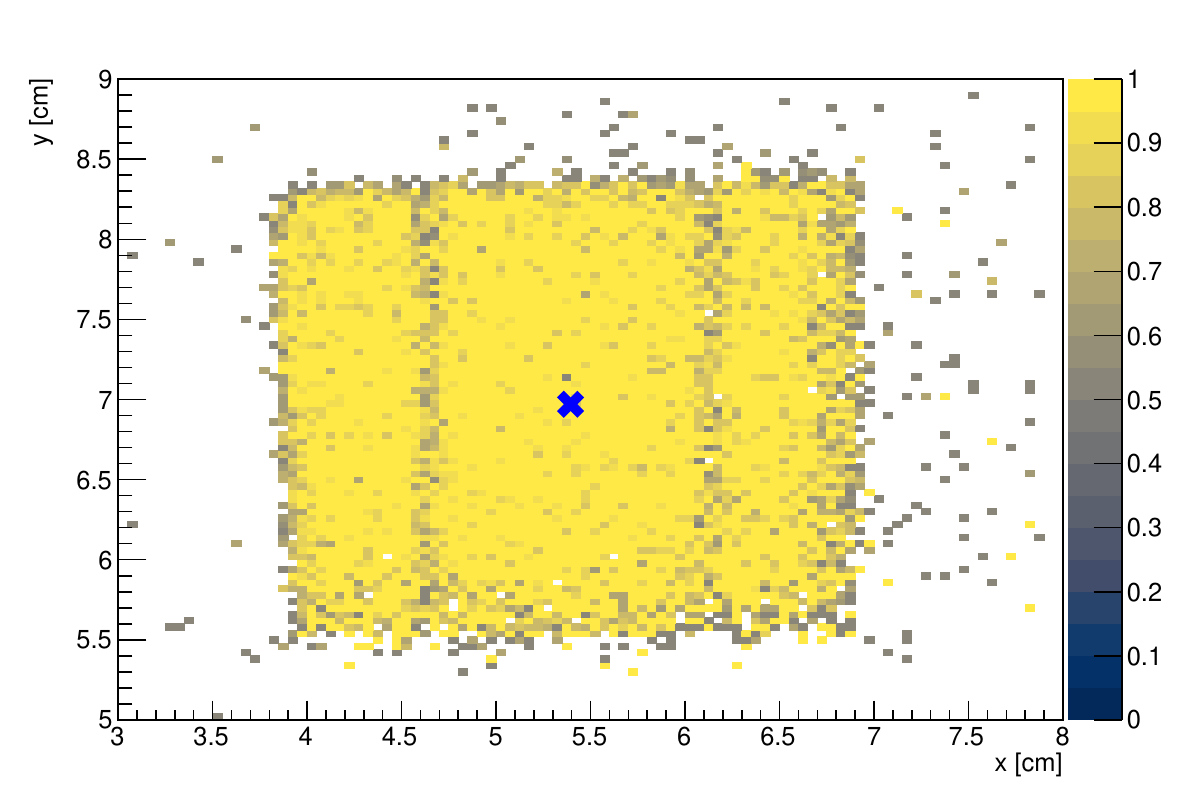}%
        \includegraphics[width=0.5\linewidth]{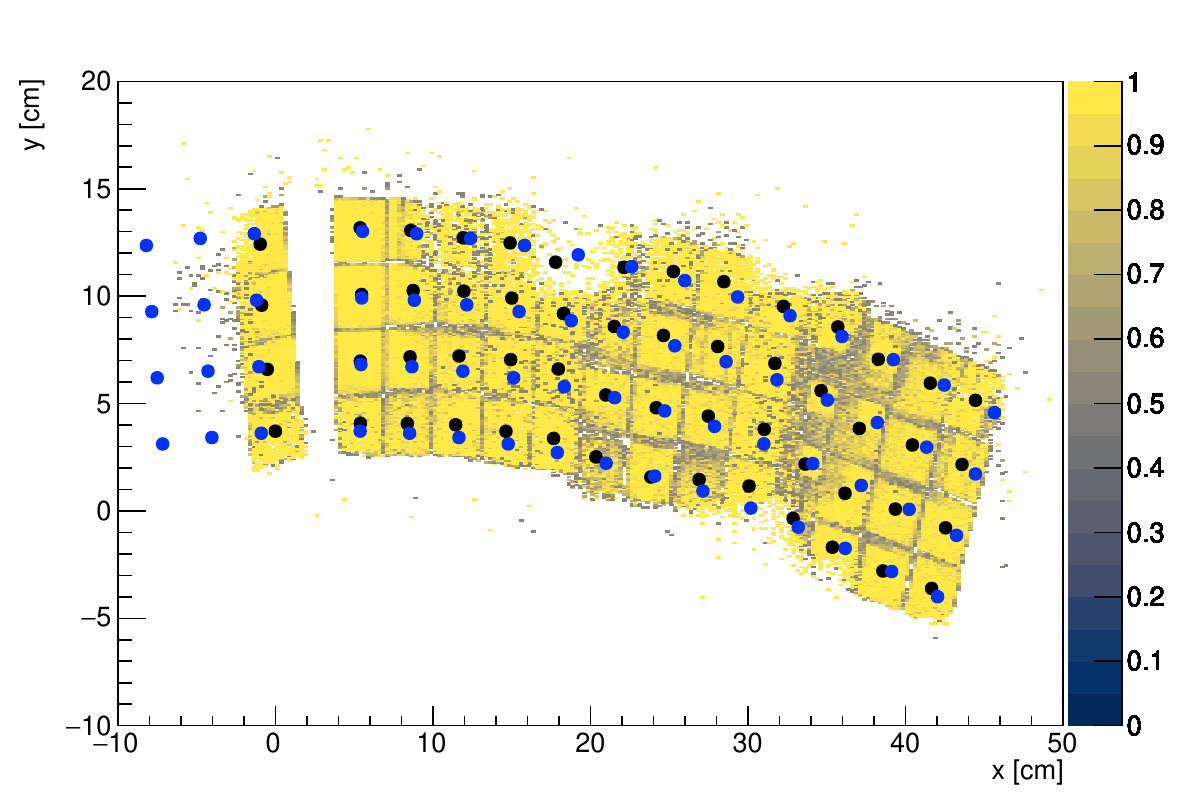}
        \caption{
            Efficiency maps and the calculated channel centers, used to define the corresponding fiducial areas.
                Left: Efficiency map of a single channel, with a marker indicating the mean of the 2D histogram used as an estimator of the channel position. Right: all efficiency maps from the 2024 beam test data. Black markers represent channel centers calculated from data, while blue markers represent channel centers as they are located in the Demonstrator simulation.}
    \label{fig:tile_center_of_gravity}
\end{figure}

The physical center ($x_c$, $y_c$) of the LCM in the $x$-$y$ plane was computed by taking the mean of the $h_e$ bins whose efficiency (bin content) exceeded 50\% (Fig.~\ref{fig:tile_center_of_gravity}).

The fiducial area -- centered in ($x_c$, $y_c$) -- was set to 1.1$\times$1.1 cm$^2$  for the first 4 $z$ layers of the Demonstrator, 1.5$\times$1.5 cm$^2$ for layers 4-8, and 2.0$\times$2.0 cm$^2$ for all subsequent layers. Some further fine tuning of the fiducial areas was adopted to compensate for the reduced statistics in the downstream Demonstrator channels and to account for the Multiple Coulomb Scattering (MCS) of MIPs within the detector. Events with primary tracks falling within this square were used as a sub-sample for the MIP energy deposit estimation. 

\begin{figure}[t]
    \centering
        \centering
        \includegraphics[width=0.5\linewidth]{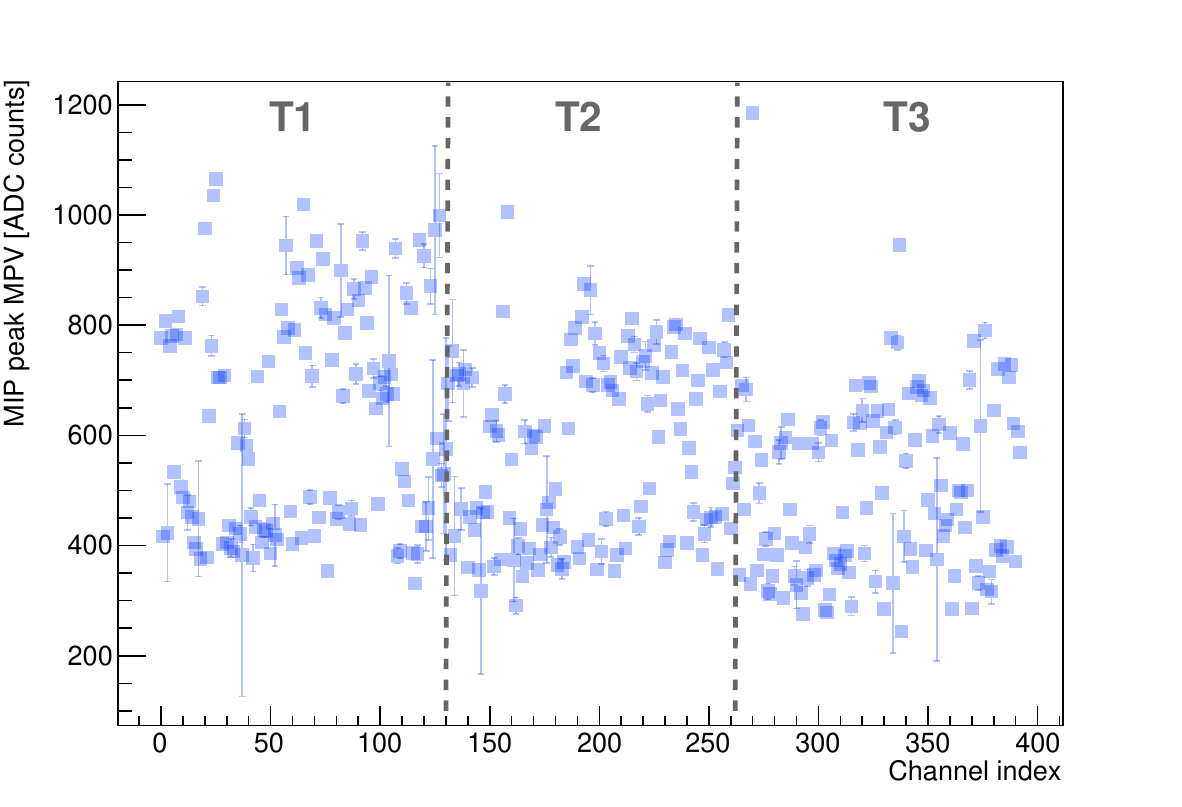}%
        \includegraphics[width=0.5\linewidth]{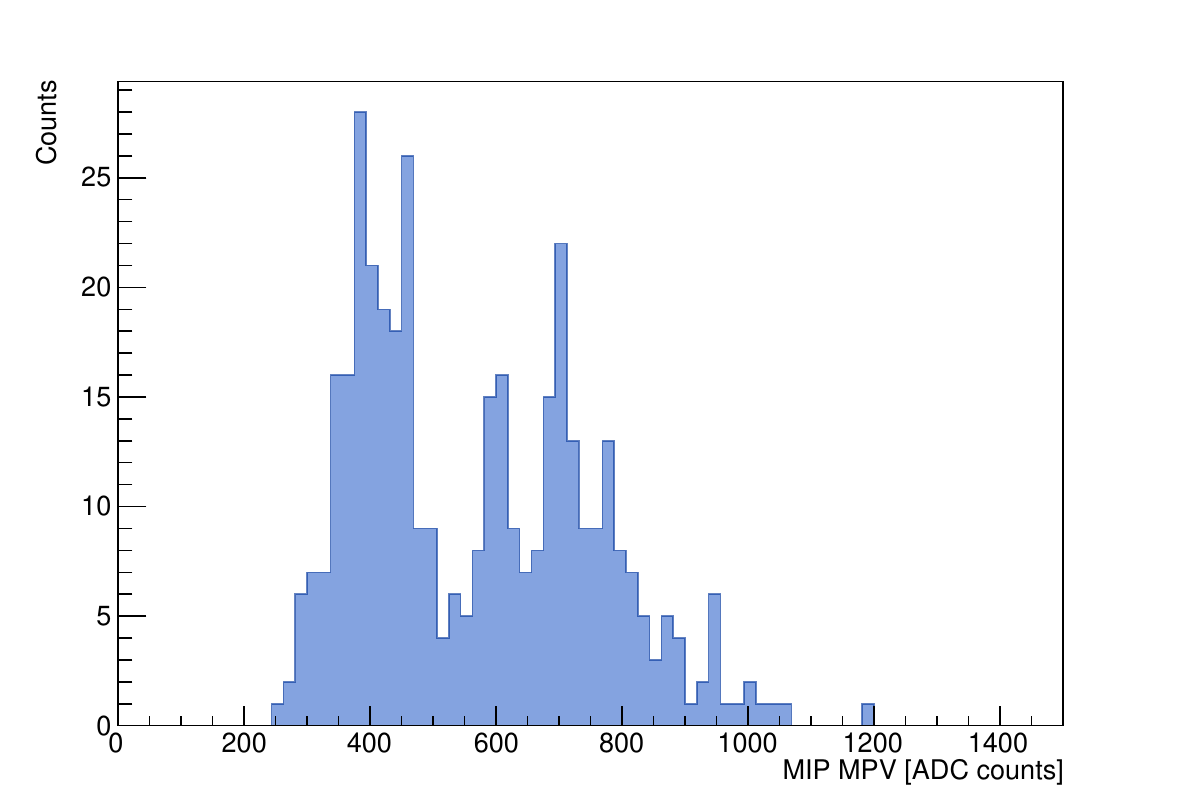}
    \caption{Most probable values of the MIP peaks for all calibrated channels. Left: MIP mean values for all calibrated channels in 2024. T1, T2, and T3 indicate different $R$-values, with T1 corresponding to the innermost and T3 to the outermost calorimetric layer. Right: distribution of the MIP mean values for all calibrated channels in 2024.}
    \label{fig:mpv_values}
\end{figure}

Given the large number of channels to be calibrated, the MIP-peak fitting was performed uniformly for all channels using a custom algorithm. After producing the channel-response histogram by applying the fiducial-area cut described above, the first step was to identify the MIP peak. To simplify this procedure for channels with lower statistics, the histogram was smoothed using the ROOT built-in \textit{TH1::Smooth} function to reduce the number of fake local minima. Once the MIP peak was located, it was fitted using a Landau-plus-linear model, and the MPV of the Landau was taken as the reference MIP energy deposit. If the fitting procedure failed for any reason (yielding a poor fit), the fallback solution was to use the highest bin within the MIP peak as the MPV. Fig.~\ref{fig:mpv_values} shows the values obtained with this procedure. The relative uncertainty on the calculated MIP mean depends on both the statistical uncertainty and the uncertainty associated with the fit model, which was estimated by varying the histogram binning and range. This uncertainty amounts to approximately 10\% for properly illuminated channels.

Fig.~\ref{fig:mpv_values} shows a slight decrease in the MIP MPV with increasing $R$ coordinate, as well as two distinct populations of MIP mean values: one around $400$ ADC counts and the other around $700$ ADC counts. These two trends are seen in both 2023 and 2024 beam test data. The observed double structure is likely due to imperfections in the optical interface caused by the optical {\color{blue}{oil}}, which exhibited a lower viscosity than initially expected, resulting in non-uniform coverage and variable performance across the Demonstrator channels. The calibration procedure effectively removes its impact in subsequent analyses.

Nevertheless it should be noted that the equalization procedure corrects the average channel response but cannot fully account for non-linear effects such as SiPM saturation.
Channel equalization largely compensates for the resulting spread in light yield, although it is clearly not an optimal solution once saturation effects become significant.

The pedestal of each channel was also determined, and its mean value was used to correct the pulse heights and MPVs for the calculation of the equalized pulse heights. To exclude spurious energy deposits from hadrons crossing nearby channels, only events with primary tracks at least one hadronic-interaction-length ($\lambda \simeq 17.6$~cm) away from ($x_c$, $y_c$) were selected. The spectrum of these events was plotted and fitted with a Gaussian function. The mean of the fit, $P$, was taken as the best estimate of the pedestal value.
This procedure was applied to most Demonstrator channels, except for dead channels and channels that were not properly illuminated by the beam during the calibration runs (see Fig.~\ref{fig:efficiency_map_procedure_sub4}).

A procedure similar to that used for the data to determine the channels response to MIPs was applied to the simulated data. The beam was generated using the transverse spatial distributions of the particles measured by the Si trackers in the calibration runs. For each LCM, a 1$\times$1~cm$^2$ fiducial cut was applied, and the spectrum of the total energy deposited in the LCM tiles was computed. The MIP peak was fitted 
and the mean deposit was obtained by averaging the MPV parameters from this fit: $MPV_{MeV} \simeq 6.12$ MeV. 

The equalized signal for each channel in the data, expressed in MeV, was then computed as:
\begin{align*}
PH_{eq} = \left( PH-P\right)\frac{MPV_{MeV}}{MPV-P}
\end{align*}
where $PH$ is the signal from the channel in ADC counts before any correction, $P$ the pedestal of that channel in ADC counts, $MPV$ is the most probable value of the Landau fit to the MIP spectrum for that channel in ADC counts and $MPV_{MeV}$, in MeV, is the most probable value of the Landau fit to the energy deposition distribution of MIPs in an LCM as predicted by the Geant4 simulation.

\subsection{Cross-talk}

\begin{figure}[h!]
\centering
\includegraphics[width=1\textwidth]{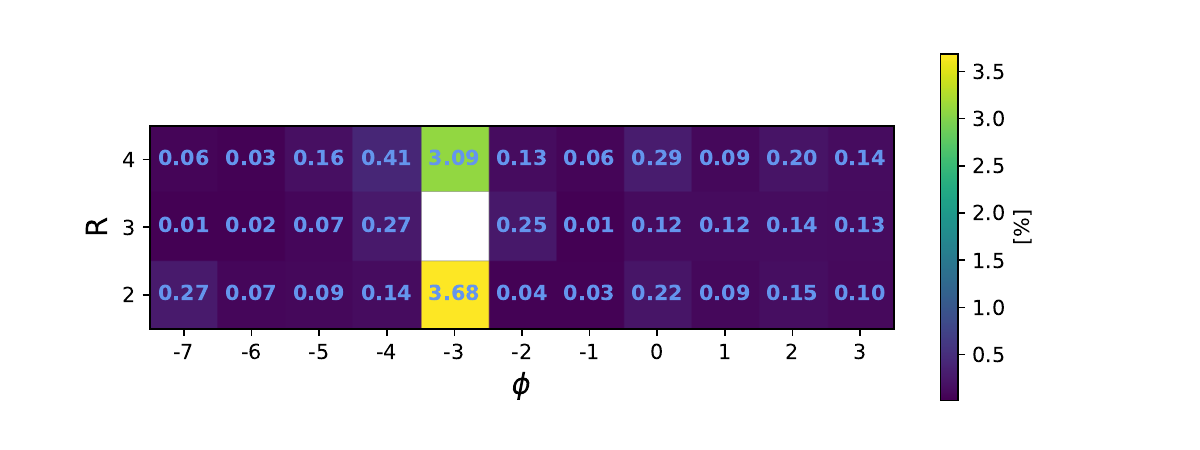}
\caption{\label{fig:CrosstalkPlot}An example of cross-talk values (in \% units) computed for the left side of the Demonstrator, using the LCM with $\left(R_0,\phi_0\right) = (3, -3)$ as reference. 
}
\end{figure}

\begin{figure}
\centering
\includegraphics[width=1.\textwidth]{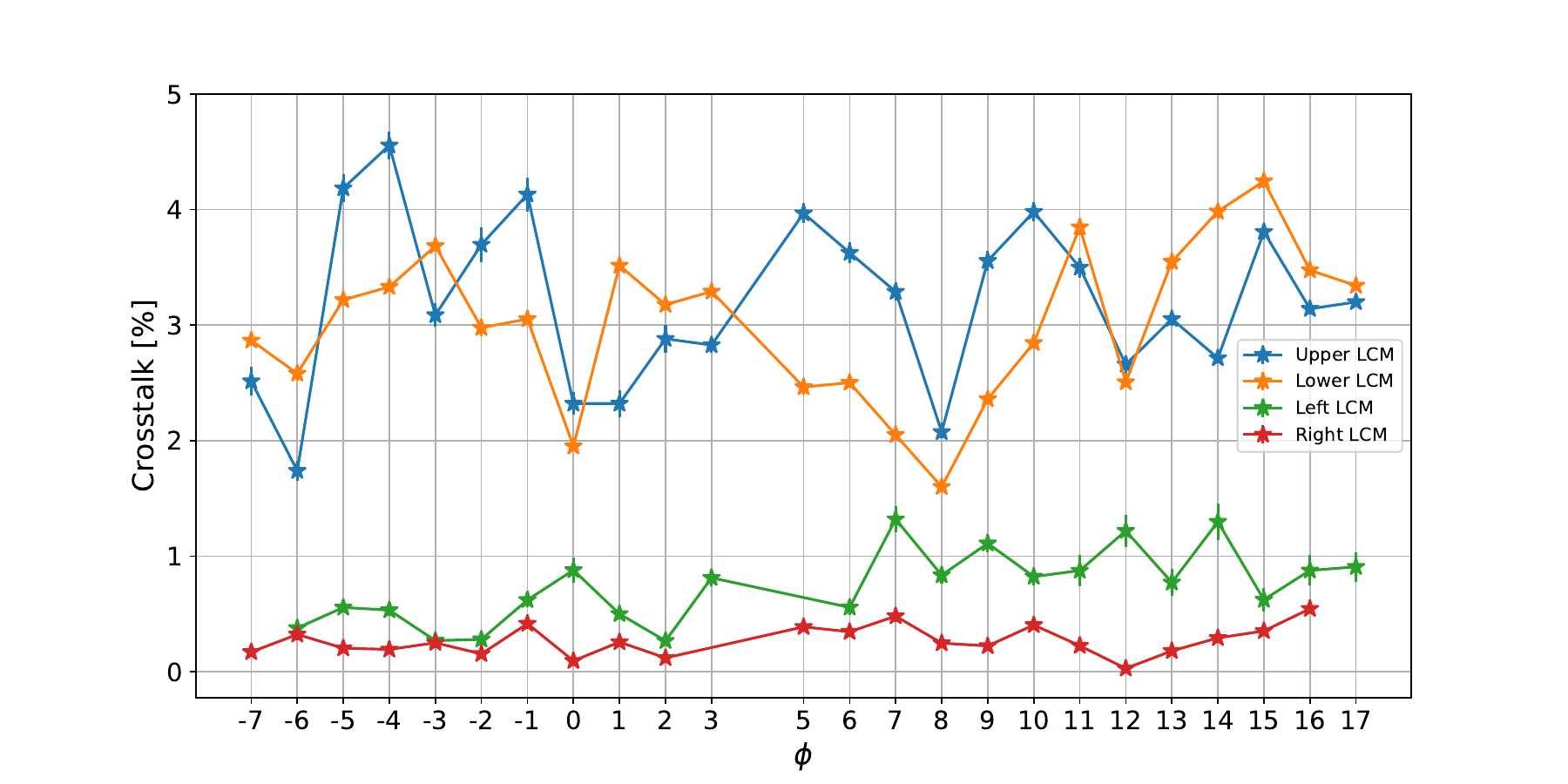}
\caption{\label{fig:CrosstalkVSphi}Cross-talk values (in \% units) as a function of $\phi$ for the first layer of the Demonstrator.}
\end{figure}

A potential source of degradation in the energy resolution and particle-identification efficiency is cross-talk between adjacent LCMs. This can occur when light from a scintillator tile escapes and enters a nearby LCM/WLS fiber, contributing to the signal of an adjacent channel. Possible causes for this effect include holes in the TiO$_2$ coating, particularly in the grooves, where achieving a uniform coating proved more difficult. Alternatively, light may escape or enter through the fibers in regions where the external cladding is degraded. Also one could envisage some level of WLS-to-WLS cross-talk as they travel close for tens of cm. To estimate cross-talk between LCMs, the calibration runs described in Sec. \ref{sec:equalization} were used. A systematic study was conducted on the most upstream $z$ layer of the Demonstrator, since in the downstream layers the actual particle track is more likely to deviate from the primary track measured by the silicon tracker due to MCS, potentially leading to errors in determining which LCM the particle impinged on. In this study, for each LCM a sub-set of events was selected, such that:
\begin{itemize}
\item the primary track falls within a  $1.5\times 1.5$ cm$^2$ fiducial cut, centered on the LCM;
\item the signal recorded in the LCM is higher than 0.75 MIPs.
\end{itemize}
Once the appropriate events were selected, the following quantity was computed for all channels in the upstream $z$ layer, identified by their $\left( R, \phi \right)$ coordinates:
\begin{align*}
\rho\left( R,\phi \right) = \frac{PH_{eq}\left(R,\phi\right) } {PH_{eq}\left(R_0,\phi_0\right) }
\end{align*}
where, for each event $PH_{eq}\left( R_0,\phi_0\right)$ is the equalized signal recorded by the LCM under consideration. An example of the ratio $\rho\left( R,\phi \right)$ for a channel in the upstream $z$ layer is shown in Fig.~\ref{fig:CrosstalkPlot}. The data show a cross-talk effect of about 3\% mostly with the upper and lower radial channels at the same $\phi$ i.e. the same fiber concentrator. In Fig.~\ref{fig:CrosstalkVSphi} we show for each tile of the central calorimetric layer as a function of their $\phi$, how the $\rho$ ratio behaves with respect to the neighboring LCMs.
Although some cross-talk appears to be present, with certain channels showing non-negligible values of $\rho\left( R,\phi \right)$, these values remain below 5\,\%. 
To cross-check the impact of cross-talk on electron energy resolution, we simulated it between neighboring radial channels at the 10\% level - approximately
three times larger than the value observed in the data. Even under this conservative assumption, its effect on the resolution was barely significant.

\subsection{Single photoelectron analysis}
\label{sec:single_photoelectron}

During the 2024 beam test, a dedicated readout system was used to measure the light yield produced by MIP particles in the demonstrator modules. Specifically, the channels of one azimuthal sector in the frontal layer were read out using an AdvanSiD SiPM Evaluation Board signal amplifier connected to a CAEN V1730 digitizer. Data were acquired with a muon beam configuration.

A sixth-order Butterworth low-pass filter with a cutoff frequency of 10 MHz was applied to the acquired waveforms as a preprocessing step. This approach proved effective in reducing high-frequency noise while enhancing the extraction of information from the measured pulses. In fact, the signal is characterized by a relatively small bandwidth, on the order of a few MHz.

\begin{figure}
\centering
\includegraphics[width=.85\textwidth]{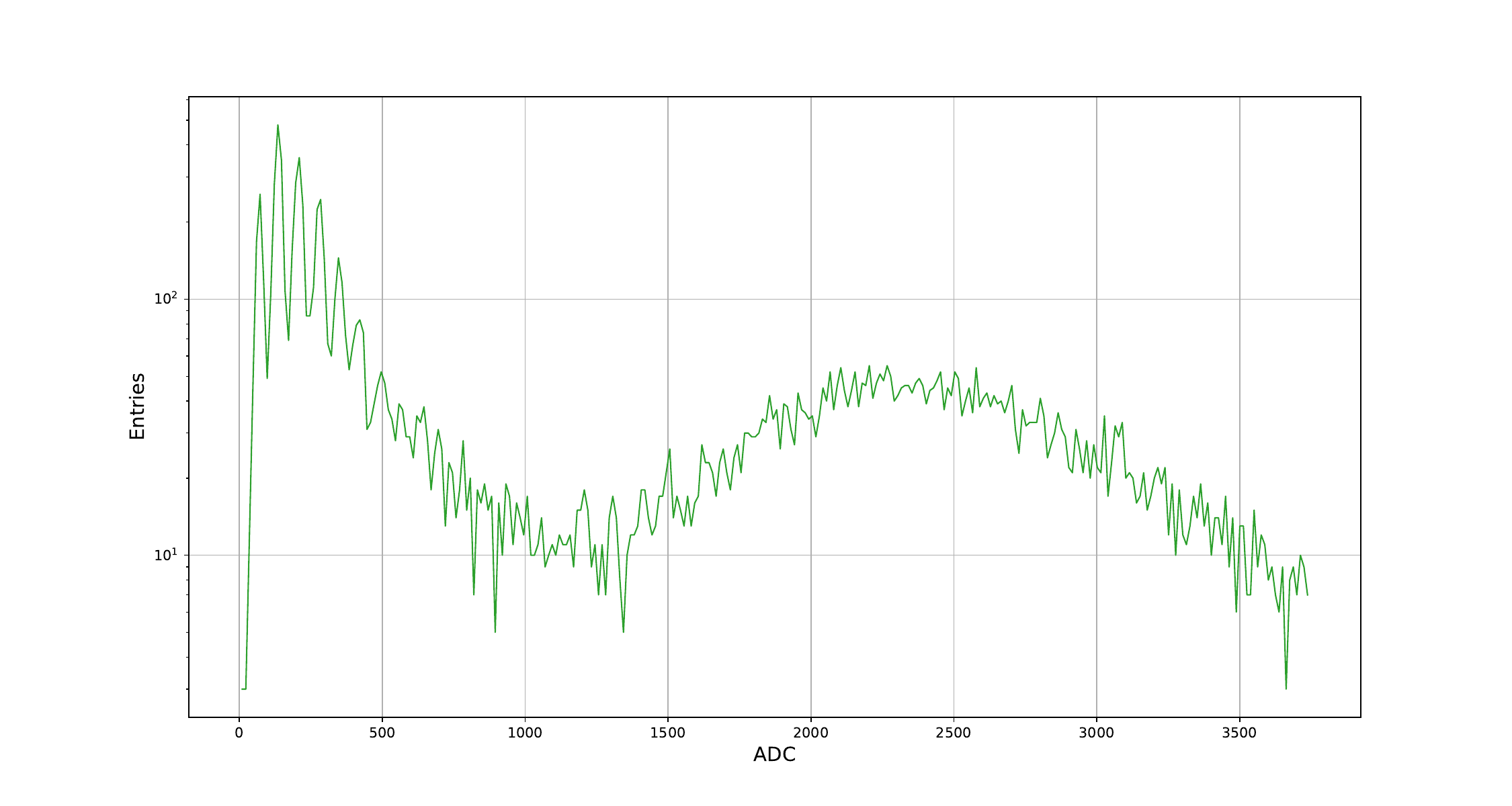}
\caption{\label{fig:pheAll}Pulse height distribution from a tile of a t0 channel of the Demonstrator. The channel output is amplified and low-pass filtered. The single- and multi-p.e. peaks from the dark noise are visible in the left part of the distribution, extending up to approximately 500 ADC counts. The peak corresponding to the MIP signal is also visible and is centered around 2200-2300 ADC counts.}
\end{figure}

\begin{figure}
\centering
\includegraphics[width=0.49\textwidth]{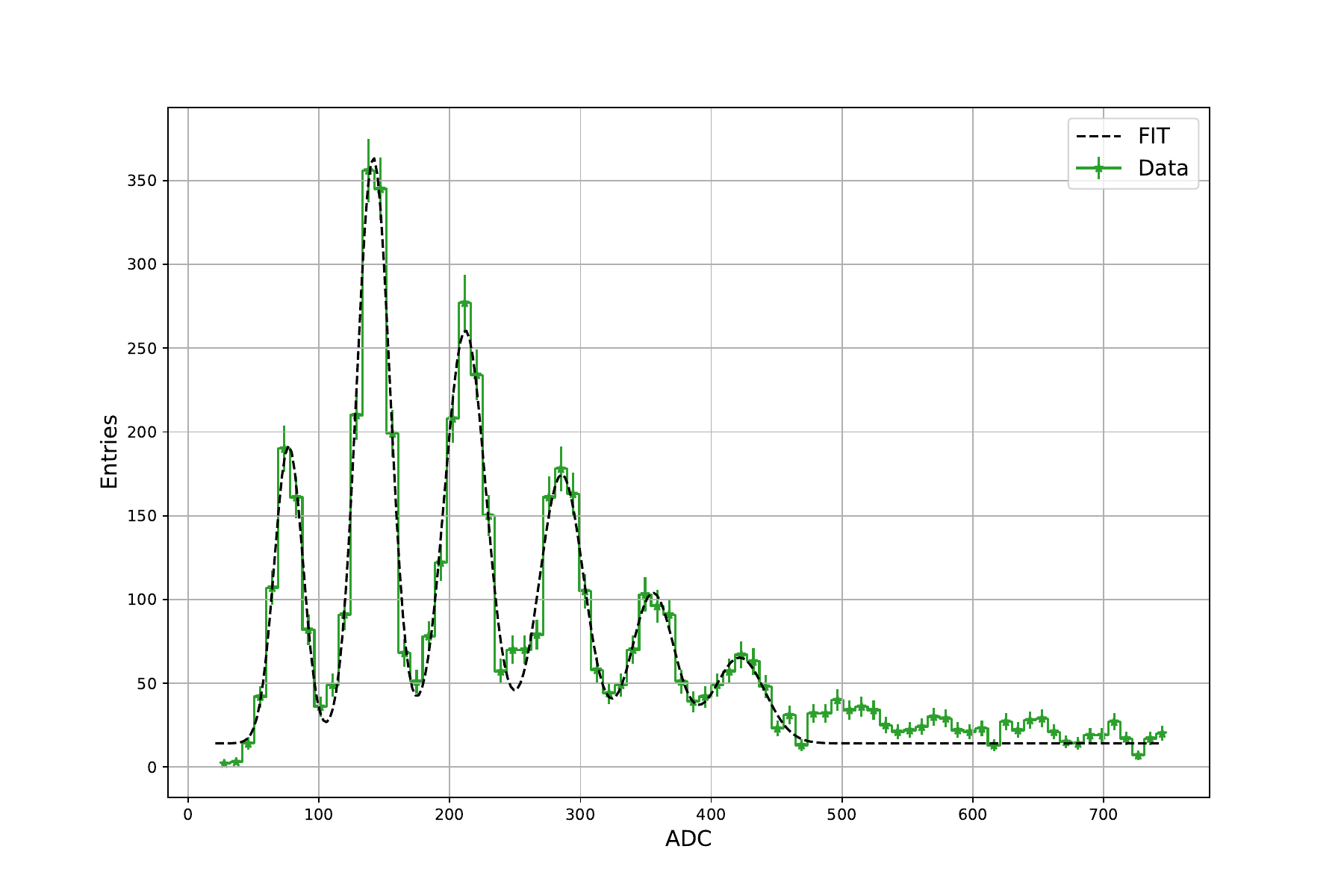} 
\includegraphics[width=0.49\textwidth]{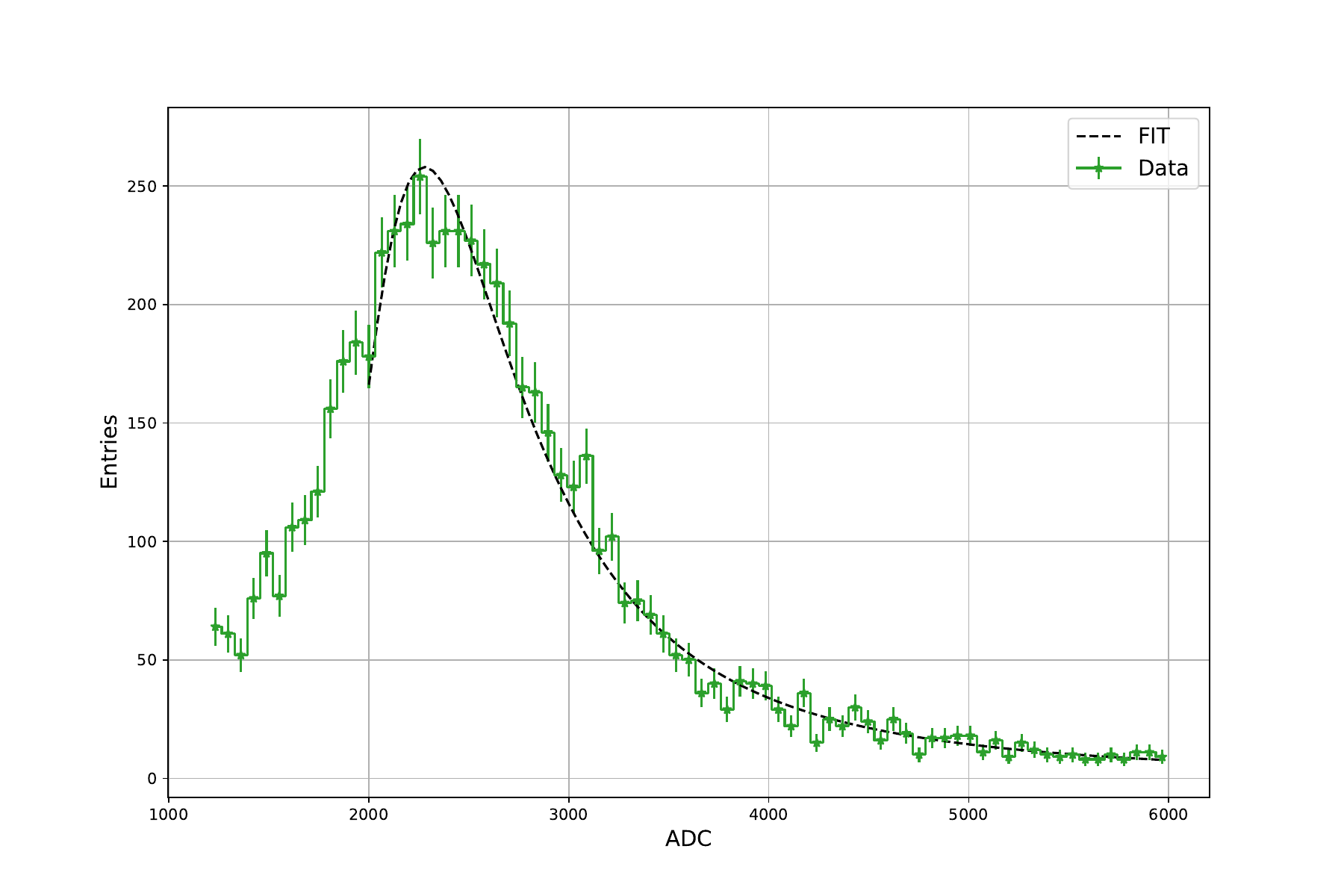} \\
\caption{(Left) Single- and multi-p.e. peaks from the dark noise, fitted with a multi-Gaussian function. (Right) Peak from the MIP signal, fitted with a Landau distribution. The data are the same as those shown in Fig.~\ref{fig:pheAll}.}
\label{fig:pheTogether}
\end{figure}

For the calorimetric channels, a fiducial cut was defined using information from the silicon trackers. This allowed the selection of a sample of MIP-like particles with tracks fully contained within the channel volumes, as well as a sample of events with no particle-induced signals. A fiducial cut was not required for the analysis of the t0 channel tiles, since their thickness ensures that all tracks traversing the channel are fully contained, while the remaining events constitute the no-signal sample.
The no-signal sample was used to characterize the dark noise of each SiPM and to estimate the amplitude of a single photo-electron (p.e.) signal.
Fig.~\ref{fig:pheAll} shows the distribution of the amplitudes of the signals recorded from a $t_0$ tile, where the single- and multiple-p.e. peaks can be clearly distinguished. A multi-Gaussian fit (Fig.~\ref{fig:pheTogether}-left) was used to estimate the single-p.e. signal amplitude from the spacing between the Gaussian peaks. The MIP-like sample was then used to estimate the MIP signal amplitude through a Landau fit to the amplitude distribution (Fig.~\ref{fig:pheTogether}-right).

The ratio of the MPV of the MIP-like signal to the single-p.e. signal provides an estimate of the number of p.e. produced by the energy deposited by MIPs. The values obtained for the channels in the analyzed sector are reported in Tab. \ref{tab:phe_analysis}. The results show improved performance with respect to the previous prototype \cite{Acerbi:2020nwd}, with a fourfold increase in the light yield in the new setup. The improvement in light yield can be attributed to the enhancements introduced in the construction of the Demonstrator, namely the adoption of the frontal-tile readout in place of the lateral readout and the more efficient optical coupling between the WLS fibers and the SiPMs.

In the current prototype, the increased light yield leads to saturation effects in the SiPMs, resulting in a distortion of the measured energy at high deposited energies, like in the case of electron events producing electromagnetic showers. The impact of these saturation effects can be evaluated by considering the expected number of photoelectrons, $N_{pe}$, produced by electron energy deposition in the absence of saturation (including the photon detection efficiency of the SiPM).
This number can be estimated from the electron energy deposition and the number of p.e. produced by the energy deposited by a MIP, $N_{pe}^{MIP}$. The estimate of $N_{pe}^{MIP}$ is obtained by averaging the measurements from the three calorimetric channels reported in Tab.~\ref{tab:phe_analysis}. The expected number of photoelectrons, $N_{pe}$, is then corrected to account for cross-talk effects as $N_{seed} \equiv (1 + P_{XT}) N_{pe}$. The SiPMs used to instrument the prototype are Hamamatsu S14160-4050HS~\cite{sipmWeb}, featuring 6331 cells and a cross-talk probability of 7\%. The area of the SiPM covered by the 10 WLS fibers from a single calorimetric module corresponds to a fraction of approximately 49\% of the total SiPM surface. Therefore, the number of illuminated cells is $N_{max} \simeq 3100$, which is smaller than the total number of available cells, $N_{cells}$. The expected number of fired SiPM cells, including saturation effects, can be approximated as~\cite{GRUBER201411}:

\begin{equation*}
N_{fired} \simeq N_{max} \left(1 - e^{-\frac{N_{seed}}{N_{max}}}\right)
\end{equation*}

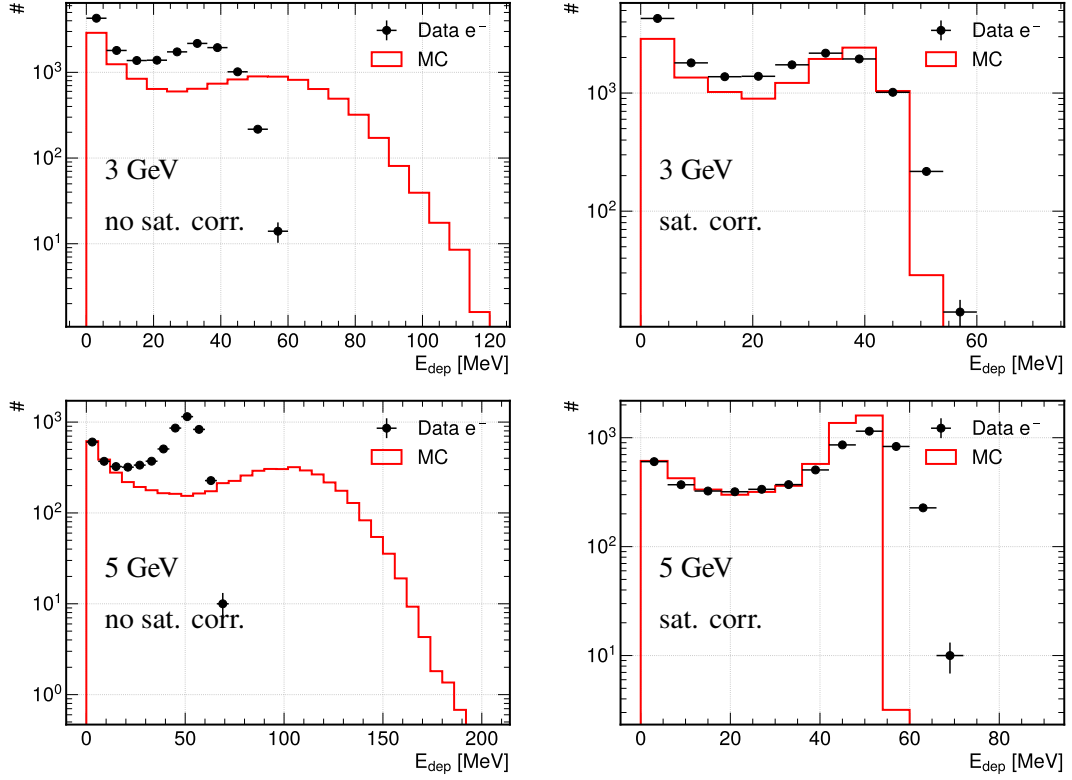
\begin{figure}[h!]
\centering
\begin{tabular}{cc}
\begin{overpic}[width=0.45\textwidth]{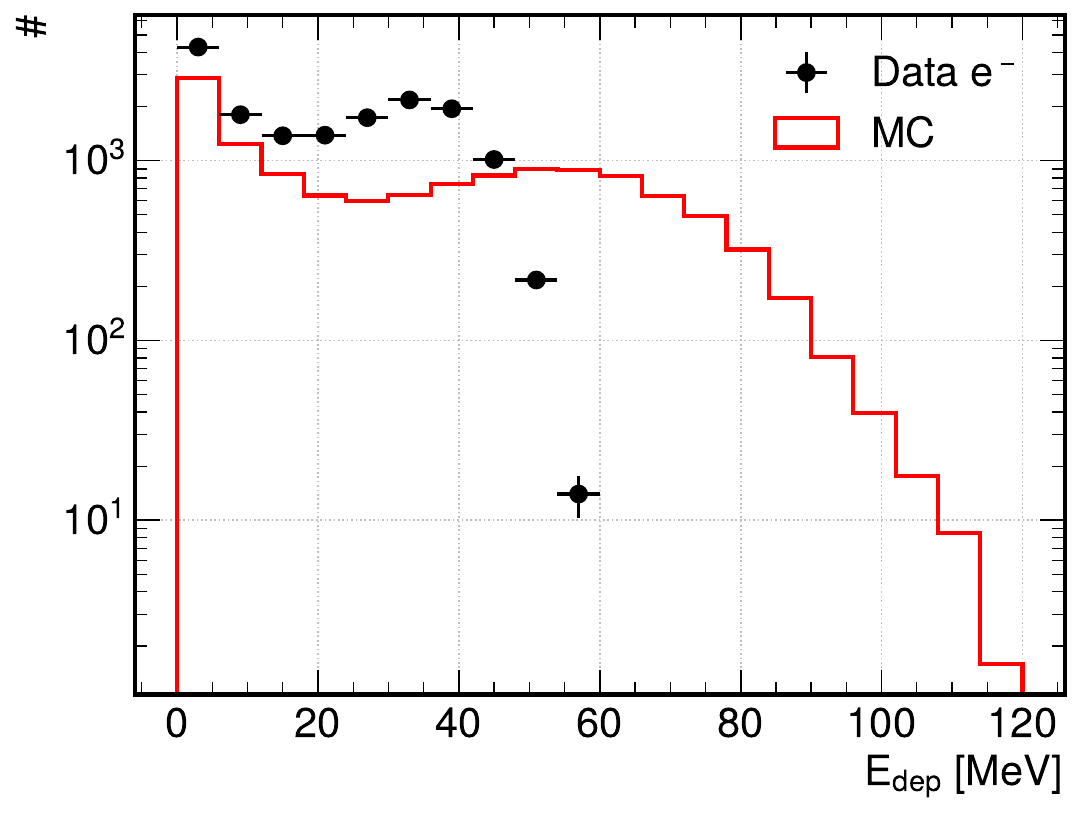}
  \put(20,40){3 GeV}
  \put(20,30){no sat. corr.}
\end{overpic}
&
\begin{overpic}[width=0.45\textwidth]{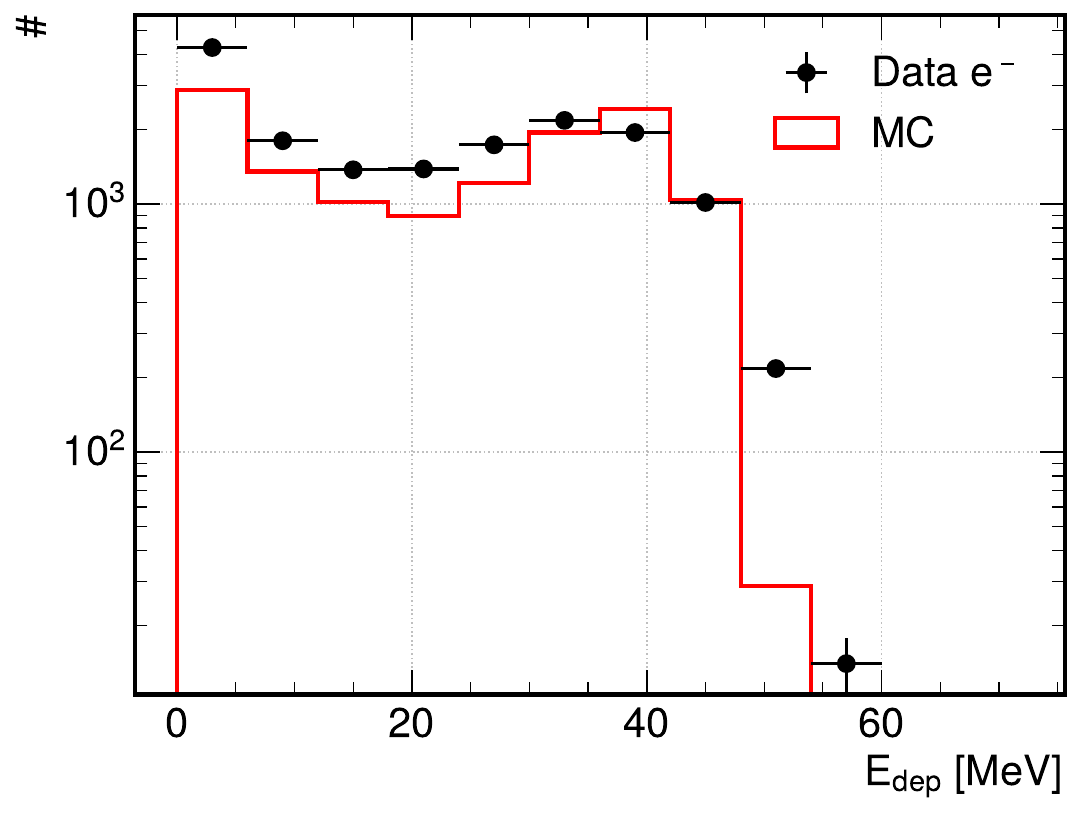}
  \put(20,40){3 GeV}
  \put(20,30){sat. corr.}
\end{overpic}
\\
\begin{overpic}[width=0.45\textwidth]{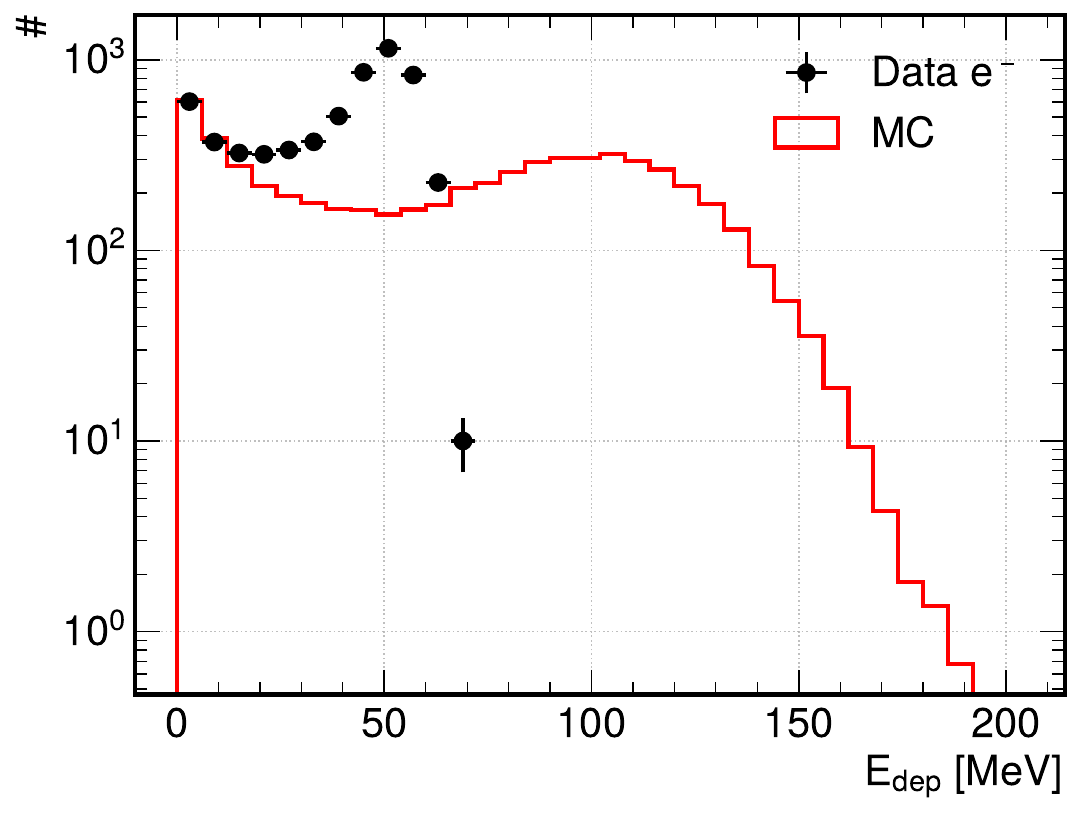}
  \put(20,40){5 GeV}
  \put(20,30){no sat. corr.}
\end{overpic}
&
\begin{overpic}[width=0.45\textwidth]{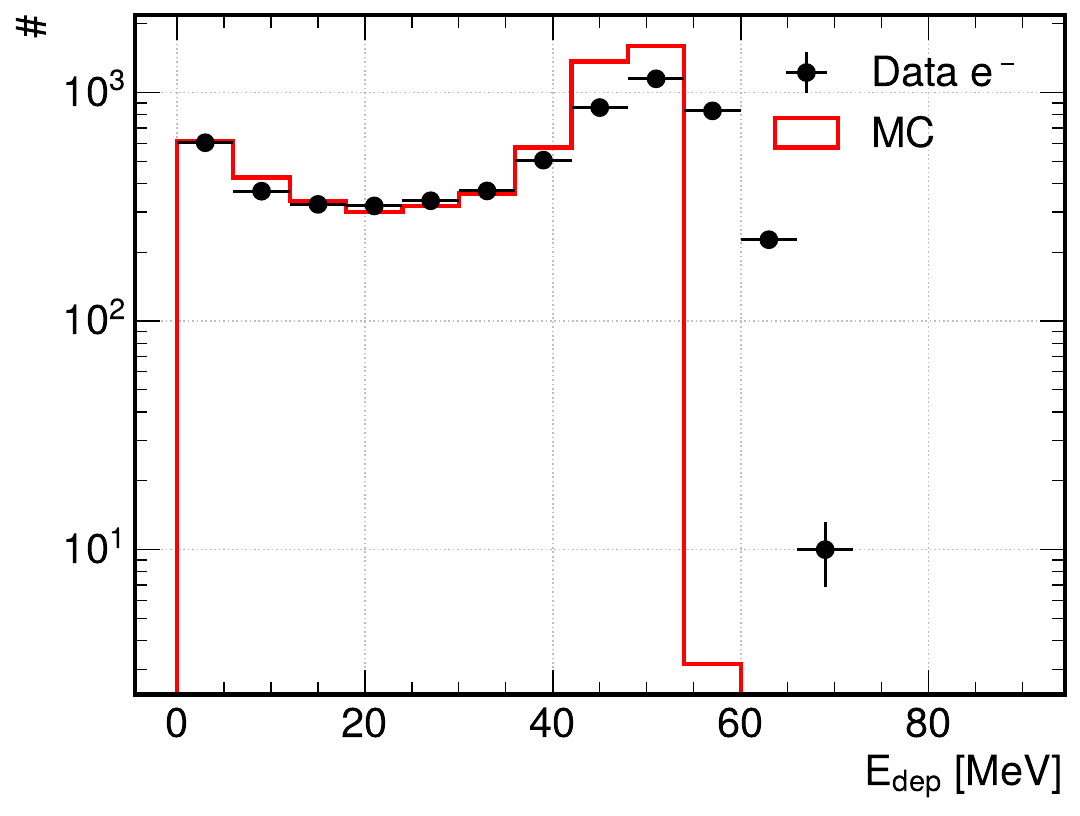}
  \put(20,40){5 GeV}
  \put(20,30){sat. corr.}
\end{overpic}

\end{tabular}

\caption{Distributions of the energy deposited in the channel located in the central radial layer, azimuth sector $\phi = 11$, and in the second layer along the beam axis, for electron runs at 3 GeV (top) and 5 GeV (bottom). This channel corresponds to the one with the largest energy deposit. The data are compared to MC simulations without SiPM saturation (left) and with saturation effects included (right).}
\label{fig:ph_saturation}
\end{figure}

Fig.~\ref{fig:ph_saturation} shows the distributions of the energy deposited in the channel with the highest deposit in the second layer for electron runs at 3 GeV and 5 GeV. The MC predictions before and after the saturation correction, as obtained from the equation above, are also shown.

As can be seen in the figure, SiPM saturation must be included in the simulation to reproduce the data distributions. From the plots, it is clear that the chosen parameter values lead to an overestimation of the saturation effects in these channels. 

We previously attempted to apply the saturation correction on a channel-by-channel basis. The number of photoelectrons could be measured directly only for a few selected channels equipped with a dedicated transimpedance amplifier coupled to a digitizer. 
We therefore estimated the number of photoelectrons collected by each channel by scaling the average measured photoelectron yield according to the ratio of the channel's MIP-peak position to the average MIP-peak position of the channels for which the light yield was measured.
As expected, this procedure did not improve the linearity and resolution. The effective SiPM area covered by the fiber is channel-dependent and determines \(N_{\mathrm{max}}\) in the saturation-correction formula. This effect cannot be disentangled from the channel-to-channel variation in light yield, \(N_{\mathrm{pe}}\). We therefore retained the average correction described above.

Saturation effects are visible in the linearity plot of the energy measured in the calorimeter for electron runs at different energies, as discussed in the following section. Moreover, SiPM saturation causes data-MC discrepancies in the electron shower profile that are larger than expected if these effects are not included in the simulation. See Sec.~\ref{sec:ene_dep_pattern} for a detailed discussion.

\begin{table}
\begin{center}
\begin{tabular}{ c c c c c c }
\hline
& $t_\text{0 up}$ & $t_\text{0 down}$ & $t_{1}$ & $t_{2}$ & $t_{3}$ \\
\hline
MIP peak & 3168 $\pm$ 18 & 2368 $\pm$ 14 & 5073 $\pm$ 26 & 4341 $\pm$ 20 & 4054 $\pm$ 35 \\
$\Delta_{pe}$ peak & 65 $\pm$ 8 & 68 $\pm$ 8 & 15 $\pm$ 2 & 14 $\pm$ 2 & 11 $\pm$ 1 \\
MIP peak / $\Delta_{pe}$ peak & 48 $\pm$ 6 & 35 $\pm$ 4 & 336 $\pm$ 45 & 306 $\pm$ 44 & 372 $\pm$ 34 \\
\hline
\end{tabular}
\caption{\label{tab:phe_analysis}Photoelectron results for calorimetric and $t_0$-layer channels. The error on the MIP peak is estimated by the Landau fit. The error on the $\Delta_{pe}$ peak is the standard deviation of the values. All units are ADC counts.}
\end{center}
\end{table}

\subsection{Response and energy resolution for electrons}
\label{sec:enereso}

To estimate the calorimeter response to electrons and its energy resolution, dedicated electron-enriched runs at different beam momenta were performed. Achievable electron purities at different beam energies are shown in Tab.~\ref{tab:ele_runs}. For a given configuration and beam momentum, the total visible energy deposit for each event was estimated by summing the equalized signals of all LCMs.
\begin{table}
\begin{center}
\begin{tabular}{ c c c c c c }
\hline
Momentum [GeV/$c$] & 1 & 2 & 3 & 4 & 5 \\
Electron purity [\%] & 93.7 & 94.7 & 91.8 & 87.0 & 74.1 \\
\hline
\end{tabular}

\caption{\label{tab:ele_runs}Run conditions for the electron-enriched runs used for the energy resolution estimation. The electron purity is measured by the \v{C}erenkov counters and contaminations are removed using cuts on the \v{C}erenkov signal (see text for details).}
\end{center}
\end{table}

\begin{figure}
\centering
\includegraphics[width=0.49\textwidth]{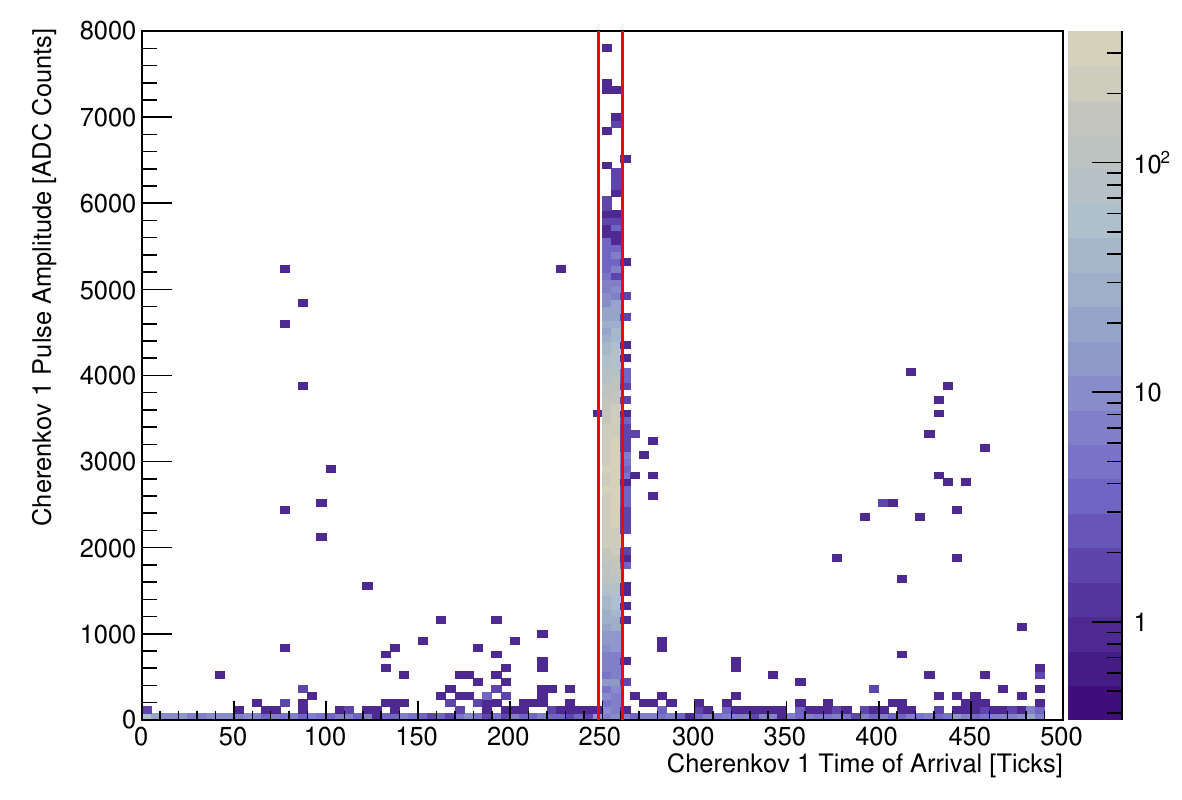} 
\includegraphics[width=0.49\textwidth]{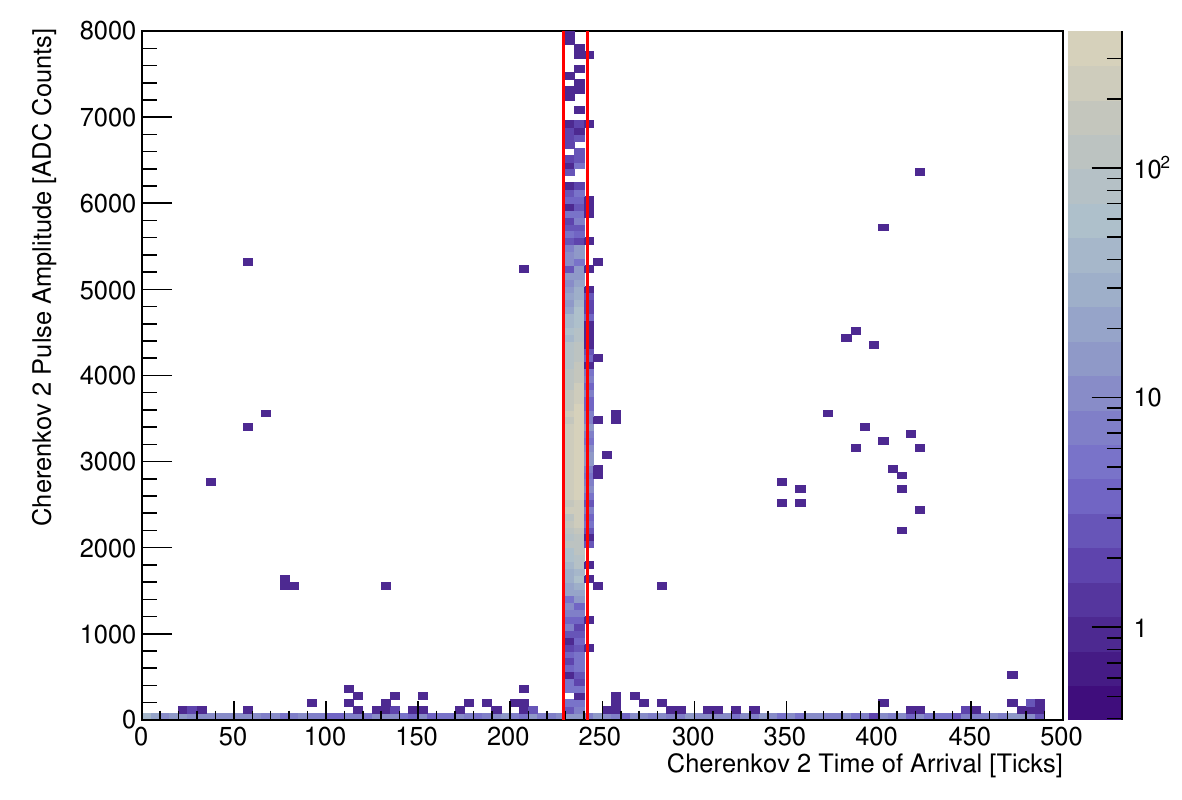} \\ 
\includegraphics[width=0.49\textwidth]{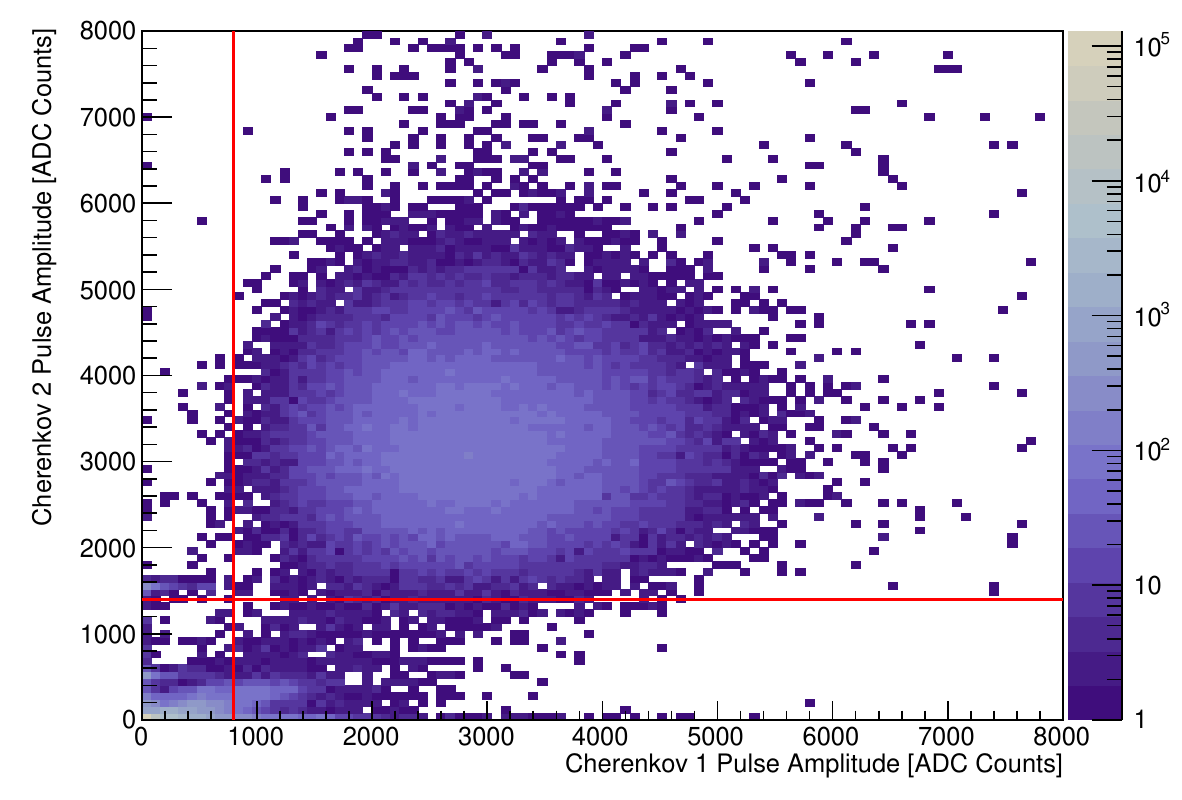} 
\includegraphics[width=0.49\textwidth]{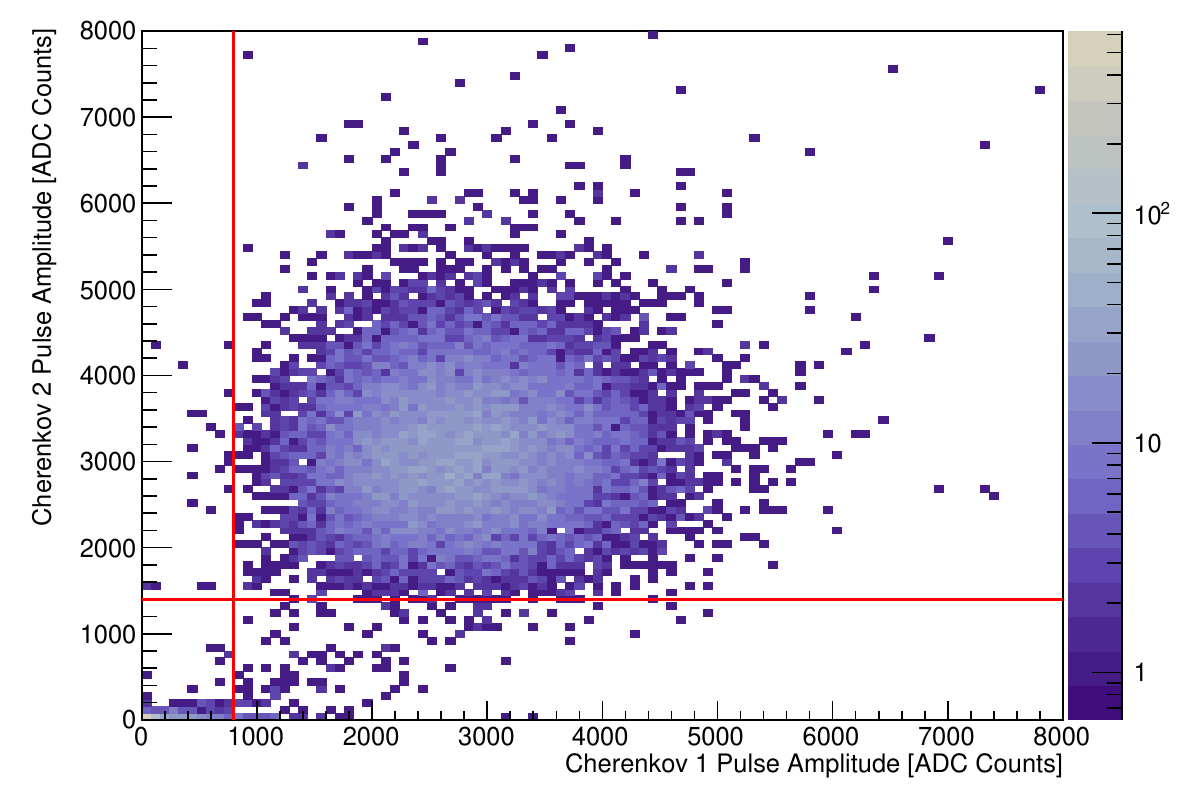} \\ 
\caption{\label{fig:cher_2dhist}Top row: correlation between pulse amplitude and time of arrival in the two \v{C}erenkov detectors. The pulse amplitude is defined as the amplitude of the highest pulse within the waveform, while the time of arrival is the waveform tick (1~Tick = 2~ns) corresponding to the maximum sample of the pulse. In a given event, if the primary particle that triggered the system produces a signal in a \v{C}erenkov detector, the time of arrival for such an event will fall within a narrow range of values (indicated by the vertical red lines in the figures). Electrons are identified by selecting this range for both \v{C}erenkov detectors. Bottom row: correlation of pulse amplitudes between the two \v{C}erenkov detectors. On the left, a 5~GeV mixed (hadrons, muons, and electrons) beam is used. On the right, an electron beam is depicted. As electrons produce a large signal in both detectors, while muons produce a lower signal in C$_1$ and no signal in C$_2$, it is possible to further isolate electrons by requiring that both \v{C}erenkov detectors register a large signal (red lines in the figures). 
}
\end{figure}

\begin{figure}[h]
\centering
\includegraphics[width=.7\textwidth]{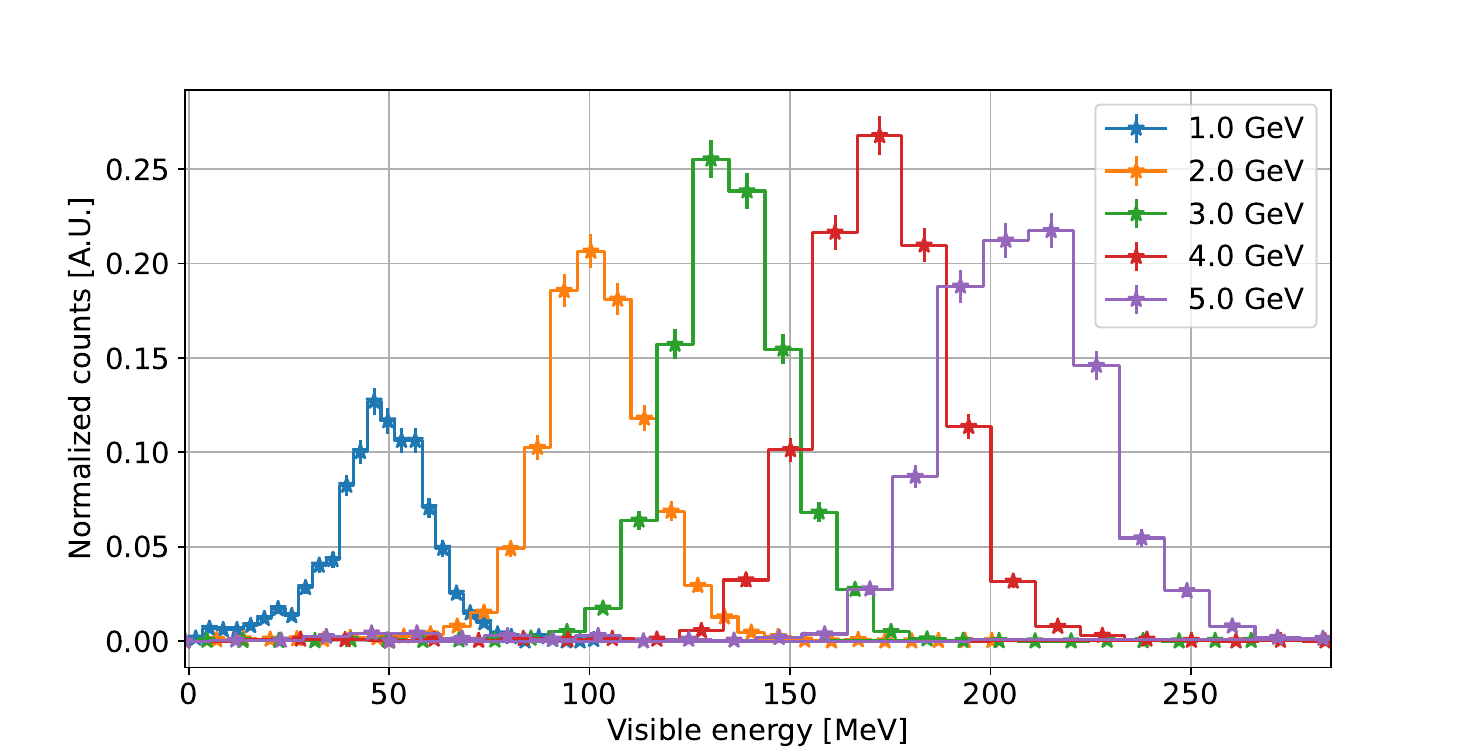}
\caption{\label{fig:EnergySpectraElectrons}Distributions of energy deposits for electron runs in data with energies of 1, 2, 3, 4,~and~5~GeV.}
\end{figure}

\begin{figure}
\centering
\includegraphics[width=.6\textwidth]{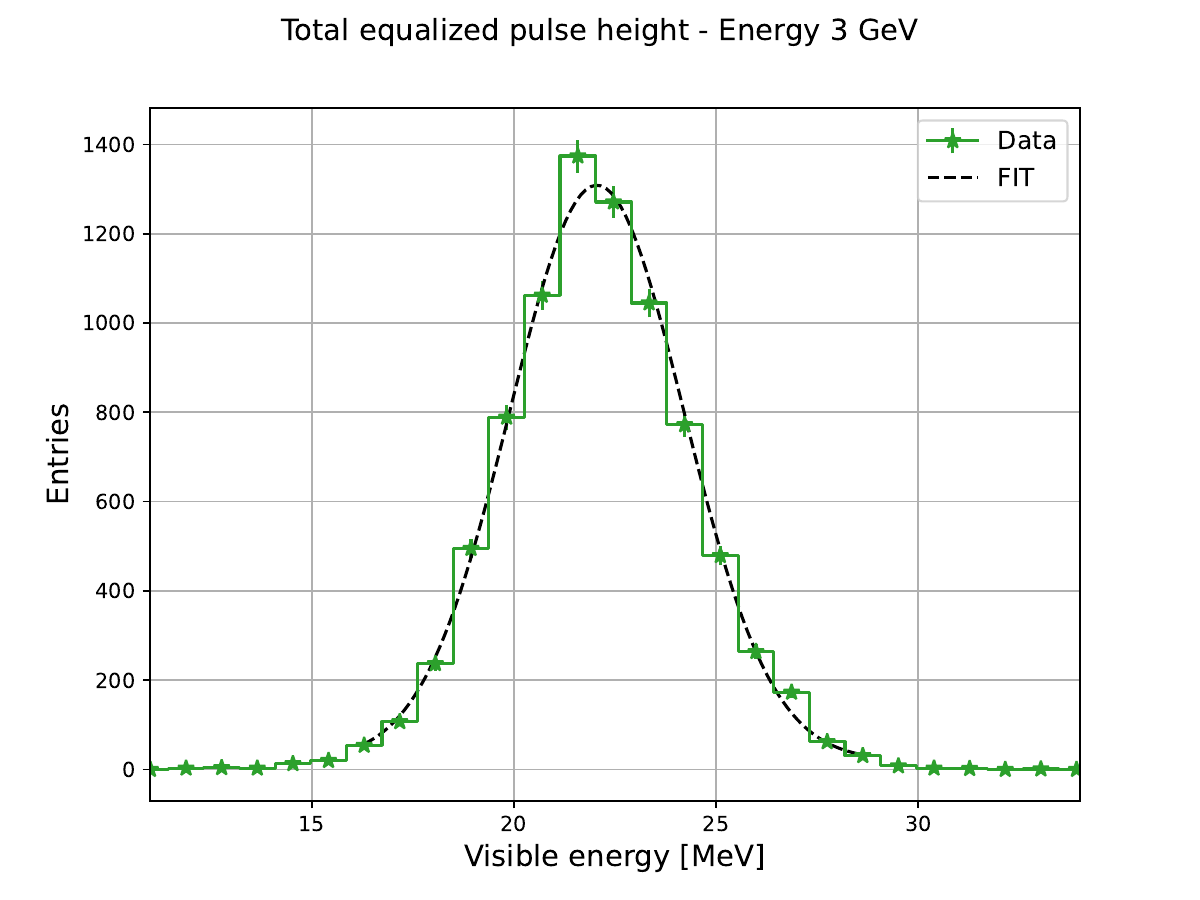}
\caption{\label{fig:Energy3GeV}Example of a Gaussian fit of the total deposited energy with a 3 GeV electron beam.}
\end{figure}

\begin{figure}
\centering
\includegraphics[width=.75\textwidth]{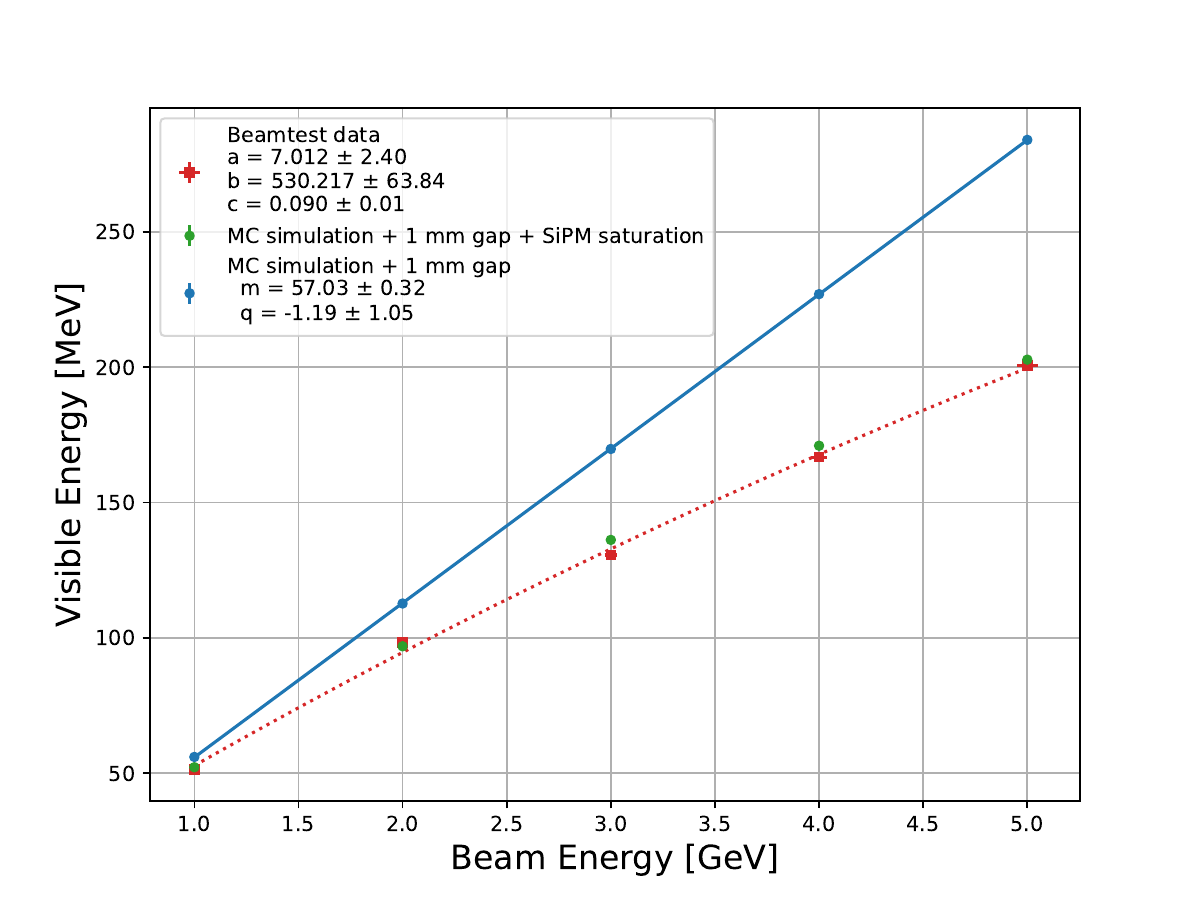}
\caption{\label{fig:Calibration}Visible energy reconstructed in the calorimeter as a function of the beam energy. Beam test data (red squares) are compared with MC simulation including (green dots) and excluding (blue dots) SiPM saturation effects. In both cases, the MC simulation includes a tile gap of 1 mm between the scintillator tiles. The linear fit (blue line) to the MC without saturation effects is also shown. The inclusion of saturation in the MC shifts the reconstructed visible energy to lower values, reproducing the behavior observed in the data. To first order, the linearity is preserved, although saturation effects introduce deviations from linearity. A fit to the data with an exponential function (dotted red line) allows this behavior to be modeled (see text).}
\end{figure}

To ensure that the distribution is not affected by spurious muons or heavier particles, the signals of the \v{C}erenkov detectors were used to select only electron events. To this end, the pulse amplitude recorded by the waveform digitizer for both \v{C}erenkov detectors, together with the corresponding pulse arrival times within the waveforms, was considered. 2D histograms of these quantities are depicted in Fig. \ref{fig:cher_2dhist}. As electrons produce a signal in both \v{C}erenkov detectors, one can isolate them by performing suitable cuts on these quantities, also shown in Fig. \ref{fig:cher_2dhist}.
To ensure full containment of the electromagnetic shower, only electrons whose primary track (projected onto the front face of the calorimeter) was impinging at least 2 Moli\`{e}re radii ($R_M \simeq 1.72$~cm in iron) away from the calorimeter edges were considered. 
The energy deposit spectra (Fig. \ref{fig:EnergySpectraElectrons}) were fitted with a Gaussian function (Fig. \ref{fig:Energy3GeV}).
To evaluate the energy linearity response of the Demonstrator, the $\mu$ fit parameters, representing the mean total visible energy in the calorimeter for different electron energies, are plotted as a function of the beam energy and shown in Fig.~\ref{fig:Calibration}. A comparison with the prediction from the MC simulation is also shown. The MC simulation provides a good description of the data once a gap in the scintillator tiles and saturation effects are included. The effect of SiPM saturation, discussed in Sec.~\ref{sec:single_photoelectron}, is already clearly visible for electrons with energies of 2 GeV. Both tile gaps and saturation reduce the linearity slope with respect to the ideal MC prediction. The MC simulation including only a 1~mm tile gap can be fitted with a first-degree polynomial  with an intercept compatible with zero, as expected for a perfectly linear response. After introducing an average correction for saturation effects in the MC, we are able to reproduce the data. To first order, the Demonstrator response remains linear; however, a linear fit does not fully reproduce the behavior observed in the data. This can be traced back to the exponential nature of the SiPMs saturation discussed in Sec.~\ref{sec:single_photoelectron}. A fit to the data with an empirical exponential function of the form $E_{vis} = a + b \cdot (1 - e^{-c x})$ is able to model the deviation from linearity introduced by saturation. The fit result is shown in Fig.~\ref{fig:Calibration}. Using the information from the fit, we estimate a deviation from pure linearity of about 12\% at a beam energy of 3~GeV.

The energy resolution at different momenta was estimated as the ratio of $\sigma$ and $\mu$ parameters obtained from the Gaussian fits. The resolution as a function of the beam energy was then fitted with the formula $\dfrac{\sigma_E}{E} = \dfrac{s}{\sqrt{E}} \oplus \dfrac{n}{E} \oplus c$ where $s$ is the stochastic term, $n$ the noise term, $c$ the constant term and $\oplus$ indicates the quadratic combination. 

The noise term in the energy-resolution fit is fixed (\(n = 1.08~\mathrm{MeV}\)). It is obtained by summing in quadrature the baseline fluctuations of the individual channels in a $\phi$ range that selects those with non negligible contributions to the total visible energy~\cite{Bramati2026}. The resulting value, initially expressed in ADC counts, is converted into MeV using the general calibration formula.

\begin{figure}
\centering
\includegraphics[width=.75\textwidth]{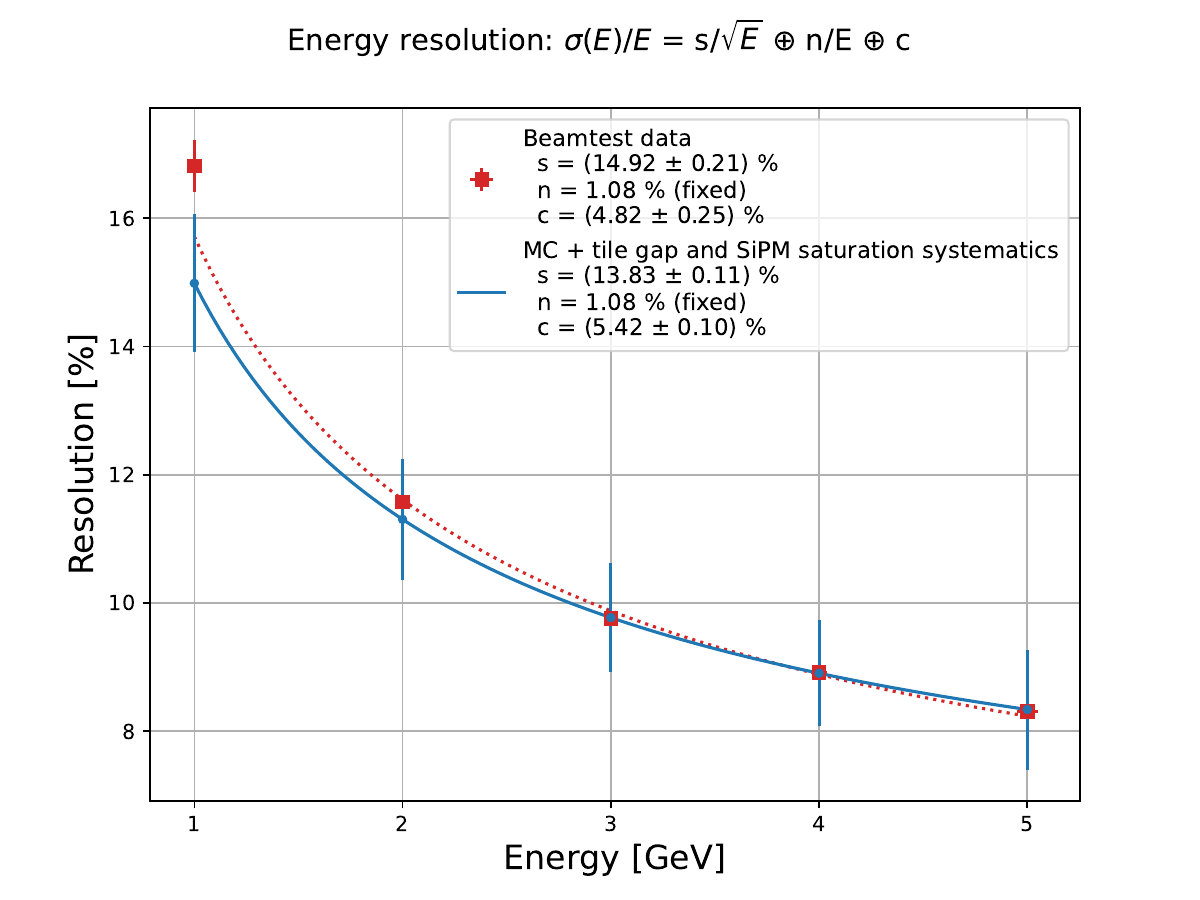}
\caption{\label{fig:EnergyResolution}Energy resolution as a function of the beam energy. Beam test data (red squares) are compared with MC simulation (blue stars). The fit results are shown by the dashed red line for the data and the solid blue line for the MC simulation; the fit parameters are reported in the plot inset. The uncertainty associated with the MC simulation represents the systematic uncertainty induced by the gaps between the scintillator tiles and the SiPM saturation effect. See text for details.}
\end{figure}

The results are presented in Fig.~\ref{fig:EnergyResolution}, where a comparison with the prediction from the MC simulation is also shown.
The achieved energy resolution is $\sim 17$\% at 1 GeV and 10\% at 3 GeV.

Given the difficulty in reproducing the resolution observed in the data when assuming average values for the tile gap size and the SiPM saturation effect, the MC result shown in Fig.~\ref{fig:EnergyResolution} is obtained by averaging the MC predictions with and without SiPM saturation, and by assigning a corresponding systematic uncertainty. A 1 mm gap between the scintillator tiles, as assumed for the linearity plot shown in Fig.~\ref{fig:Calibration}, is consistently adopted.

The 1~mm estimate for the tiles gap central value is based on a quantitative study of the tile thickness in the $\phi$ direction, measured before and after the application of the TiO$_2$ coating. 
The average difference between painted and not-painted tiles is approximately 400~$\mu$m. This estimate represents a lower bound on the actual gap between neighbouring $\phi$ sectors. Because the three (or four) tiles are aligned by the WLS fibers, the largest displacement tends to determine the effective thickness of the assembly. For the outermost radial layer (type 7), we observe an average value of approximately 0.6~mm.

In practice, the actual gaps are larger because of the combined effect of irregularities introduced during the final assembly, as can be seen from visual inspection. The value of 1~mm adopted in the simulation therefore appears to be well supported by these experimental observations.

The standard deviation of the painted-tile measurements is larger than that of the unpainted tiles. This is due not only to the somewhat smaller available samples of painted-tiles used in the measurement, but also to variations in the coating thickness, often caused by the formation of paint droplets during drying under gravity. This issue could be avoided by using chemical etching, which was the solution originally envisaged.

The systematic uncertainty is estimated from the differences between the MC results obtained assuming two conservative values for the tile gap, namely 0.5 mm and 1.5 mm, as well as between the MC results obtained with and without SiPM saturation, where the saturation is computed as discussed in Sec.~\ref{sec:single_photoelectron}. The two contributions are summed in quadrature to obtain the final uncertainty estimate on the MC predictions shown with error bars in the plot.
The values of the fit parameters for the stochastic and the constant term for data are $s=(14.92 \pm 0.21)\%$ and $c=(4.82 \pm 0.25)\%$ respectively, while for the MC simulation the obtained values are $s=(13.83 \pm 0.11)\%$ and $c=(5.42 \pm 0.10)\%$.

\subsection{Energy deposition pattern}
\label{sec:ene_dep_pattern}

A key feature of the ENUBET calorimeter is its capability for electron/hadron identification, which is essential for distinguishing positrons from charged pions in the few-GeV energy range. This separation is enabled by the longitudinal segmentation of the calorimeter. In this section, we report the total energy deposition produced by $e^-$, $\pi^-$, and $\mu^-$ , together with the shower-profile results for $e^-$ and $\pi^-$. The data are equalized following the procedure discussed in Sec.~\ref{sec:equalization}, where a factor $MPV_{MeV} \simeq 6.12$ MeV is used for the conversion to MeV. A validation of the calorimeter simulation is also presented by comparing the data with the simulation of the prototype. The MC includes a 1 mm gap between tiles together with the average SiPM saturation effects, as discussed in the previous section. 
An energy-scale correction (see Tab.~\ref{tab:ene_scale}) is required to account for residual data–MC differences after including the average effects of SiPM saturation and tile gaps in the simulation, as well as for limitations in the modeling of hadronic showers in the $\pi^-$ sample. The correction factors are derived from the total energy-deposition distribution and are then applied to all other distributions presented in this section.

Discrepancies of approximately 10--20\% for the pion energy are not entirely unexpected and have already been reported in the literature. For example, the CALICE Collaboration observed discrepancies of a similar magnitude in the reconstructed pion energy between 2 and 10~GeV, consistently across several GEANT4 physics lists (Fig. 7 in ~\cite{CALICE2015_HadronicModels_SiW}).

\begin{table}
\begin{center}
\begin{tabular}{ c c c c }
\hline
 & $e^{-}$ & $\pi^-$ & $\mu^-$ \\
\hline
 $\epsilon_{scale}$ & 0.97 & 0.82 & 1 \\
\hline
\end{tabular}
\caption{Energy scale for the energy of electrons, pions and muons in the MC simulation.}
\label{tab:ene_scale}
\end{center}
\end{table}

\begin{figure}[h!]
\centering
\includegraphics[width=\textwidth]{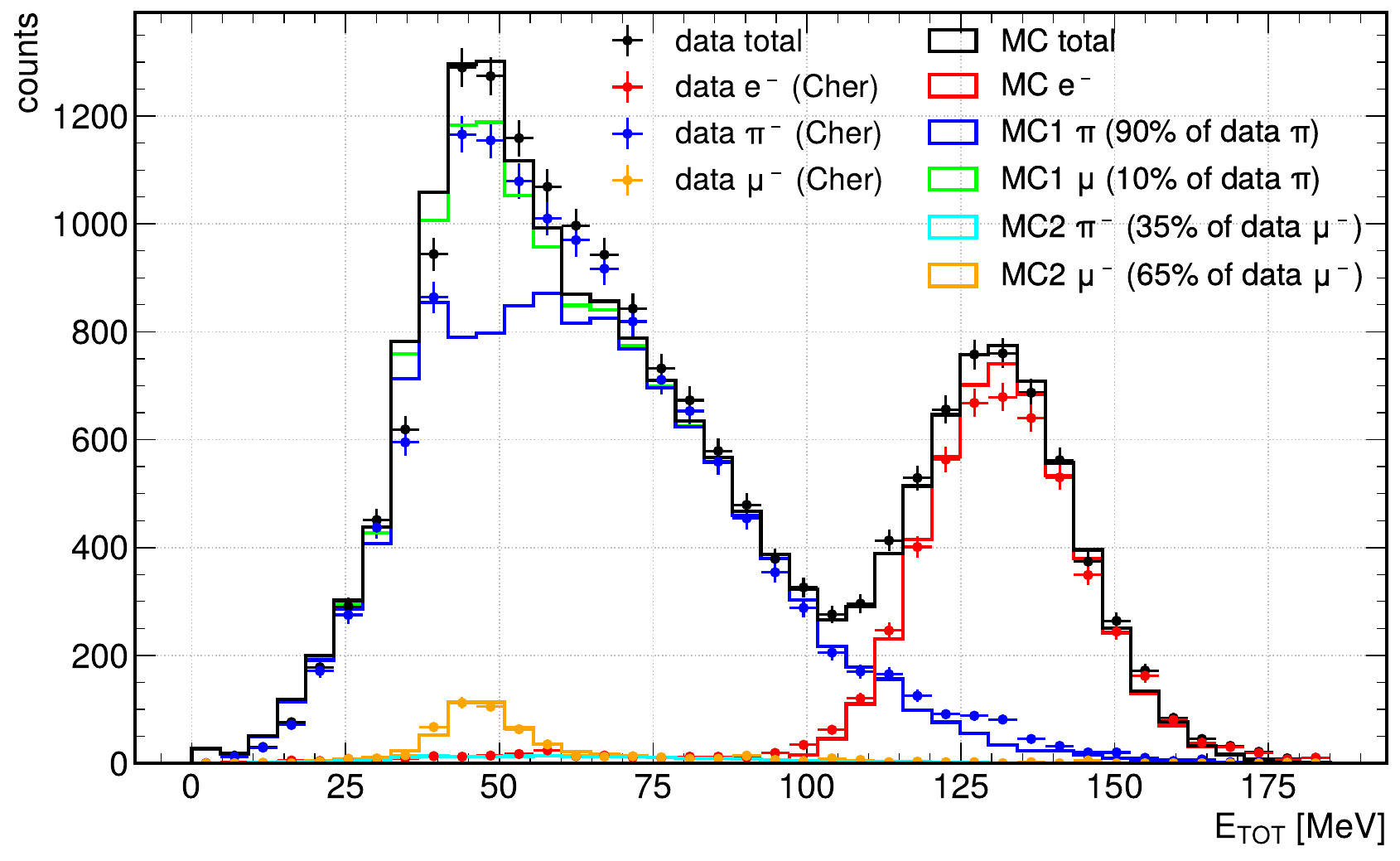} 
\caption{Distribution of the total energy deposited in the scintillator for electrons, pions, and muons in a 3 GeV hadron run. The different particle species are selected using \v{C}erenkov cuts. The hadronic component ($\pi$ data) is dominated by pions (90\%, MC1 $\pi$ template) with a small muon contamination (10\%, MC1 $\mu$ template); while the muon component ($\mu$ data) consists of 65\% true muons (MC2 $\mu$ template) and 35\% pion contamination (MC2 $\pi$ template). In the hadronic component, the MC1 $\pi$ and MC1 $\mu$ templates are stacked; the same applies to the muon component, where the MC2 $\mu$ and MC2 $\pi$ templates are stacked. The black dots and the black line represent the sum of all particle contributions in data and MC, respectively.}
\label{fig:tot_energy_by_particle}
\end{figure}

\begin{figure}[h!]
\centering
\includegraphics[width=0.49\textwidth]{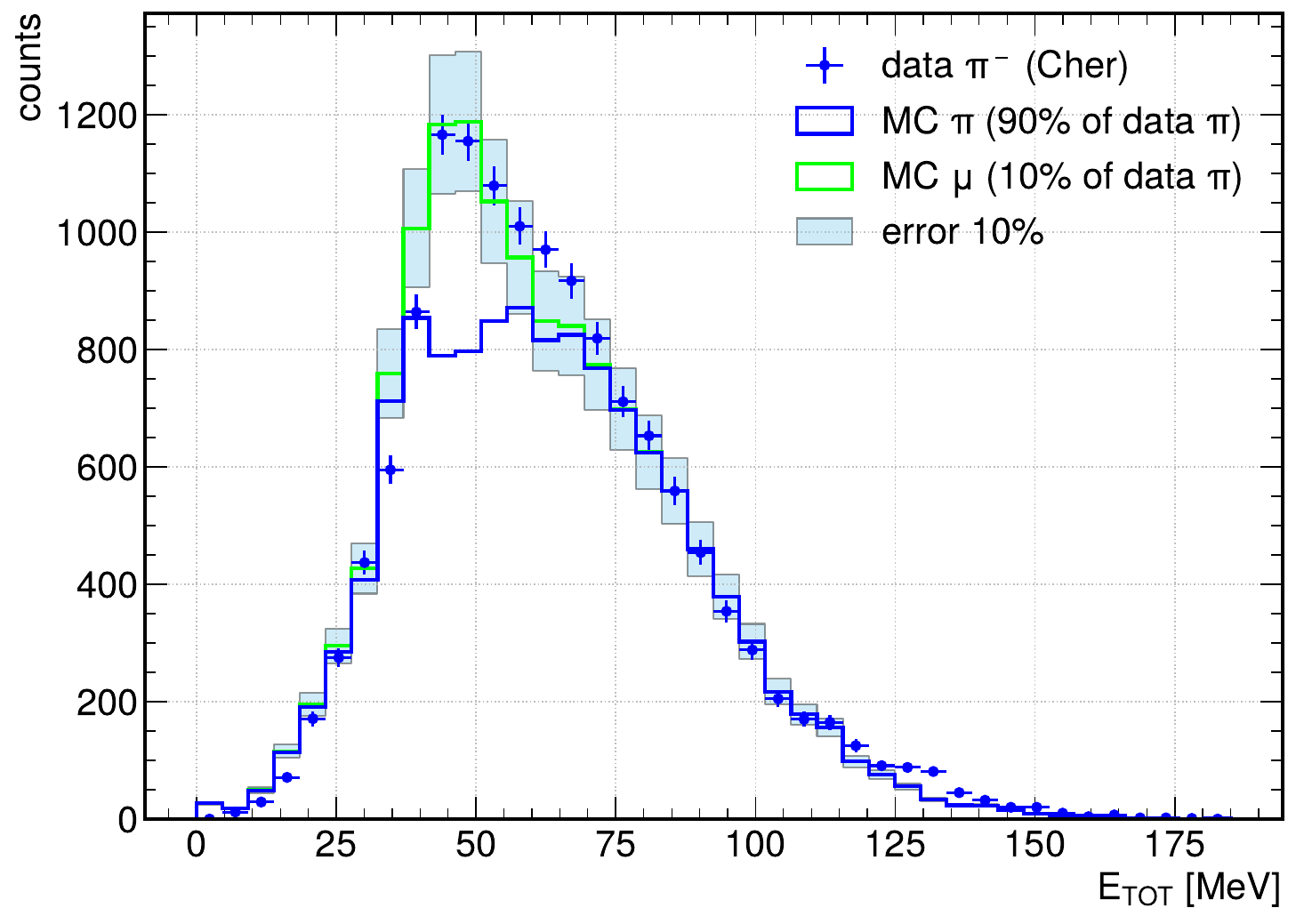} 
\includegraphics[width=0.49\textwidth]{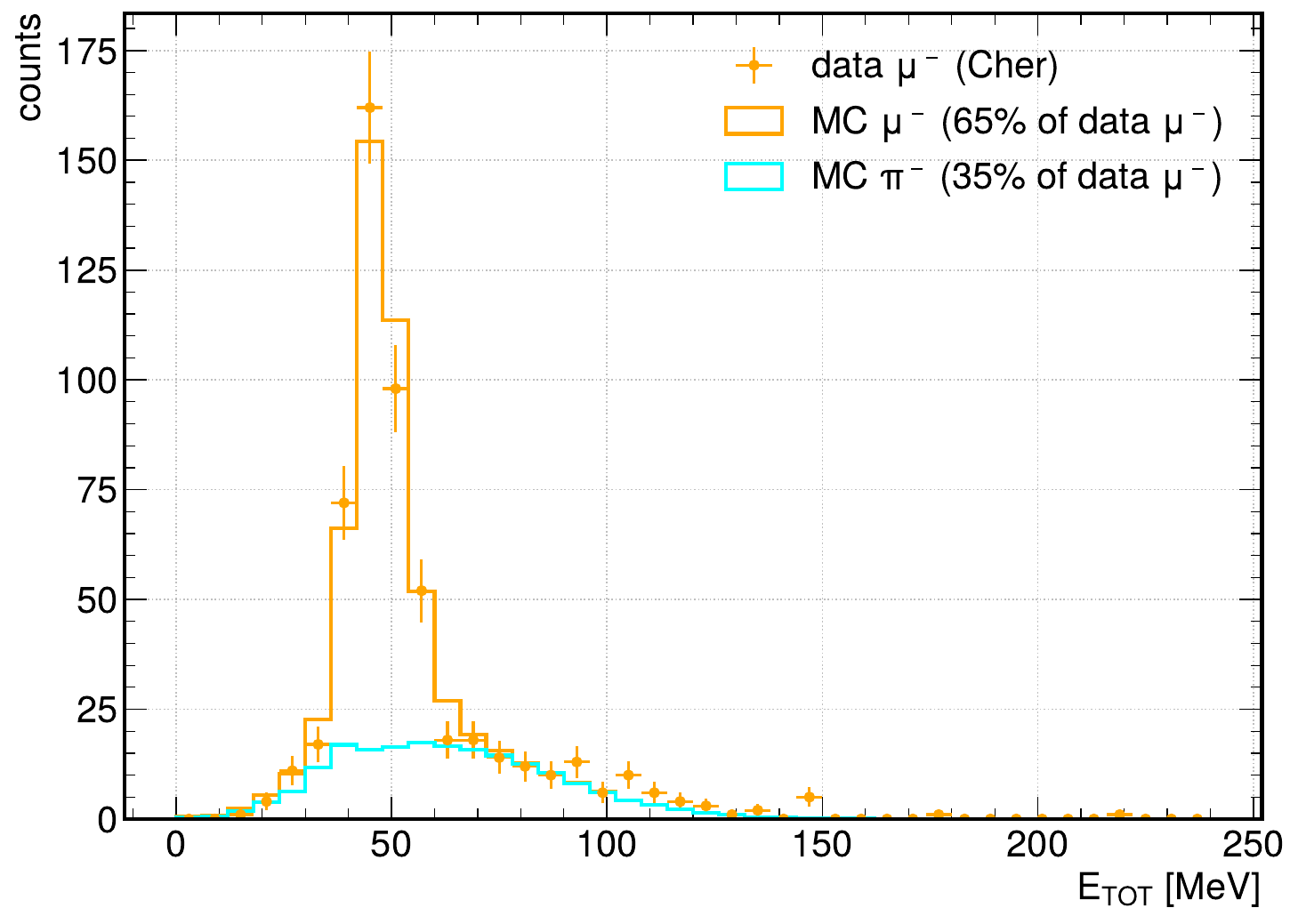} \\
\caption{The total energy distributions from Fig.~\ref{fig:tot_energy_by_particle} are shown separately for hadrons (left) and muons (right). For the hadronic component, a 10\% uncertainty band is added to the MC prediction. In the hadronic component, the MC1 $\pi$ and MC1 $\mu$ templates are stacked; the same applies to the muon component, where the MC2 $\mu$ and MC2 $\pi$ templates are stacked.}
\label{fig:pi_only_mu_only}
\end{figure}

The total energy response of the detector, summed over all channels, was computed and compared to the MC prediction for 3 GeV $e^-$, $\pi^-$, and $\mu^-$. The corresponding distributions are shown in Fig.~\ref{fig:tot_energy_by_particle}. The individual $\pi^-$ and $\mu^-$ contributions are shown separately in Fig.~\ref{fig:pi_only_mu_only}. For the $\pi^-$ sample, a 10\% uncertainty band is included in the MC prediction. 
Both the $\pi^-$ and $\mu^-$ distributions are affected by contamination from misidentified particles of the other type, arising from the limited purity of the \v{C}erenkov-based particle selection.
In order to reproduce the data, a 10\% contamination of muons has to be added to the pion MC sample for the hadronic component, while for the muon one a 35\% pion contamination is required. On the other hand the electron component shows a pretty high purity and no additional contributions are needed in the MC to account for the data.

\begin{figure}[h!]
\centering
\includegraphics[width=0.49\textwidth]{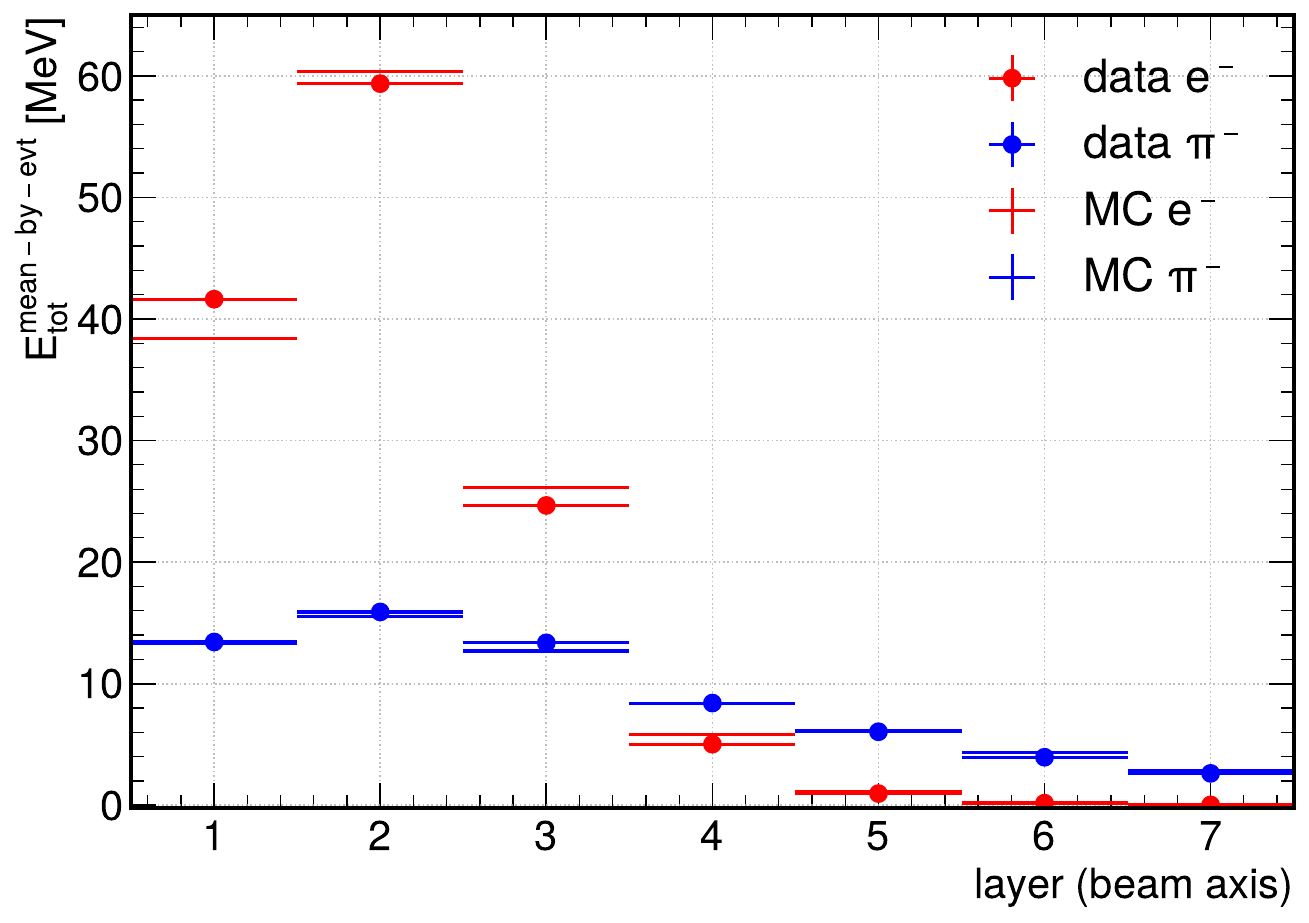} 
\includegraphics[width=0.49\textwidth]{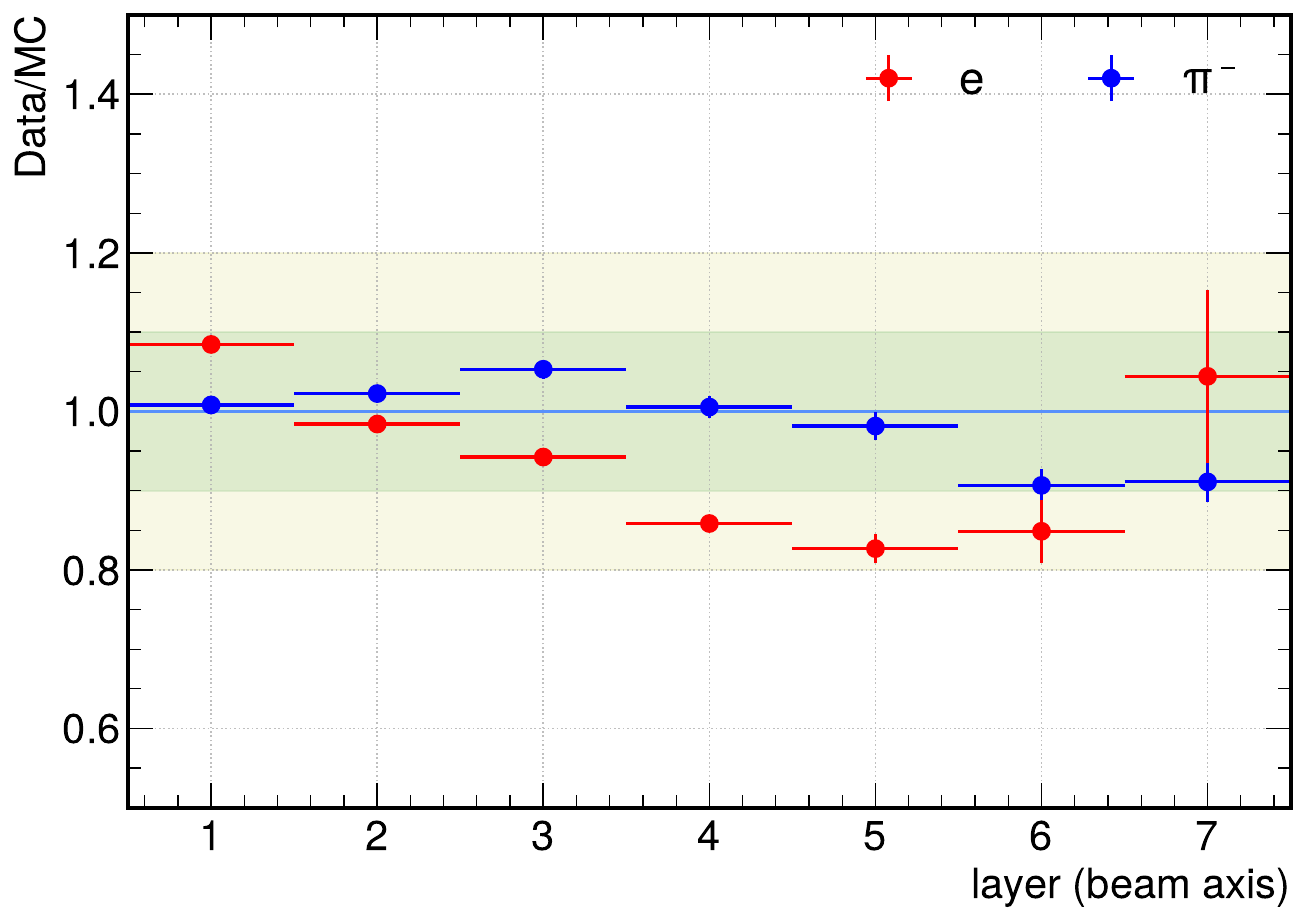} \\ 
\caption{(Left) Average energy deposited in the scintillator as a function of the shower depth for 3 GeV pions and electrons. The depth is expressed as the number of channel layers along the beam axis. (Right) Energy ratio between data and MC.}
\label{fig:shower_profile}
\end{figure}

Fig.~\ref{fig:shower_profile}-left shows the average total energy deposited per event in the channels of each layer along the beam axis for 3 GeV $e^-$ and $\pi^-$, while Fig.~\ref{fig:shower_profile}-right shows the corresponding data/MC ratio. The distributions in Fig.~\ref{fig:shower_profile}-left provide a proxy for the longitudinal shower profile in the calorimeter. Layer 1 is the front layer, facing the beam head-on. Layer 7 corresponds to a calorimeter depth of $7\times4~X_0 = 28~X_0$ and $7\times0.42~\lambda_0 = 2.94~\lambda_0$.

The particle type is selected using \v{C}erenkov counters, and for the $e^-$ run, a fiducial area is defined on the calorimeter front face as described in Sec.~\ref{sec:enereso}. Due to a failure of one of the silicon trackers during the acquisition of the $\pi^-$ run, the definition of the fiducial area in this case relies on the selection of MIP-like particles in the central channels of the front layer.\footnote{Because of time constraints during the beam test, it was not possible to acquire additional $\pi^-$ runs with fully operational silicon trackers.}

From the ratio plot in Fig.~\ref{fig:shower_profile}-right, we observe that the MC reproduces the data within 10-20\% for both $e^-$ and $\pi^-$.
Limitations in the detector description within the Geant4 simulation, in particular the scintillation photon production and transport, and channel-by-channel variations in the SiPM-fiber optical coupling, not included in the simulation, contribute to the observed discrepancies for both $e^-$ and $\pi^-$. For $\pi^-$, an additional contribution to the discrepancy originates from known limitations in hadronic-shower modeling~\cite{Kiryunin:2007zz}. The global scaling factors in Tab.~\ref{tab:ene_scale} provide good agreement between data and MC for the total deposited energy and for the normalization of the shower profiles, but are less effective in reproducing the layer-by-layer dependence of the shower profile, as they cannot account for variations in the detector response between layers.

\begin{figure}[h!]
\centering
\includegraphics[width=0.45\textwidth]{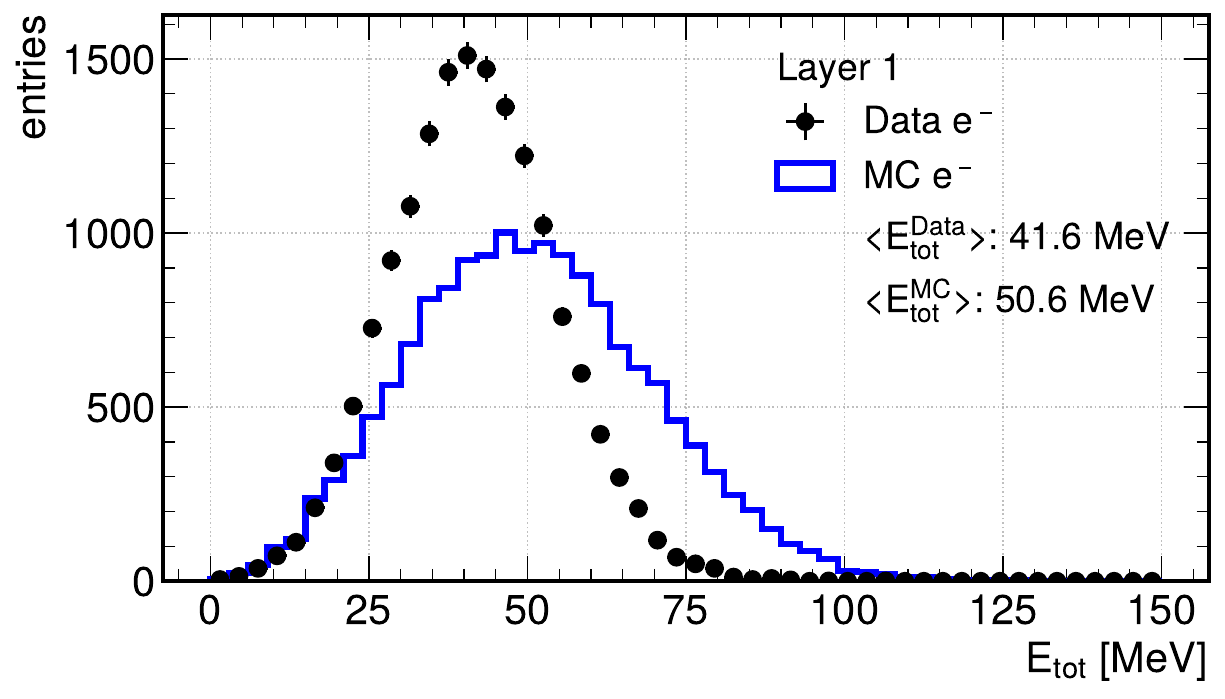} 
\includegraphics[width=0.45\textwidth]{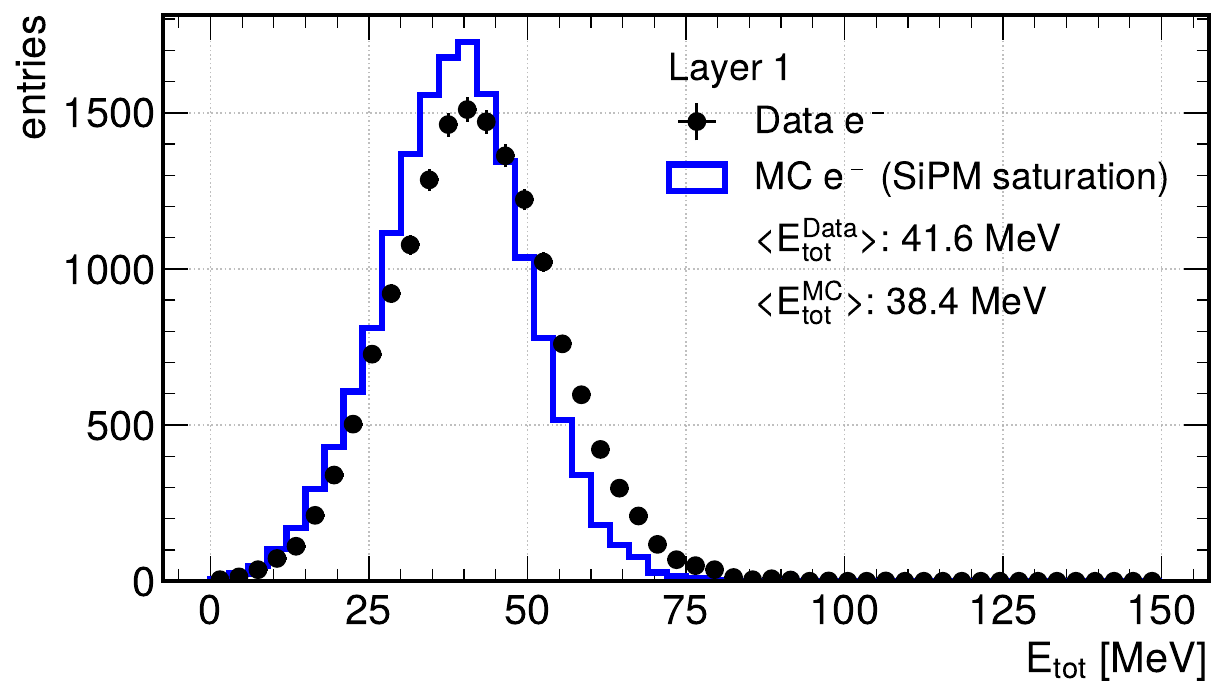} \\
\includegraphics[width=0.45\textwidth]{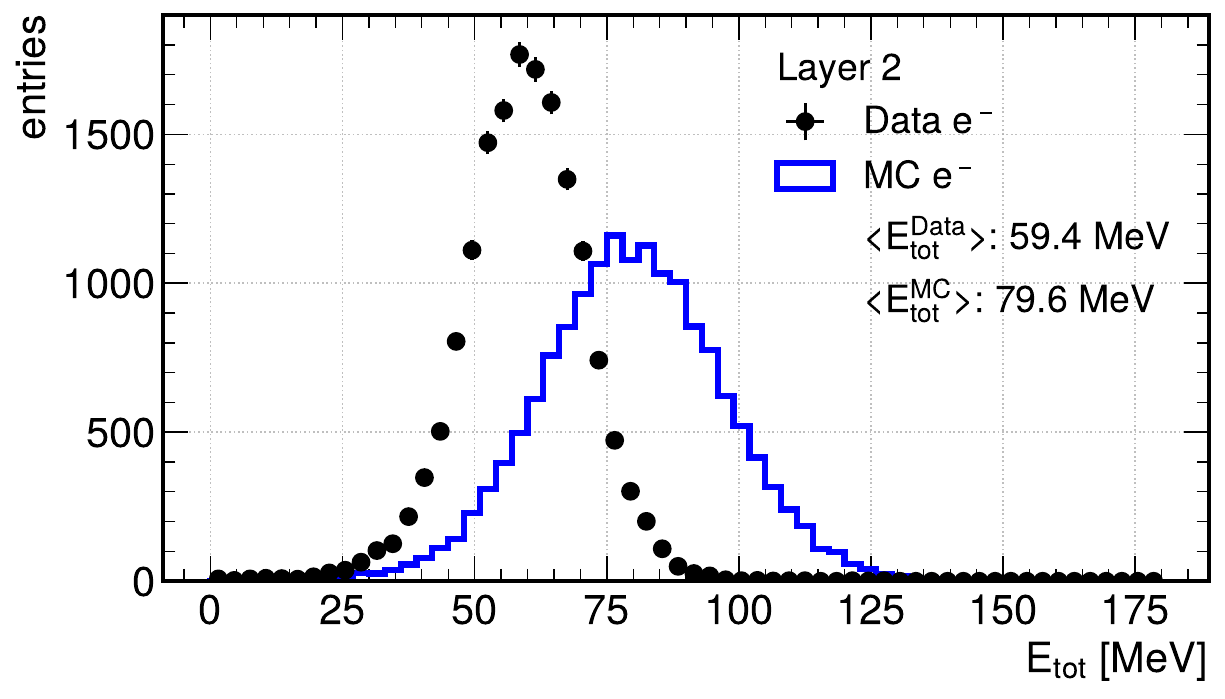} 
\includegraphics[width=0.45\textwidth]{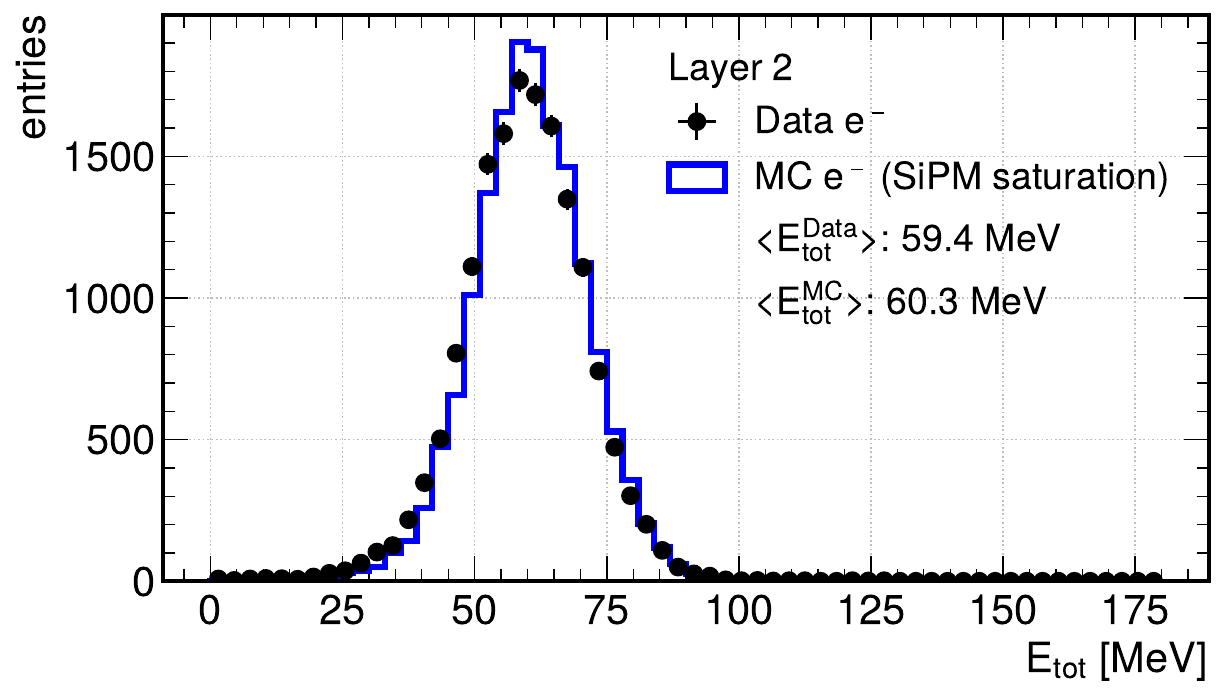} 
\caption{Distributions of the total energy deposited by 3 GeV electrons in the first (layer 1) and second (layer 2) calorimeter layers. The data are compared to MC simulations without SiPM saturation (left) and with saturation effects included (right).}
\label{fig:etot_layer2_sipm_saturation}
\end{figure}

In principle, a better agreement is expected for $e^-$. However, part of the discrepancies can be attributed to the channel-by-channel variations in SiPM saturation effects, which can lead to the differences observed layer by layer. Such channel-dependent variations are expected, as the differences in optical coupling between the WLS fibers and the SiPMs lead to channel-dependent light yields. The effect of SiPM saturation was studied for the different calorimetric layers. Figure~\ref{fig:etot_layer2_sipm_saturation} shows the distributions of the total energy deposited by $e^-$ in layers 1 and 2, where the energy deposition is largest, comparing data with MC simulations with and without SiPM saturation effects. As shown in the figure, the inclusion of saturation results in improved agreement between data and MC. However, this correction is layer dependent due to saturation non-uniformities.

\section{Conclusions}

The Demonstrator represents a full-size prototype of the ENUBET instrumented tunnel and has provided a robust experimental validation of lepton identification in monitored neutrino beams.

The experimental campaign carried out in 2022–24 has demonstrated that the technology of segmented iron–scintillator calorimeters read out by WLS fibers and SiPMs performs reliably over the entire energy range relevant for ENUBET. The component production and installation procedures described in Sec.~\ref{sec:assembly} are fully scalable to the final experiment. Compared to previous prototypes, the adopted installation strategy resulted in a significant improvement in light yield. Residual non-uniformities associated with the deposition of the diffusive layer and optical cement, as well as with tile positioning tolerances, have been identified, providing clear guidance for further optimization.

The achieved energy resolution ($\sim 17$\% at 1 GeV and 10\% at 3 GeV) fully satisfies the requirements for positron identification. In addition, the measured MIP and hadron responses are in agreement with expectations and validate the ENUBET full simulation and its particle-identification performance \cite{ENUBET:2023hgu}.

Nevertheless, the increased light yield highlights some limitations of the current prototype. In particular, SiPM saturation effects are already observed for energy measurements within the range of interest for ENUBET. Although this does not compromise the target energy resolution of the calorimeter, it could limit particle identification performance based on energy deposition patterns. A solution to mitigate SiPM saturation will be identified and implemented in future detector developments.
In the final detector, better control of the optical coupling and of the tile--fiber assembly, together with the use of SiPMs containing a larger number of cells and therefore providing a wider dynamic range, will substantially reduce both the channel-to-channel variations and the associated saturation effects.

Overall, the Demonstrator confirms the results of the end-to-end ENUBET simulation and demonstrates that the proposed detector technology can achieve the monitoring sensitivity required to reach a 1\% precision in the determination of the neutrino flux for next-generation neutrino cross-section experiments \cite{Acerbi:2025wzo}.

\acknowledgments{
This project has received funding from the European Union's Horizon 2020 Research and Innovation programme under Grant Agreement no. 681647 and the Italian Ministry for Education and Research (MIUR, bando FARE, project: NUTECH and PRIN2022 project: GIGANU). It is also supported by the Agence Nationale de la Recherche (ANR, France) through the PIMENT project (ANR-21-CE31-0027), by the Ministry of Science Education and Youth of Republic of Croatia Grant No.PK.1.1.10.0002, and by Swiss National Science Foundation (SNSF) and Croatian Science Foundation (HRZZ) under grant MAPS IZ11Z0\_230193.
This project has received funding from the European Union's Horizon Europe research and innovation programme under grant agreement no. 101057511 (project: EuroLabs).

We would like to thank the Mechanical workshop of INFN Padova and LNL for the support during the demonstrator construction and all the people who participated in the assembly phase of the detector:  M.L.~Fasolo, F.~Longhin, L.~Ranjbarlari, A. Olivito, and L.~Ramina. We are grateful to CERN BE-EA for invaluable support during the beamtest data taking and, in particular, to D. Banerjee, J. Bernhard, L. Gatignon, and A.E. Rahmoun.
}

\bibliographystyle{JHEP}
\bibliography{bibliography.bib}
\end{document}